\documentclass[useAMS,usegraphicx]{mn2e}

\newcommand {\apgt} {\ {\raise-.5ex\hbox{$\buildrel>\over\sim$}}\ }
\def\simlt{\mathrel{\hbox{\rlap{\hbox{\lower4pt\hbox{$\sim$}}}\hbox{$<$}}}}
\def\simgt{\mathrel{\hbox{\rlap{\hbox{\lower4pt\hbox{$\sim$}}}\hbox{$>$}}}}

\def\ale{\mathrel{\hbox{\rlap{\hbox{\lower4pt\hbox{$\sim$}}}\hbox{$<$}}}}
\def\age{\mathrel{\hbox{\rlap{\hbox{\lower4pt\hbox{$\sim$}}}\hbox{$>$}}}}

\def\nodata{---}

\title[Nearby Supernova Rates from LOSS: III. Rates]{Nearby Supernova
  Rates from the Lick Observatory Supernova Search.  III. The
  Rate-Size Relation, and the Rates as a Function of Galaxy Hubble
  Type and Colour}

\author[Li et al.]{Weidong Li$^{1}$\thanks{Email: wli@astro.berkeley.edu}, 
Ryan Chornock$^{1,2}$, 
Jesse Leaman$^{1,3}$, Alexei V. Filippenko$^{1}$,
\newauthor
Dovi Poznanski$^{1,4,5}$, Xiaofeng Wang$^{1,6,7}$, Mohan
Ganeshalingam$^{1}$, and Filippo Mannucci$^{8}$\\ 
$^{1}$Department of Astronomy, University of California,
Berkeley, CA 94720-3411, USA \\
$^{2}$Harvard-Smithsonian Center for Astrophysics, 60
Garden Street, Cambridge, MA 02138, USA\\
$^{3}$NASA Ames Research Center, Mountain View, CA 94043, USA \\
$^{4}$Computational Cosmology Center, Lawrence Berkeley National
Laboratory, 1 Cyclotron Road, Berkeley, CA 94720, USA \\
$^{5}$Einstein Fellow \\
$^{6}${Department of Physics, Texas A\&M University,
College Station, TX 77843-4242, USA} \\
$^{7}$Physics Department and Tsinghua Center for Astrophysics (THCA),
Tsinghua University, Beijing, 100084, China \\ 
$^{8}$INAF -- Osservatorio Astrofisico di Arcetri, Largo E. Fermi 5, I-50125, 
Firenze, Italy
}

\begin{document}
\maketitle

\begin{abstract}

This is the third paper of a series in which we present new
measurements of the observed rates of supernovae (SNe) in the local
Universe, determined from the Lick Observatory Supernova Search
(LOSS).  We have considered a sample of $\sim 1000$ SNe and used an
optimal subsample of 726 SNe (274 SNe~Ia, 116 SNe~Ibc, and 324 SNe~II)
to determine our rates.  We study the trend of the rates as a function
of a few quantities available for our galaxy sample, such as
luminosity in the $B$ and $K$ bands, stellar mass, and morphological
class. We discuss different choices (SN samples, input SN luminosity
functions, inclination correction factors) and their effect on the
rates and their uncertainties.  A comparison between our SN rates and
the published measurements shows that they are consistent with each
other to within the uncertainties when the rate calculations are done in
the same manner. Nevertheless, our data demonstrate that the rates
cannot be adequately described by a single parameter using either
galaxy Hubble types or $B - K$ colours.  A secondary parameter in
galaxy ``size," expressed by luminosity or stellar mass, is needed to
adequately describe the rates in the rate-size relation: the galaxies
of smaller sizes have higher SN rates per unit mass or per unit
luminosity. The trends of the SN rates in galaxies of different Hubble
types and colours are discussed. We examine possible causes for the
rate-size relation. Physically, such a relation for the core-collapse
SNe is probably linked to the correlation between the specific
star-formation rate and the galaxy sizes, but it is not clear whether
the same link can be established for SNe~Ia. We discuss the
two-component (``tardy" and ``prompt") model for SN~Ia rates, and find
that the SN~Ia rates in young stellar populations might have a strong
correlation with the core-collapse SN rates.  We derive volumetric
rates for the different SN types (e.g., for SNe~Ia, a rate of $(0.301
\pm 0.062) \times 10^{-4}$ SN Mpc$^{-3}$ yr$^{-1}$ at redshift 0) and
compare them to the measurements at different redshifts.  Finally, we
estimate the SN rate for the Milky Way Galaxy to be $2.84 \pm 0.60$ SNe
per century (with a systematic uncertainty of a factor of $\sim 2$),
consistent with published SN rates based on several different
techniques.

\end{abstract}


\section{Introduction}

The Lick Observatory Supernova Search (LOSS; Li et al. 2000;
Filippenko et al. 2001; Filippenko et al. 2011) has been the most
successful nearby supernova search engine in the past 12 years.
During the period from March 1998 through the end of 2008 (on which
the data from this study are based), LOSS found 732 SNe, easily
exceeding any other searches for nearby SNe and accounting for more
than 40\% of all SNe with redshift $z < 0.05$ reported to the Central
Bureau for Astronomical Telegrams. It found an even larger fraction of
the reported young SNe, discovered close to or before maximum
brightness.  One major goal of LOSS is to improve our understanding of
the statistics of SNe --- in particular, the SN rates in galaxies of
different types and colours. Here, in Paper III of this series on LOSS
SN rates in the local Universe, the goal is to put all of the
ingredients together to derive the SN rates.

In this section, we first summarise what we have learned from Paper I
(Leaman et al. 2011) and Paper II (Li et al. 2011) {\it that is
  relevant to the rate calculations}, and then discuss the details of
the control-time and rate calculations.  The rest of the paper is
organised as follows. Section 2 discusses an observed correlation
between the normalised SN rates and the host-galaxy
sizes\footnote{Hereafter, ``the galaxy size" refers to the magnitude
  of both the luminosity and stellar mass, unless otherwise specified,
  because the mass is directly calculated from the luminosity, with a
  small dependence on $B - K$ colour (Paper I; Mannucci et al. 2005).}
(the ``rate-size relation''), while \S 3 reports the SN rates for a
fiducial galaxy size. Section 4 discusses comparisons with the
published SN rates, the possible causes of the rate-size relation, the
two-component model for SN~Ia rates, and the volumetric rates; it also
provides an estimate of the SN rate in the Milky Way Galaxy. Our
conclusions and possible future improvements are summarised in \S
5. In the Appendix, we offer additional discussion of the rate-size
relation, including its discovery and an alternative description using
the rate-colour relation. We adopt a Hubble constant of $H_0 = 73$ km
s$^{-1}$ Mpc$^{-1}$ (Spergel et al.  2007) in our study, consistent
with the recent direct determination based on Cepheid variables and
SNe~Ia by Riess et al. (2009).

\subsection{Summary of Papers I and II}

Paper I of this series discusses the construction of the galaxy and SN
samples. Two galaxy samples are heavily used here, in Paper III: the
``full" sample with a total of 14,882 galaxies, and the ``optimal"
sample with a total of 10,121 galaxies.  The ``optimal'' sample excludes
all of the small (major axis $< 1\arcmin$) E and S0 galaxies, as well
as highly inclined ($i > 75^\circ$) spirals, in the ``full'' sample to
avoid the uncertainties in the detection efficiencies and inclination
correction factors (see Paper I for details).  Four SN subsamples (out
of the 7 discussed in Paper I) are used throughout this paper,

\begin{enumerate}

\item{The ``full" SN sample with a total of 929 SNe, which include all
  of the SNe that occurred in the ``full" galaxy sample. }

\item{The ``full-optimal" SN sample with a total of 726 SNe, which are
  all of the SNe that occurred in the ``optimal" galaxy sample.}

\item{The ``season" SN sample with a total of 656 SNe, which include
  all of the SNe discovered ``in season''\footnote{An ``in-season" SN
    is one that explodes during the active monitoring period of its
    host galaxy. The active monitoring period refers to the time when
    the galaxy emerges from being too close to the Sun in the sky to
    the time when it once again becomes unobservable, a period during
    which the galaxy is actively monitored in our survey with a short
    observation interval. In other words, a SN discovered in the first
    image of a galaxy after a long break when the galaxy was too close
    to the Sun was not counted as an ``in-season SN.'' See Paper I for
    further discussion.} and that occurred in the ``full" galaxy
  sample.}

\item{The ``season-optimal" sample with a total of 499 SNe, which are
  all of the SNe discovered ``in season" and that occurred in the
  ``optimal" galaxy sample.}

\end{enumerate}

Paper I also shows that for each individual epoch of imaging in our
database, the limiting magnitude can be calculated from several
parameters (flux ratio, seeing conditions, and sky background) recorded
in the log files, to a precision of 0.2--0.3 mag. The detection
efficiency (DE) for SNe with different significance is also determined
through Monte Carlo simulations, and reaches a limit of about 90\%
because some of the SNe are missed near the centres of galaxies.

Paper II discusses the construction of a complete SN sample.  A total
of 175 SNe are selected from the ``season" SN sample with a cutoff
distance of 80~Mpc for SNe~Ia and 60~Mpc for SNe~Ibc and
II. Photometry is collected for every SN to derive the light-curve
shape and peak absolute magnitude, and the incompleteness of each SN
is studied and corrected.\footnote{The completeness of a SN is defined
  as the ratio between the total control time for the SN and the total
  season time. In other words, for a SN that in our survey has 100\%
  completeness, we should have discovered all such SNe during our
  monitoring peroid.} The peak absolute magnitudes are corrected only
for the Galactic extinction in the direction of each SN.  Because of
this, we do not need to consider the host-galaxy extinction (which is
poorly known) toward each SN, as it is naturally included in these
``pseudo-observed" luminosity functions (LFs). The LFs also show
significant dependence on the host-galaxy Hubble types. To alleviate
the effect of small-number statistics, the LFs are constructed in two
broad Hubble-type bins: E--Sa and Sb--Irr for SNe~Ia, and S0--Sbc and
Sc--Irr for SNe~Ibc and II.

We note that for the control-time calculations, it is important to
match the different SN subsamples with the proper subsets of SNe in
the LFs, as long as there is a sufficient number of objects in the
LFs. As discussed in Paper II, the LFs do not change significantly
regardless of whether the SNe discovered in small (major axis $<
1\arcmin$) early-type (E/S0) galaxies are considered, so the ``full,"
``full-nosmall," ``season," and ``season-nosmall" SN samples can use
the full set of SNe in the LFs. On the other hand, the ``full-optimal"
and ``season-optimal" SN samples exclude all of the SNe that occurred
in highly inclined spiral galaxies, so the rate calculations for these
samples should compute the control times using the subset of SNe in
the LFs that are not in highly inclined spiral galaxies as
well. Fortunately, only 40 out of the 175 SNe (23\%) in the full LF
sample occurred in highly inclined spiral galaxies, leaving a
reasonable number of SNe in the LFs when they are excluded.

\subsection{The Control-Time Calculation}

Section 3 and the Appendix of Paper I provide the mathematical details
of the control-time method. Here we provide the numerical details
regarding how the control-time calculation is performed.

As discussed in Paper II, each SN in the LF sample is a discrete
point, with its own light-curve shape and peak absolute magnitude,
and a fractional contribution proportional to the completeness
correction factor.  We first calculate the control time for a single
SN from the LF.  The uncertainty of the peak absolute magnitude is
used to generate a random correction (according to Gaussian
statistics). This correction, together with the Galactic extinction
toward a specific galaxy in a galaxy sample, is applied to the peak
absolute magnitude of the SN.  The light curve with the derived peak
absolute magnitude is then converted to the apparent light curve
according to the distance of the galaxy.

For a single image recorded in the log files of this galaxy, the
limiting magnitude is calculated from the parameters in the log files,
as detailed in Paper I. The difference between the apparent light
curve and the limiting magnitude is then converted to a control-time
curve using the DE curves for the Hubble type of the galaxy, as
reported in Paper I. This process is demonstrated in Figure 1; SN
2002fk is used as an example, in a galaxy with a distance of 100~Mpc
and an extinction of $A_V$(Galactic) = 0 mag. The apparent light curve
is shown in the upper panel, with the limiting mag also marked
(assumed to be 19; dashed line). The middle panel shows the DE curve
(as derived in Paper I) for the Sb--Sbc bin, the Hubble type we assume
for the galaxy. The offset between the apparent light curve and the
limiting magnitude can then be converted to the control-time curve
shown in the lower panel.  Depending on the apparent light-curve shape
and the offset between the peak and the limiting magnitudes, the
control-time curve can have different shapes, but generally has a
rising, a constant, and a declining portion, and is different from a
step function (i.e., 0 when the SN is fainter than the limiting
magnitude, and 1 when it is brighter). The total integration for the
curve marks the maximum possible contribution to the control time from
this single epoch.

Next, the total control time for all of the epochs of images for this
galaxy is computed. As each epoch can have different control-time
curves, it is difficult to compute the total control time
analytically.  Instead, the problem is solved numerically. We use a
large array with each cell corresponding to a single day in the survey
period.  The maximum allowable control time for each day is the limit
of the DE at the bright end plus a random correction according to the
uncertainty of the DE. For any given epoch of image, the control-time
curve is calculated, and is allowed to shift along the time axis to
compute the total contribution to the control time at or before the
epoch (as the image cannot ``control'' any SNe that occurred after the
observation). The shift that gives the maximum contribution is then
used. Our control-time algorithm follows the simple philosophy of
maximizing the contribution to the control time from any given epoch,
which is the principle of the control-time method.  

We also note that because of the small observation intervals in our
survey, the contribution to the total control time from each epoch is
typically the DE multiplied by the observation interval; in other
words, the constant portion of the control-time curve is used most of
the time. Consequently, our rates are relatively insensitive to the
input SN LF (see more discussion in \S 3.2), especially for SNe~Ia
which are very luminous.

Following the same procedure, the control time is calculated for the
SN for all of the galaxies in the galaxy sample, and then for all of
the SNe in the LF sample. For each galaxy, the total control time for
each SN type (Ia, Ibc, and II) is then calculated according to
Equation (A11) of Paper I --- that is, the sum of the control time of
each SN component weighted by its fractional contribution to the
luminosity function.

The adopted light-curve shapes are important for the control-time
calculation, so in Figure 2 we compare our light curves as constructed
in Paper II with those used by Cappellaro et al. (1999; hereafter,
C99). The differences between the two sets of light curves are
significant, with the C99 light curves in general evolving faster than
our light curves. This is not surprising, since the C99 light curves
are in the $B$ band while ours are in the $R$ band.

\subsection{The Rate Calculation} 

The total control time calculated for the galaxy sample can be
normalised by a chosen factor to generate the total normalised control
time, which is then used to calculate the rate, as described by
Equations (A3) and (A4) of Paper I.  The normalisation factors we
choose to use are the $B$-band luminosity ($L_B$), the $K$-band
luminosity ($L_K$), and the stellar mass. The resulting rates are
labeled SNuB, SNuK, and SNuM, which have units of one SN per 100 yr
per $10^{10}\,{\rm L}_\odot(B)$, $10^{10}\,{\rm L}_\odot(K)$, and 
$10^{10}\,{\rm M}_\odot$, respectively.

However, before we proceed with the rate calculations, we need to
discuss a strong observed correlation between the SN rates and the
sizes of the host galaxies, which fundamentally changes the way our SN
rates are determined.  The details are described in the next section.

\section{The Rate-Size Relation}

In this section, to compute the SN rates we use the ``full-optimal" SN
sample with 726 objects that occurred in the ``optimal" galaxy
sample. As discussed later in the paper, we also adopt this
combination of the SN and galaxy samples for the final rate
calculations.

An important step in the rate calculations is to find an ideal method
to divide the galaxies into different groups so that galaxies within
each group have the same rate. As the specific star-formation rate
(SSFR, the star-formation rate per unit mass) is often considered to
be intimately connected to the (mass-normalised) SN rates, especially
for core-collapse supernovae (CC~SNe), a galaxy sequence that also
represents a SSFR sequence could be used to describe the SN
rates. Historically, the SN rates have been published in galaxies of
different Hubble types or $B - K$ colour. The underlying assumption is
that the galaxy Hubble type or $B - K$ colour is a good proxy for
SSFR, and the SN rate is a constant in galaxies having the same Hubble
type or $B - K$ colour.

In the process of checking the robustness of our rate-calculation
pipeline, however, we found that the SN rates {\it cannot} be
adequately described by a single parameter, either the galaxy Hubble
type or galaxy $B - K$ colour (see \S A of the Appendix for more
details). Instead, a secondary parameter of galaxy size, expressed in
either luminosity (in the $B$ or $K$ bands) or stellar mass, is needed
to quantify the rates.  One would normally expect the SN rates to be
constant for galaxies of different sizes, since the rates have been
{\it linearly} normalised by the galaxy size as indicated by Equations
(A3) and (A4) in Paper I.  But in fact, below we show that there is a
strong correlation between SN rates and galaxy sizes, for the rates in
galaxies of different Hubble types (\S 2.1) or $B - K$ colours (\S
2.2).

\subsection{The Rate-Size Relation for the Hubble-Type Rates}

We first consider whether there is a correlation between galaxy sizes
and CC~SN rates in galaxies of different Hubble types. The results for
the SNuM rates are shown in Figure 3. Only galaxy Hubble types 3--7
(Sab--Scd) are considered because of the small number of CC~SNe
discoveries in E, S0, and Irr galaxies. For each Hubble type, the
galaxies are sorted in order of their masses, and then divided into
several bins from the least massive to the most massive, with roughly
the same number of discovered SNe in each bin because small-number
statistics are often the dominant source of uncertainty. The SN~Ibc
and SN~II rates are then calculated for each mass bin.  Only the
statistical errors are considered here.

  For the SN~II rates (Figure 3, top panel), we find that {\it there is
a strong correlation between SNuM and galaxy mass, with smaller
galaxies having a higher SNuM}.  A $\chi^2$-minimizing technique
is used to fit a power law of SNuM $\propto M^{-0.55}$ (solid line,
the final adopted relation in our calculations), using the rates in
the Sbc galaxy bins as the anchor points and scaling the rates in
the other Hubble types by a multiplicative constant (with proper
error propagation). The reduced $\chi^2$ (i.e., $\chi^2$/DOF [degree
of freedom]) of the fit is $\sim 0.7$, suggesting a good fit to
the data.

For the SN~Ibc rates (Figure 3, bottom panel), there is more scatter due
to small-number statistics, but they can be well fit ($\chi^2$/DOF
$\approx 1.0$) by the same relation as determined for the SN~II rates
after scaling the rates in each Hubble type.  The power-law indexes
between the rates and masses, which we call the rate-size slopes
(RSSs) hereafter, are also measured for the SN~II rates in each
Hubble-type bin and are listed in the third column of Table 1. The
RSSs in different Hubble types have individual statistical
significances of 3--5$\sigma$ and are consistent with each other. The
significance of the RSS after combining the rates in Sab--Scd galaxies
(i.e., the linear fit in the top panel of Figure 3) is $\sim
10.7\sigma$.

The results for the SNuM rates of SNe~Ia are shown in Figure 4 and the
RSSs are listed in the second column of Table 1.  Here the rates in
the Sb galaxies are used as the anchor points, and the power-law index
of $-0.50$ as plotted is the final adopted value in our rate
calculations. The reduced $\chi^2$ of the fit is $\sim 0.6$,
suggesting a good fit to the data. The RSSs in different Hubble-type
bins are significant at the 2--3$\sigma$ level and are generally
consistent with each other.  The combined significance (i.e., the
linear fit in Figure 4) is $\sim 7.4\sigma$.  More discussion of
the RSSs and their significance can be found in \S 4.2.

We investigate the dependence of the RSSs on the normalisation ($L_B$,
$L_K$, or mass), different SN types (Ia, Ibc, and II), various Hubble
types, and distinct SN samples.  The RSSs have a relatively strong
dependence on the normalisation, increasing from SNuB to SNuK to
SNuM. The RSSs for the two types of CC~SNe are generally consistent
with each other and are thus not discriminated from each other
hereafter.  There are some differences (with low significance due to
uncertainties) between the RSSs for SNe~Ia and those for CC~SNe.
Different SN samples yield consistent RSSs for the same SN type and
normalisation.  For each type of SN, no significant difference is
found in the RSSs in various galaxy Hubble types, though the
uncertainties for some RSSs are relatively large (e.g., as shown in
Table 1).

The power-law correlation between the SN rate and the sizes of the
galaxies is called the {\it rate-size relation} hereafter. The
rate-size relation can be explicitly expressed as

\begin{equation}
 {\rm SNuB}(L_B) = {\rm SNuB}(L_{B0}) \Bigl(\frac{L_B}{L_{B0}}\Bigr)^{{\rm RSS}_B},
\end{equation} 

\begin{equation}
 {\rm SNuK}(L_K) = {\rm SNuK}(L_{K0}) \Bigl(\frac{L_K}{L_{K0}}\Bigr)^{{\rm RSS}_K},~{\rm and}
\end{equation} 

\begin{equation}
 {\rm SNuM}(M) = {\rm SNuM}(M_{0}) \Bigl(\frac{M}{M_{0}}\Bigr)^{{\rm RSS}_M}, 
\end{equation} 
\noindent
where $L_{B0}$, $L_{K0}$, and $M_{0}$ are the fiducial galaxy sizes,
and RSS$_B$, RSS$_K$, and RSS$_M$ are the rate-size slopes for the
different normalisations.

\subsection{The Rate-Size Relation for the $B - K$ Colour  Rates}

Historically, SN rates have also been parameterised by the $B - K$
colours of the host galaxies (e.g., Mannucci et al. 2005; hereafter
M05). Unlike the galaxy Hubble types, which are discrete points in
parameter space, the galaxy $B - K$ colours span a wide range and
follow a continuous distribution, so it is impractical to group the
galaxies in {\it constant} $B - K$ colours and then study the SN rates
in different galaxy sizes. 

We adopt the following procedure to investigate whether there is a
rate-size relation (i.e., Eqs. 1--3) in the $B - K$ SN rates.  The
SNuK rates of SNe~II are used as an example (Figure 5; the SNuB and
SNuM rates of SNe~II, and the SN~Ia rates, all show a similar
relation).  As illustrated in the top-left panel, the galaxies are
first sorted according to their $B - K $ colours, and then divided
into four colour groups from the bluest to the reddest (the same
symbol, left to right). For each colour group, the galaxies are
subsequently sorted by $L_K$, and divided into seven $L_K$ bins (the
different symbols). For clarity, only three size bins are shown: the
smallest (size bin 1, open circles), the intermediate (size bin 4,
half-solid circles), and the largest (size bin 7, solid
circles). Next, for each bin the SNuK rate, the average $B - K$
colour, and the average $L_K$ are calculated and plotted. The size of
the symbol is proportional to the logarithm of the average $L_K$. The
dashed line is the average rate for the different $B - K$ groups
(i.e., all galaxies are used in the rate calculations without
considering the differences in $L_K$). A systematic trend is observed
in this panel: the rates for the bins with the intermediate $L_K$
(half-solid circles) closely follow the average rates (dashed line),
while the bins with the smallest $L_K$ (open circles) are higher, and
the bins with the largest $L_K$ (solid circles) are lower than the
average rates. This trend becomes more obvious after the rates are
normalised by the average curve (the bottom-left panel).

At face value, this trend suggests that there is a rate-size relation
for the $B - K$ SN rates. To further investigate this, we study the
SNuK $-$ $L_K$ correlation in two narrow ranges of galaxy $B - K$
colours. As can be seen in the bottom-left panel of Figure 5, there
are only minimal colour changes in the different $L_K$ bins for the
groups of galaxies at $B - K \approx 2.9$ and 3.3 mag. Thus, for each
of these two colour groups, any correlation between SNuK and $L_K$
(i.e., the rate-size relation) is not significantly affected by the
rate changes due to colour variation within the group. The results are
shown in Figure 6, using the rates for the $B - K \approx 2.9$ mag
galaxies as the anchor points and scaling the rates for the $B - K
\approx 3.2$ mag galaxies.  The linear fit has a power-law index of
$-0.38$, the final adopted RSS in our analysis. The existence of a
rate-size relation is verified at $\sim 3.5\sigma$ using these two
colour groups alone.

To quantify the RSSs for the rate-size relation for the $B - K$ SN
rates, we use two numerical methods. The first employs a multi-variate
linear regression model to fit the rates as a function of both galaxy
$B - K$ colours and $L_K$, so that

\begin{equation}
 {\rm log (SNuK)} = c_1 + c_2 \, {\rm log} \Bigl(\frac{L_K}{L_{K0}}\Bigr) + 
  c_3 \, (B - K) + c_4 \, (B - K)^2,
\end{equation} 
\noindent
where $c_1$, $c_2$, $c_3$, and $c_4$ are the coefficients to be evaluated
during the fitting process. It can be seen that $c_2 = {\rm RSS}_K$ in this
equation.  Here we also assume that the logarithm of the rates for a
fiducial galaxy can be adequately fit by a second-order polynomial
function of $B - K$ colour, an assumption that is verified by the
discussion in \S 3.6.

The second method employs a $\chi^2$-minimizing technique and is
demonstrated by the right-hand panels in Figure 5. A wide range of RSS
values is tested to convert the rates in all of the $L_K$ bins as
well as the average rates to a fiducial galaxy size using Equation
(2), and the optimal RSS is the one that yields the minimum $\chi^2$
when the rates in different $L_K$ bins are compared to the average
rates.  As the right-hand panels of Figure 5 show, after the rate-size
relation is considered and all of the rates are converted to the same
fiducial galaxy size, the systematic trend presented in the left
panels is gone, and the rates in different $L_K$ bins are consistent
with the average rates to within $\sim 1\sigma$.

The RSSs derived from these two methods are fully consistent with each
other, so we average them as our adopted values. We derived the RSSs
for the SN~Ia and SN~II rates, but not for the SN~Ibc rates due to the
relatively large uncertainties. Instead, we assume that the SN~Ibc
rates have the same RSSs as the SN~II rates.\footnote{We tested this
  assumption by adopting the RSSs from the SN~II rates in the SN~Ibc
  rate calculations, and found that these RSSs adequately removed any
  rate-size relation in the SN~Ibc rates.} Unlike the RSSs for the SN
rates in galaxies of different Hubble types, which exhibit a
significant dependence on the normalisations ($L_B$, $L_K$, or mass),
the RSSs for the rates in galaxies of different $B - K$ colours are
within a narrow range and consistent with each other for the different
normalisations (for the same SN type), so only two RSSs are needed.

Our final adopted RSSs, which are the averages for the different SN
types and normalisations, are reported in Table 2.  We adopt an
uncertainty of 0.10 for most RSSs, roughly the value of adding the
scatter of the RSSs from different SN samples and the uncertainty of
an individual RSS measurement in quadrature.  Somewhat larger errors
of 0.20 and 0.15 are adopted for the SN~Ia SNuB Hubble-type rates and
all $B - K$ rates due to larger RSS measurement scatter or
uncertainties.

\subsection{The Effect of the Rate-Size Relation on the Rate Calculations}

The existence of the rate-size relation has two implications. First,
the SN rate before the normalisation by the sizes of the galaxies
(i.e., the SN frequency, or number of SNe per year) is not linearly
proportional to the galaxy size, but to a power law of size$^{(1 +
  {\rm RSS})}$, where size can be $L_B$, $L_K$, or mass.  For example,
for the $B$-band normalisation, the SN frequency for SNe~II is
proportional to $L_B^{0.73}$ instead of to $L_B^{1.00}$.  Second,
since the rate varies with galaxy size, we need to choose a fiducial
galaxy size to compute the rate, so that the rates for the other
galaxy sizes can be evaluated using the RSSs. As the exact value of
the fiducial size is not of great importance, we use a value that is
close to the average galaxy size in each normalisation for this
purpose: $L_{B0} = 2 \times 10^{10}\, {\rm L}_\odot$ for SNuB, $L_{K0}
= 7 \times 10^{10}\, {\rm L}_\odot$ for SNuK, and $M_0 = 4 \times
10^{10}\, {\rm M}_\odot$ for SNuM. These values are listed in the last
column of Table 2.

Using SNuM as an example, here we show how the rates are computed for
a fiducial galaxy size.  Let $M_0$ be the fiducial galaxy size. Then the
rate-size relation can be written as

\begin{equation}
 {\rm SNuM}(M) =\frac{N{\rm (SN)}}{MC} = {\rm SNuM}(M_0)
 \Bigl(\frac{M}{M_0}\Bigr)^{{\rm RSS}_M},
\end{equation} 
\noindent
where $C$ is the control time. This can be rewritten as 

\begin{equation}
{\rm SNuM}(M_0) = \frac{N{\rm (SN)}}{MC(M/M_0)^{{\rm RSS}_M}}.
\end{equation}
\noindent
In other words, the rate for each galaxy can be effectively converted
to the rate for the galaxy with the fiducial galaxy size (hereafter,
the fiducial galaxy) when the control time $C$ is scaled by a factor
of $(M/M_0)^{\rm RSS}$.  This is the main modification to the rate
calculations discussed in \S 3 of Paper I and in \S 1.3 here.

We note that a nonlinear proportionality between the SN frequency and
the host-galaxy size has been reported for SN~Ia rates in star-forming
galaxies by Sullivan et al. (2006), although our results are somewhat
different. More detailed discussion of this and the possible causes of
the rate-size relation can be found in \S 4.2.

An alternative parameterisation of the SN rates using Hubble types
and colour as the two independent variables is discussed in \S B
of the Appendix. 

\section{The SN Rate in a Fiducial Galaxy} 

\subsection{The SN Rates in Different SN Samples}

As discussed in Paper I and summarised in \S 1.1, there are several SN
subsamples with different associated galaxy samples. One test to
investigate the robustness of our rate-calculation pipeline is to
compute the rates using different SN subsamples, and check for their
consistency, as shown in Figure 7. Here SNuM for a fiducial galaxy is
calculated for SNe~Ia, Ibc, and II in different galaxy Hubble
types. Only the statistical errors are shown. The solid circles are
for the rates of the 929 SNe in the ``full" sample, the triangles are
for the 726 SNe in the ``full-optimal" sample, the open squares are
for the 656 SNe in the ``season" sample, the solid squares are for the
499 SNe in the ``season-optimal" sample, and the open circles are for
the 583 SNe in the ``full-optimal" sample but only using SNe
discovered before the end of the year 2006. As discussed in \S 1.1,
the full set of SNe in the LFs is used to calculate the control times
for the galaxy samples for the ``full" and ``season" SN samples, while
the LFs without the SNe occurring in highly inclined spiral galaxies
are used for the galaxy samples for the ``optimal'' SN samples.

Inspection of Figure 7 reveals that the rates from different SN
subsamples are consistent with each other to within 1$\sigma$.  For
each Hubble-type bin, we calculate the average and root-mean square
(RMS) of the rates, and find that the RMS is about 6\% of the average
for SNe~Ia, 12\% for SNe~Ibc, and 11\% for SNe~II.  The rates using
the SNe in the ``full-optimal" sample before the end of the year 2006
are consistent with the rates using the whole ``full-optimal"
sample. This suggests that our rate-calculation pipeline is robust in
terms of the cutoff period for the SN sample.

Our final rates use the 726 SNe in the ``full-optimal" sample, which
provides a good balance between improving small-number statistics and
avoiding systematic biases. We emphasize, however, that using a
different SN sample does not significantly affect our discussion in
the subsequent sections of this paper. As the different SN samples are
not independent of each other, a straight average or median of the
rates is not the proper way to proceed.

\subsection{The SN Rates with Different LFs}

In this section, we investigate how our rates are affected by the
choices of the input LFs for the SNe. Three sets of LFs are
considered. The first set of LFs splits the LF SNe into two broad
Hubble-type bins (hereafter 2LF), which is our choice for the final
rate calculations. The second set of LFs combines all of the SNe into
a single LF for each SN type (hereafter 1LF). The third set of LFs is
actually not a LF at all, but a single light curve with a single peak
absolute magnitude as adopted in the C99 rate calculations (hereafter
C99-LF).\footnote{The C99 rate calculation was performed with a Gaussian
LF and $BVR$ light curves depending on the specific search. Here only
the $B$-band light curve and the average peak absolute magnitude are 
used.} As shown in Figure 2, the light curves adopted by the C99
study are quite different from those used in our rate calculations,
and are only suitable for surveys done in the $B$ band.  Since our
unfiltered survey is more closely matched to the $R$ band, the
calculations using the C99-LF are not very meaningful except to
demonstrate the effect of an extreme choice of the input LF.

The results are shown in Figure 8 for the ``full-optimal" sample of
SNe. For SNe~Ia, the rates are remarkably stable with different input
LFs, even when the extreme choice of the C99-LF is used. When all
three rates are used to calculate the average for each Hubble-type
bin, the RMS is about 7\% of the average, similar to the scatter found
in the previous section for the different SN samples. The reason the
SN~Ia rates are insensitive to the choice of the input LF is
simple. Due to the depth of our SN survey, the short observational
intervals, and the luminous nature of SNe~Ia, our survey is largely
volume-limited for SNe~Ia, so the control time is close to the season
time for any reasonable choice of the input LFs.

For the CC~SNe, the rates are more sensitive to the choice of the
input LFs.  This is not unexpected, as the CC~SNe already suffer some
incompleteness within 60~Mpc, as discussed in Paper II.  When the 2LF
and 1LF rates are used to calculate the average for each Hubble-type
bin, the RMS is about 16\% of the average for SNe~Ibc and 12\% for
SNe~II. Compared to the 2LF rates of SNe~Ibc, the 1LF rates are
smaller in early-type spirals and bigger in late-type spirals, while
it is the opposite for SNe~II. This is consistent with the
expectations from the LF study of the SNe~Ibc and II in Paper II. The
average luminosity of SNe~Ibc in early-type spirals is fainter than
that of SNe~Ibc in late-type spirals. Consequently, using a separate
LF for the SNe~Ibc in the early-type spirals will enhance the rates in
these galaxies.  The same logic can be applied to the SNe~Ibc in
late-type spirals and the SNe~II.

The C99-LF rates are dramatically different from the other rates for
the SNe~Ibc and II.  These rates, though not very meaningful, do
provide information on how our rates and the published C99 results
compare when the same sets of light curves and peak absolute
magnitudes are used. When the C99-LF is used, our rates are depressed
for SNe~Ibc (by $\sim 50$\% and $\sim 20$\% for the early-type and
late-type spirals, respectively).  This is because the C99 SN~Ibc
light curve has a peak absolute magnitude of $-17.0$, brighter than
more than 60\% of the SNe in our SN~Ibc LF. As a result, the control
time is increased, yielding a lower rate.  The SN~II rates, on the
other hand, are enhanced (by $\sim 60$\% and $\sim 10$\% for the
early-type and late-type spirals, respectively).  This is likely to be
mainly caused by the differences in the adopted light-curve
shapes. The C99 SN~II light curves have a much narrower peak than
ours, resulting in a smaller control time and a higher SN rate.

We note that the effect of the light-curve shape is dramatically
reduced for the rates in the ``season" and ``season-optimal" SN
samples, as these calculations do not include the control time for the
first epoch of each season, which is often the only epoch when the
control time from light-curve shape is needed [the other epochs mostly
  use DE $\times$ (observation interval)].  Accordingly, the SN~II
rates in the ``season" and ``season-optimal" SN samples using the C99
light curves do not show a significant difference from those obtained
with the 2LF and 1LF LFs.

\subsection{The Inclination Correction Factor} 

The presence of a strong bias in the discovery of SNe in inclined 
{\it spiral} galaxies was first reported by Tammann (1974), and
subsequently discussed by van den Bergh \& Tammann (1991), Cappellaro
et al. (1997, hereafter, C97), and C99.  Historically, researchers have
used an inclination correction factor (ICF), which is the ratio of the
SN rate in a face-on galaxy to that in an inclined galaxy, to account
for the bias. A significant ICF (on the order of 2--3) has been
reported in searches conducted visually or with photographic plates
(e.g., C97, C99).

Since our search is conducted with a red-sensitive CCD camera and our
SNe are discovered via image subtraction to deal with the bright
central regions of galaxies, the ICF is expected to be relatively
small in our rates compared with previous studies.  To verify this, we
divide the ``full-nosmall" SN sample into three inclination bins
($0^\circ-40^\circ, 40^\circ-75^\circ, 75^\circ-90^\circ$), and
calculate the respective rates ($r0$, $r1$, $r2$) for different SN
types and normalisations. We also divide the spiral galaxies into
early-type (Sa--Sbc) and late-type (Sc--Scd) bins (as mentioned
previously, the inclination angle is not meaningful for the elliptical
or irregular galaxies). Since the goal is to address how our rates are
affected by a possible ICF, we used the subset of SNe in the LFs that
are not in highly inclined spiral galaxies, which are adopted in the
final rate calculations, to calculate the control times. The results
are listed in Table 3 and plotted in Figure 9.  Only the statistical
errors\footnote{To simplify the rate ratio calculations, the upper and
  lower uncertainties due to Poisson statistics are averaged to
  generate the statistical errors reported in Table 3.} are considered
here.

Inspection of Table 3 and Figure 9 reveals that there may be a sizable
ICF in our rates. In particular, for the SN~II rates in the late-type
spirals, the ICFs between the face-on galaxies ($r0$;
$0^\circ-40^\circ$) and the highly inclined galaxies ($r2$;
$75^\circ-90^\circ$) are 3.2--4.7 (the column labeled as $r0/r2 - 1$),
although their uncertainties are relatively large due to the rate
uncertainties.

For our final rate calculations, we elect not to adopt an ICF.
Instead, the highly inclined galaxies ($75^\circ-90^\circ$) and the
SNe that occurred in them are not considered. These rates are reported
as $r3$(0--75) in Table 3, together with the ratios when compared to
the rates in the face-on galaxies ($r0/r3 - 1$). We avoid using an ICF
for two main reasons, as follows.

(1) The significance of the ICFs for the two bins with small and
medium inclinations ($0^\circ-40^\circ$ and $40^\circ-75^\circ$) is
low.  As shown by $r0/r1 - 1$ in Table 3, the SN~Ia and Ibc rates do
not have a significant ICF for all of the normalisations. The SN~II
rates display differences in the two bins with a significance level of
only $\sim 2\sigma$ for all normalisations.

(2) We fail to explain the presence of an ICF for the SN~II rates,
but not for the SN~Ia and SN~Ibc rates.  Historically, the presence of
an ICF is attributed to greater extinction toward the SNe in more
highly inclined galaxies. Consequently, the SNe in inclined galaxies
are, on average, dimmer than those in face-on galaxies. Using the
average LF without considering the inclinations thus overestimates the
control time for the inclined galaxies and underestimate the rates.
However, as discussed in \S 5.2 of Paper II, when the LF SNe are
considered, only SNe~Ibc are consistent with greater extinction in
more highly inclined galaxies, with small-number statistics. Thus, an
ICF for the SN~II rates, if real, cannot be easily explained by
greater extinction in more highly inclined galaxies.

As a further test to investigate whether the differences in the SN~II
rates are caused by an ICF, we calculate the rates for the SNe~II in
the late-type spirals in two distance bins, and plot the results in
Figure 10. The open circles are for the rates in the galaxies with
distance $D < 75$~Mpc, while the solid circles are for the galaxies
with $D \ge 75$~Mpc.  In theory, the control times for the more nearby
galaxies should be less affected by additional extinction in more
highly inclined galaxies, because a large fraction of the galaxies are
in the volume-limited regime.  As Paper II discussed, the SNe~II in
the LF sample (with $D < 60$~Mpc) have only a small ($\sim 10$\%)
incompleteness in our search; hence, a smaller ICF for the rates is
expected for the more nearby galaxy bin.  Figure 10 does not support
such a conclusion, but it does not eliminate the conclusion either
because of the relatively large uncertainties.

As described in Paper II, we have host-galaxy inclination information
for all of the LF SNe. To investigate whether the discrepancies in the
rates are caused by the differences in the LFs in the various
inclination bins, we calculate the rate for each inclination bin using
the subset of SNe with the same inclination range in the LFs to
calculate the control times.  The results are shown in Figure 11. The
SN~Ia rates do not show a significant ICF. The SN~Ibc rates, on the
other hand, show a negative ICF due to the strong dependence (with
small-number statistics) of the SN~Ibc LFs on the inclinations. The
SN~II rates still exhibit a significant ICF for the late-type spiral
galaxies. Thus, inclination-dependent LFs, at least with the
small-number statistics in our LFs, do not solve the problem for the
SN~II rates in late-type spiral galaxies.

We have also investigated whether the rates in galaxies having
different $B - K$ colours show significant differences at various
inclinations.  The spiral galaxies (Sa--Scd) are split into two bins
with $B - K < 3.1$ mag and $B - K \ge 3.1$ mag. No significant
difference is found for the SN~Ia and Ibc rates in the smallest and
medium-inclination bins, but a 2--3$\sigma$ difference is found for
the SN~II rates in both colour bins.

We note that the inclination effect for the SN~II rates in late-type
spiral galaxies appears stronger in SNuB than in SNuK or SNuM. As
discussed in Paper I, the galaxy $B$ luminosities are corrected for
internal extinction due to inclination using the prescription by
Bottinelli et al. (1995).  It is possible that this prescription
overestimated the galaxy luminosity correction, and thus the SNuB
rates in the edge-on galaxies are underestimated.  However, incorrect
internal extinction will not explain the strong inclination effect for
the SN~II rates in late-type spirals in SNuK, as the $K$-band
luminosities of the galaxies have {\it not} been corrected for any
internal extinction (see Paper I for more details). It should also be
noted that any attempt to remove the inclination effect for the SN~II
rates in late-type spirals by changing the galaxy luminosities will
also result in a negative inclination effect for the SN~Ibc rates, as
the same set of galaxies is used to calculate the rates for both types
of SNe.

We conclude that invoking extinction to explain the differences in the
SN~II rates does not present a coherent picture when all of the
observational evidence is considered. Rather, the differences may be
caused by a combination of several factors, such as small-number
statistics, systematic errors (\S 3.4), errors in the control-time
calculation (due to the limitation of the LF and the light-curve
shape, as discussed in Paper II), and the presence of an ICF.

Regardless of the reasons for the differences in the SN~II rates in
the different inclination bins, the differences themselves may be
real. If true, our neglect of a correction factor will result in an
underestimate of the SN~II rates. As the values of $r0/r3 - 1$ in
Table 3 show, the average SN~II rates in the $0^\circ - 75^\circ$ bin
for the late-type spirals are underestimated by about 40--50\% when
compared to the rates in the face-on galaxy bin.  For the galaxies
with different $B - K$ colours, the average rates are underestimated
by about 30--70\%.  The ICF (or lack thereof) thus becomes the largest
uncertainty in our treatment of the SN rates, especially for SNe~II,
as discussed in the next section. We also note that for the SN~Ia and
Ibc rates, the presence of an ICF cannot be completely ruled out on
statistical grounds because of the relatively large uncertainties in
the rate ratios. It is thus important to substantially enlarge the
sample size in future SN rate calculations, to further evaluate the
rate dependence on the galaxy inclinations.

\subsection{Error Budget}

It is important to have a reasonable uncertainty estimate for the SN
rates before discussing any trends or biases. Here we describe the
error budget for our rates, considering the statistical and systematic
errors separately.

We emphasise that it is nearly impossible to account for every
possible source of uncertainty in the rate calculations because of the
large amount of involved data and the complexity of the pipeline. Even
though we tried to make use of the best available data in the current
astronomical database (see the discussion in Paper I of how our galaxy
and SN databases were constructed), many measurements are ultimately
limited by our knowledge and/or the precision of the existing
astronomical quantities. For example, for the galaxies, the $B$ and
$K$ photometry suffers from relatively large uncertainties due to the
difficulty of cleanly measuring fluxes of extended objects. For the
SNe, the LFs (Paper II) were measured from a sample of nearby objects,
whose distances derived from the Hubble law suffer from relatively
large uncertainties due to peculiar motions in the local Universe. For
the rate-calculation pipeline, the choice of the RSSs and whether an
inclination correction factor is adopted have significant effects on
the final derived rates.

One positive aspect of the uncertainties, resulting from the sheer
number of galaxies and SNe involved in the calculations, is that the
uncertainty is determined by the sample as a whole; the effect of
the uncertainty for a single galaxy or SN becomes relatively small.

For the statistical errors, we use Poisson statistics.  The upper and
lower Poisson 1$\sigma$ uncertainties of the number of SNe involved in
a rate calculation are computed and used to derive the errors (Gehrels
1986).  For the rates in the Irr galaxies, or the CC~SNe in early-type
galaxies (E--S0), the statistical errors can be as large as $\sim
100$\% of the measurements due to small-number statistics.  For the
other rates with significant numbers of SNe involved, this value is
$\sim 10$--30\% (see, e.g., the rates listed in Tables 4 and 5,
discussed below).

For the systematic errors, we adopt the following methodology to
calculate the contribution from each likely source except those from
the ICF.  The rates from the ``full-optimal" sample are used as the
``anchor points." For a new set of rates with a different choice of
parameters, the difference is calculated as a percentage of the anchor
point, and its absolute value is used as both the upper and lower
uncertainties. For the ICF, an asymmetric error matrix is used, as
discussed in detail below. The final upper and lower uncertainties (as
percentages of the anchor points) are calculated with the individual
components added in quadrature, and then converted to errors by
multiplying the values of the anchor points.

We consider the following sources for the systematic errors.

\begin{enumerate}

\item{Scatter from using different SN samples. While \S 3.1 provides a
  detailed discussion of how the rates are affected by using five
  different SN samples, the scatter in the rates are not all
  independent of the other uncertainties discussed below.  For
  example, part of the difference between the ``full" sample and the
  ``full-optimal" sample may be caused by an inclination correction
  factor. For this reason, the contribution to the systematic errors
  due to the sample selection is calculated from the ``full-optimal"
  and ``season-optimal" samples. The sample selection causes an
  uncertainty in the range $\sim 5$--20\%, with a median at $\sim
  10$\%.}

\item{Scatter from using different input LFs. The change in the rates
  when using 1LF or 2LF demonstrates the effect of the input LFs.
  Ideally, a LF should be constructed for each galaxy Hubble type, but
  our small-number statistics preclude such an exercise. While it is
  difficult to predict how the rates would change from 2LF to multiple
  LFs, we can use the differences between the 1LF and 2LF rates as a
  reasonable estimate of the uncertainty caused by the inadequate
  precision in the input LFs. The choice of the input LFs causes an
  uncertainty in the range $\sim 5$--30\%, with a median of $\sim
  10$\%.}

\item{Uncertainty caused by the errors in the RSSs. The errors in the
  RSSs as reported in Table 2 are used to calculate the resulting
  uncertainty in the rates, and the two errors from the upper and
  lower uncertainty of the RSSs are averaged. The RSS errors cause an
  uncertainty in the rates in the range $\sim 5$--25\%, with a median
  of $\sim 10$\%.}

\item{Uncertainty caused by the treatment of the ICF. As discussed in
  the previous section, the SN~II rates show a potential ICF for the
  late-type spirals or galaxies with different $B - K$ colours. As the
  adoption of an ICF will only increase the rates, we use the
  following asymmetric error matrix. For the upper uncertainty, the
  percentage that the average rate in the $0^\circ - 75^\circ$ bin is
  underestimated relative to the face-on bin is adopted (40--50\% for
  the late-type spirals, 30--70\% for the galaxies with different $B -
  K$ colours). For the lower uncertainty, a global 10\% is
  assumed. For all of the other rates, a global $\pm$10\% uncertainty
  is adopted.}

\item{Uncertainty caused by miscellaneous small factors. As mentioned
  earlier, it is very difficult to fully assess the uncertainties
  caused by the errors of the various measurements (such as
    photometry, hubble types, inclination, and distance) for a large
  number of galaxies and SNe.  Since the previous several sources all
  contribute roughly 10\% each toward the total systematic error
  budget, we adopt a global uncertainty of $\pm 10$\% for all
  remaining miscellaneous factors.}

\end{enumerate}

As discussed in the next several sections, for most of the rates the
systematic errors are roughly the same size as the statistical errors.
For the SN~II rates in the late-type spirals and in galaxies of
different $B - K$ colours, the systematic errors are a factor of $\sim
1$--4 times that of the statistical errors, and can reach $\sim 80$\%
of the measurements.  We emphasise that our final systematic errors
are quite uncertain due to the rough estimates from several
components. Fortunately, for most discussions of the internal trends
and comparisons based on one set of chosen parameters, only the
statistical errors need to be considered (as we have done in \S 3.1
and 3.2).  The systematic errors become relevant when our rates are
compared with other published results, or when the rate-size relation
and/or ICF play a significant role.  We shall discuss the
uncertainties and their significance in the following sections on a
case-by-case base.

\subsection{The SN Rates as a Function of Galaxy Hubble Type}

In the previous sections, we have shown that our rates are stable for
different SN subsamples but are sensitive to the choice of the input
SN LFs.  For the final rate calculations, we elect to use 2LF (for
more detailed LFs) and the 726 SNe in the ``full-optimal" sample (for
a good balance between statistical and systematic uncertainties). The
rates for a galaxy with the fiducial size are computed according to
the RSSs in Table 2 (also listed in Table 4), and reported in Table 4
for different Hubble types. The statistical errors are given together
with the systematic errors (in parentheses). To calculate the rate for
a specific galaxy, one simply needs to apply Equations (1) through (3)
(assuming the size of the galaxy is known).  Table 4 shows that our
rates are derived from significant numbers of SNe for most of the SN
types and galaxy Hubble types, except for the Irr galaxies (not enough
galaxies) and for CC~SNe in E--S0 galaxies (CC~SNe are intrinsically 
rare in such galaxies).

These rates, together with the statistical errors, are plotted in
Figure 12. To illustrate the effect of adopting a fiducial galaxy
size for each normalisation, we also evaluate the rates at the 
median galaxy size for each Hubble type and plot them as open 
circles in Figure 12. Inspection of the figure reveals the following.

\begin{enumerate}

\item{The SNuK and SNuM rates for the same SN type display very
  similar trends, so we choose to discuss only SNuM in this
  section. The results on SNuM generally apply to SNuK, unless
  explicitly expressed otherwise.}

\item{The SNuB of SNe~Ia declines from the early- to the late-type
  galaxies, with only an upper limit derived for the Irr galaxies.
  Since the $B$-band luminosity of a galaxy is heavily influenced by
  the amount of blue, young, massive stars, $L_B$ is not a good
  indicator of the total amount of mass that is responsible for the
  production of SNe~Ia, which arise from white dwarfs. This is
  particularly true for the late-type galaxies having abundant massive
  stars from recent star formation.  Consequently, the SNuB rates in
  the late-type galaxies are depressed because their $L_B$ are
  significantly contaminated by massive stars. }

\item{The SNuM rates of SNe~Ia are consistent with being 
  {\it constant} for the different Hubble-type bins.  Without considering
  the upper limit in the Irr galaxies, the rest of the rates can be
  fit as a constant (SNuM = 0.136 $\pm$ 0.018) with a reduced $\chi^2
  \approx 0.8$. }

\item{The rates of the CC~SNe in the early-type galaxies (E and
  S0) are close to 0 for all of the normalisations. These small rates
  provide a strong constraint on the amount of recent star formation
  and/or the delay-time distribution (DTD; a distribution of the delay
  time between the formation of the progenitor star and the explosion
  of the SN) in these galaxies, as discussed later in this paper.}

\item{The CC~SN rates generally {\it increase} from early- to
  late-type spiral galaxies for all of the normalisations (except
  perhaps for the SNuB rates of SNe~Ibc which are nearly
  constant). The SN~II rates have a more dramatic change than the
  SN~Ibc rates, especially considering the fact that the SN~II rates
  in Sc/Scd galaxies may be underestimated due to the presence of an
  ICF.  There might be a declining trend from the Sc to the Irr
  galaxies, but the significance of such a trend hinges on the
  uncertain rates in the Irr galaxies. For example, when the rates for
  the Irr galaxies are not considered, such a trend would have a low
  significance for the SN~II rates. For the SN~Ibc rates, the decline
  from the Sc galaxies to the Scd galaxies is more obvious, but the
  difference is still within 2$\sigma$ of the uncertainties.  We need
  more SNe to reduce the statistical uncertainties of the rates and
  verify the presence of such a trend.}

\item{The rates evaluated at the median galaxy size for each Hubble
  type, nearly identical to those found when not adopting the
  rate-size relation in the rate calculations (see \S 4.1 for more
  discussion), show that the biggest differences from the rates using
  a single fiducial galaxy size are presented in E (for SNe~Ia) and
  Scd/Irr (for all SN types) galaxies.  Not surprisingly, these
  galaxies are also at the two extreme ends of the luminosity/mass
  size distribution (most luminous/massive for E, and least
  luminous/massive for Scd/Irr galaxies).}

\item{Understanding the observed trends in the SN rates requires
  knowledge of the SSFR and the initial mass function (IMF) in the
  different Hubble types, DTDs for stars with different masses, and
  the link between stars of different masses and the different SN
  types (see Smith et al. 2011). }

\end{enumerate} 

\subsection{The SN Rates as a Function of Galaxy $B-K$ Colour}

It is well known that the Hubble-type sequence from E to Irr
corresponds to a sequence in the star-formation rate (SFR). The SFR is
virtually zero in ellipticals and becomes increasingly larger toward
late-type spirals. An alternative indicator of the SFR are the
broad-band colours (in particular the optical to near-infrared), with
bluer galaxies hosting a younger stellar population having stars that
are more massive than those in redder galaxies. For this reason and
following the work of M05, we calculate the SN rates for galaxies with
different $B - K$ colours. As discussed in Paper I, we have secured
the $B - K$ colour measurements for a majority of the LOSS sample
galaxies.

We divide the galaxies into different $B - K$ colour bins, calculate
the SN rates for the fiducial galaxy, and report the results in Table
5. To evaluate the rate for a specific galaxy, one needs to know the
galaxy size and apply Equations (1) through (3) (with the RSSs listed
in both Tables 2 and 5). The rates in Table 5, together with their
statistical errors, are plotted in Figure 13. The rates are also
evaluated at the median galaxy size for each colour bin, and plotted
as open circles.  The dashed lines shown in Figure 13 represent the
second-order polynomial fits (as a function of $B - K$ colour) for the
logarithm of the rates as determined during the multi-variate linear
regression model analysis using Equation (4). As mentioned in \S 2.2,
this analysis is not applied to the SN~Ibc rates due to their
relatively large uncertainties\footnote{We actually performed the
  analysis for the SN~Ibc rates, and the model provides a reasonable
  fit to the data. The fits are not shown in Figure 13 in order to be
  consistent with the discussion in \S 2.2.}. Inspection of the figure
reveals the following.

\begin{enumerate}

\item{As in Figure 12, the SNuK and SNuM rates for the same SN type
  display very similar trends, and we choose to discuss only SNuM as
  an example.}

\item{The SNuB rate of SNe~Ia increases from blue to red galaxies,
  likely due to the increasing influence of massive stars in the total
  $B$-band luminosity in the bluer galaxies. The SNuB rate of SNe~II,
  on the other hand, is consistent with a constant for the several
  bins at the blue colour end, and then declines toward the red
  colours. This is likely caused by the increasing influence of an old
  stellar population in the redder galaxies.  The SNuB rate of SNe~Ibc
  rises from the bluest galaxies to $B - K = $ 3.0 mag, then declines
  thereafter. }

\item{The SNuM rate of SNe~Ia increases dramatically from red to blue
  galaxies (by a factor of $\sim 6.5$). This is different from the
  Hubble-type rates where the SN~Ia rates are consistent with being a
  constant in different Hubble types for SNuK and SNuM.} 

\item{The CC~SN rates are small (but not zero) in the reddest
  galaxies, and in general become progressively higher for bluer
  galaxies. However, the SN~Ibc rate becomes smaller for the bluest
  galaxy bin. Aside from small-number statistics, other possible
  reasons for this change are the metallicity effect on the binary
  progenitor evolution of SNe~Ibc, the progenitor-star mass range,
  and/or the DTD.  More detailed discussion is beyond the scope of the
  current analysis. }

\item{ The dashed lines provide excellent fits to the SN~Ia and SN~II
  rates, indicating that we have adopted a reasonable functional form
  during the multi-variate linear regression model analysis in \S 2.2.}

\item{The rates evaluated at the median galaxy size for each colour
  bin show that the biggest differences from the rates using a single
  fiducial galaxy size are present in the bluest galaxies, which have
  the lowest luminosity per unit mass among all of the galaxies. }

\end{enumerate}

\section{Discussion}

\subsection{Comparison with Historical Results}

In this section, we compare our SN rates with the published results,
in particular to the benchmark work of C99 and M05. There are many
differences in the calculations, as detailed in Papers I and II and
the previous sections of this paper, such as the total number of SNe,
the survey method, the treatment of the LFs, the light-curve shapes,
the host-galaxy extinction, and the ICFs.  The biggest difference,
however, is our adoption of the rate-size relation and the use of the
RSSs.  Accordingly, our rates are calculated for a fiducial galaxy
size. Since the C99 and M05 results do not consider a RSS, their rates
are for the average galaxy sizes. To mimic the calculations performed
by C99 and M05, there are two options. One is to evaluate our rates
(for the fiducial galaxies) at the average galaxy size for different
Hubble types or colours, while the other is to calculate the rates
without using the RSSs in the rate-calculation pipeline.  The two
options are not exactly the same, as the rates without using the RSSs
in the pipeline are the average of the rates for the galaxies weighted
by their control times.  In practice, however, the rates from the two
approaches are nearly identical, as there are numerous galaxies
involved in the calculations and the effect of the control time is
averaged out.

We elect to calculate the rate for the average KAIT galaxies without
using the RSSs in the pipeline, exactly the same way the rates were
calculated by C99 and M05. The rates are listed in Tables 6 (for
different Hubble types) and 7 (for different $B - K$ colours), and
they are plotted in Figures 14 and 15.  As no RSSs are used to
calculate the average SN rates, the systematic errors reported in the
tables are the combination of the remaining components discussed in \S
3.4. The total uncertainties (the statistical and systematic errors
added in quadrature) are also plotted in Figures 14 and 15; since our
rates are compared to the measurements from another analysis, we need
to show the full error matrix.

For the rates as a function of galaxy Hubble type (Figure 14), our
results and those published by C99 and M05 are generally in good
agreement within the uncertainties, even though nominally our fiducial
SN~Ibc rates are higher (by a factor of $\sim 2$), and our fiducial
SN~II rates are lower (by a factor of $\sim 1.5$).  The only
significant difference is the rates in the Irr galaxy bin.  As
discussed earlier, there is a deficit of Irr galaxies in the LOSS
galaxy sample, and only 11 out of the 929 SNe considered in the rate
calculations were discovered in the Irr galaxies. Consequently, the SN
rates for the Irr galaxies are quite uncertain in our calculations,
but we are in the process of remedying this by monitoring more Irr
galaxies in our search. Nevertheless, our rates in the Irr galaxies,
derived from a small number of SNe for SNe~Ibc and II, and the upper
limit of our rate for SNe~Ia, do not support the dramatic increase of
the rates in the Irr galaxies suggested by C99 and M05. We suspect
that the true SN rates in the Irr galaxies are in between our rates
(or limits) and the C99/M05 results. Obviously, better constraints
will be obtained once more SNe are discovered in the galaxies and
incorporated into future rate calculations.

For the rates as a function of galaxy $B - K$ colour (Figure 15), the
SN~Ia rates show good agreement, and exhibit a dramatic increase from
the red to the blue galaxies, much more so than the rates for the
fiducial galaxies (Figure 13).  This is caused by the differences in
the average masses of the galaxies with different colours, as
discussed in Paper I. Bluer galaxies tend to have smaller masses, and
their SNuM becomes higher as indicated by the rate-size
relation. Since the CC~SN rates are combined together by M05, we also
combine our CC~SN rates, giving the comparison in the lower panel of
Figure 15. Again, our rates agree with the M05 results to within the
uncertainties. We also plot the SN~Ibc rates (dashed line) and SN~II
rates (dash-dotted line). The SN~II rates show a more dramatic
increase from the red to the blue galaxies than the SN~Ibc rates.

The good agreement between our rates and these reported by C99 and
M05, though with different approaches to treat the various aspects of
the rate calculations, suggests that both analyses employed reasonable
assumptions and corrections to deal with the observational biases and
uncertainties involved. However, we note that the agreement is only
achieved when the rates are calculated in the same manner, without
considering the important rate-size relation that we discovered
during the course of our research.

\subsection{The Rate-Size Relation}

In this section, we offer more discussion of the rate-size relation.
We emphasise that this relation is empirically derived from the data;
finding the exact causes of the relation is not critical for the rate
calculations, but may shed light on the correlation between the SFR
and the galaxy properties, and on the DTD for the various types of
SNe. As also discussed in the next section, the rate-size relation has
a significant effect on the study of the two-component model fit to
the SN~Ia rates.

We first attempt to quantify the effect of adopting the rate-size
relation in our rate calculations. Figures 12 and 14 show our rates in
different Hubble types with and without the adoption of the rate-size
relation, respectively. The differences are significant. For example,
the SNuM rate of SNe~Ia exhibits only a weak increasing trend from the
early-type to the late-type galaxies, and is consistent with a
constant in Figure 12, but a much more prominent increasing trend is
seen in Figure 14. The ratio of the rates between Figures 12 and 14
for the same Hubble-type bin reflects the corrections caused by the
rate-size relation.

Numerically, the existence of the rate-size relation indicates that
the rates cannot be adequately described by a single parameter using
either galaxy Hubble type or $B - K$ colour. The galaxy size ($L_B$,
$L_K$, mass) is thus used as a second parameter to quantify the rates
(in the form of the rate-size relation).  We have considered other
combinations of parameters to describe the rates --- that is, to
replace the rate-size relation with some other empirical
correlations. One combination that merits more discussion is to
parameterise the rates as a function of both galaxy Hubble type and
$B - K$ colour; see \S B of the Appendix.

Physically, what could possibly cause the SN rates to be sensitive to
the sizes of the galaxies? For the CC~SNe, which come from massive
stars and are intimately connected to the recent SFR, the rate-size
relation might be explained by the correlation between the SSFR and
the galaxy mass recently reported by Noeske et al. (2007a, 2007b),
Salim et al. (2007), and Schiminovich et al. (2007). Using the
ultraviolet-optical colour-magnitude diagram in conjunction with
spectroscopic and photometric measurements derived from the Sloan
Digital Sky Survey spectroscopic sample, Schiminovich et al. (2007)
studied the physical properties of the galaxies as a function of SSFR
and stellar mass. As demonstrated in the rightmost panel of their
Figure 7, the SSFR of the galaxies has an apparent dependence on the
stellar mass of the galaxies, with SFR/$M \propto M^{-0.36}$ for
star-forming galaxies, and SFR/$M \propto M^{-0.16}$ for
non-star-forming galaxies. The main cause of this correlation is
likely the higher gas mass fractions and surface densities in the
low-mass galaxies.  

The average SSFR for all of the galaxies, weighted by the intensity of
the contour map (Table 3 of Schiminovich et al.), is shown in Figure
16 as a function of galaxy mass. Due to the mix of the star-forming
and non-star-forming galaxies and their loci on the SFR/$M$ vs. $M$
diagram, the average SSFR for all of the galaxies is proportional to
$M^{-0.55 \pm 0.09}$, as shown by the solid line in Figure 16. Note
that our CC~SN SNuM rate is proportional to $M^{-0.55 \pm 0.10}$
(Table 2).  The correlations thus have an essentially identical
dependence on galaxy mass, indicating the consistency of these two
tracers of star-formation activity.

While the SN~Ia SNuM rate shows a dependence of $M^{-0.50 \pm 0.10}$,
similar to the correlation between the SSFR and galaxy mass, a link
between the two correlations is more difficult to understand. First,
SNe~Ia are believed to come from the thermonuclear explosion of a
white dwarf in a binary system, so they are often associated with the
old population of their host galaxies, although recently a component
of SNe~Ia that is associated with the intermediate-age population, or
perhaps even the young/star-forming population, has been proposed
(i.e., the ``prompt" component in the SN~Ia rates; see, however, the
discussion in the next section). Still, a direct link between the
SN~Ia rate and the SSFR is not to be expected. Rather, the DTD needs
to be considered. 

Perhaps more troubling is the fact that the SN~Ia rates in the E--S0
galaxies, or even in the E galaxies, show the same rate-size
correlation as in the spiral galaxies. While there is some
observational evidence for a widespread, low-level presence of star
formation in the early-type E and S0 galaxies (see Mannucci et
al. 2008, and the references therein), the SNe~Ia in these early-type
galaxies should be dominated by the ``tardy" component (the component
that is associated with the old population), as demonstrated in the
next section. We further argue against the influence of the SSFR in
the early-type galaxies as being the main cause of the rate-size
relation, because the near-zero rate of CC~SNe in these galaxies
suggests that their SSFR is low.


Possible reasons for the rate-size relation for the SN~Ia rates are as
follows. (a) The DTD and the age of the stellar populations.  Maoz et
al. (2011, Paper IV in this series) developed a method to recover the
DTD for SNe~Ia, and found that the SN~Ia rate decreases monotonically
with the age of the stellar population, with the relatively ``young''
(age $< 420$~Myr) stellar populations having a rate that is at least
an order of magnitude higher than the ``old'' (age $> 2.4$~Gyr)
stellar populations. If smaller galaxies have a younger average age
for the stellar populations, they would have a higher rate. (b) The
probability of a white dwarf in a binary system exploding as a
SN~Ia. If the less massive galaxies affect the binary evolution of the
white dwarf in such a way as to boost the probability of a SN~Ia
explosion (due to metallicity or other factors), the SN~Ia rate can be
enhanced.  We consider reason (a) to be more likely, and reason (b) to
be a secondary, more speculative possibility.

We emphasise that the above discussion of the rate-size relation of
SNe~Ia hinges on the existence of the rate-size relation for the
galaxies having different Hubble types or $B - K$ colours. We note the
relatively large uncertainties in some of our RSS measurements due to
small-number statistics. For example, the significance of the
rate-size relation is only 1.6$\sigma$ for the SN~Ia SNuM rate in the
E galaxies. It is thus conceivable that the SN~Ia rates do {\it not}
depend on the mass of these galaxies, and that the rate-size relation
of the SN~Ia rates in the star-forming galaxies is indeed related to
the dependence of the SSFR on the galaxy mass. Sullivan et al.  (2006)
reported a nonlinear proportionality between the SN~Ia frequency and
the galaxy mass for the star-forming galaxies, with a RSS of $\sim
-0.30 \pm 0.08$, while for the non-star-forming (``passive") galaxies,
the SN~Ia frequency is consistent with a linear relation with the
galaxy mass (i.e., no RSS is required). While the discrepency between
our results and those reported by Sullivan et al. (2006) does not have
high significance due to the large uncertainties involved in both
studies, the different results nonetheless highlight the need to
further increase the sample sizes and reduce the uncertainties of the
RSSs.

We note that Sullivan et al. used the SFR to split the galaxies into
different bins, while we use the galaxy Hubble types and colours.  Even
though the galaxy Hubble type or colour sequence reflects a sequence in
the SFR, there is not a one-to-one association.  As discussed in \S 2.1 
and \S 2.2, the rate-size relation shows a dependence on how the
galaxies are grouped to calculate the rates: the RSSs depend on the
normalisation for the Hubble-type rates, while they are insensitive to
the normalisation for the $B - K$ colour rates. The different behaviour
of the rate-size relation with the two different grouping methods for
the galaxies leaves the possibility that the rate-size relation may
not be needed for certain galaxy grouping methods.  We plan to perform
a rate calculation using the SFR for the galaxies in a future paper,
when the SFRs for the LOSS sample galaxies are derived.  One test, for
example, is to investigate whether there is a rate-size relation when
the galaxies are binned according to the SSFR.  The expectation is 
that the rate-size relation should still be present if it is universal 
and not related to the SSFR and galaxy-mass relation.  Otherwise, no 
such rate-size relation should be present.

We also note that while studying the SN rates in galaxies of different
SFRs is a different and valuable approach, our measurements for
galaxies of different Hubble types and $B - K$ colours have their own
merits.  In particular, the Hubble type and $B - K$ colour of a galaxy
are {\it observed} quantities and widely available for the nearby
galaxies, while the SFR of a galaxy is an {\it inferred} quantity
based on synthetic models of the integrated broad-band fluxes or
spectra. Moreover, when SFR measurements are derived from spectral
energy distribution (SED) fitting, a strong degeneracy with dust
extinction (which, in general, is relatively poorly known) is usually
found, reducing the precision of the derived values. The SFR
measurements of the nearby galaxies also suffer from the difficulty of
properly measuring fluxes of extended objects (especially when there
are Galactic stars along their lines of sight), and could introduce
systematic uncertainties into the rate calculations. The situation is
improved at moderate to high redshift, where galaxies become more like
a point source and Galactic contamination is minimal, so photometry
can be more accurately conducted and modeled.

\subsection{The Two-Component Model for the SN Ia Rates} 

Based on the fact that the SN~Ia rate per unit mass (SNuM) in
late-type or blue galaxies is approximately an order of magnitude
higher than in early-type or red galaxies, a trend similar to that
seen for CC~SNe, M05 and Scannapieco \& Bildsten (2005) suggested that
the overall SN~Ia rate could be described as the sum of two
components.  One, denoted the ``young" or ``prompt" component, is
proportional to the ongoing SFR (and thus to the CC~SN SNuM) and has a
relatively short DTD. The other, called the ``old" or ``tardy"
component, is proportional to the mass of the galaxies (and thus a
constant SNuM) and has a relatively long DTD.  The so-called
``two-component model" for the SN~Ia rates, with its limitation as a
simplified analytic model, has been discussed in numerous subsequent
studies of SN rates (e.g., Neill et al. 2006; Sullivan et al. 2006;
Mannucci et al. 2006; Dahlen et al. 2008; Pritchet et al.  2008; see
also Bartunov, Tsvetkov, \& Filimonova 1994, who over a decade earlier
found that SNe~Ia occur in spiral arms with a frequency similar to
that of SNe~II).\footnote{Note that the idea that some SNe~I come from
  a relatively young stellar population was first proposed long ago by
  Dallaporta (1973) and Oemler \& Tinsley (1979). However, at the time
  these papers were published, SNe~Ib and SNe~Ic were still not
  recognized as separate classes from SNe~Ia, so there was potential
  contamination of the SN~Ia sample by SNe~Ibc. }

As discussed in \S 4.2, the rates used to derive the two-component
model by M05 and Scannapieco \& Bildsten (2005) have not been
corrected for the rate-size relation, and thus are for the average
galaxy sizes. We perform a similar analysis and display the results in
the left panel of Figure 17. We also apply the model to the rates for
the fiducial galaxy (i.e., after the rate-size relation is
considered); the results are shown in the right-hand panel. For both
cases, we confirm that the SN~Ia rates in galaxies of different $B -
K$ colours (the solid circles) can be well fit by a constant plus a
fraction of the CC~SN rate (the dashed line; the error bars of the fit
are not shown but are comparable to those of the SN~Ia rates), as
follows:

\begin{equation}
\begin{array}{ll}
 {\rm SNuM(Ia)} = &(0.036 \pm 0.022) \\
  & + \,\, (0.220 \pm 0.067)\, {\rm SNuM(CC)},~~{\rm and}
 \end{array}
\end{equation} 

\begin{equation}
\begin{array}{ll}
 {\rm SNuM(Ia,} M_0) = &(0.046 \pm 0.019) \\
  & +\,\,  (0.248 \pm 0.071) \, {\rm SNuM(CC,} M_0).
 \end{array}
\end{equation} 
\noindent
Compared with the fit parameters reported by M05, the constant (the
tardy component) is in good agreement, while the fraction of the
CC~SN rate (the prompt component) is somewhat different.  Our
fractions (0.220 $\pm$ 0.067, 0.248 $\pm$ 0.071) are smaller than
those reported by M05 (0.35 $\pm$ 0.08), but the differences are not
significant once the uncertainties are considered.

To further investigate the correlation between the SN~Ia and CC~SN
rates, we adopt an approach to visualise the correlation without using
the galaxy Hubble type or colour as the platform.  We first create
$(X,Y)$ = (CC~SN rate, SN~Ia rate) pairs for the galaxies with the
same Hubble type or colour range, and then fit a linear correlation $Y
= a + bX$ to quantify the coefficients and the significance of the
correlation. For each correlation, we also calculate the $\chi^2$/DOF
for a constant fit to the SN~Ia rates (i.e., no correlation with the
CC~SN rate).

We demonstrate how the two-component model for the SN~Ia rates is
affected by the choices of the RSSs and the sizes of the galaxies in
Table 8 and Figure 18. The rates in galaxies of different $B - K$
colours are used to construct the $(X,Y)$ pairs. The first three
entries of Table 8 and the top panel of Figure 18 show the results for
the different RSSs for the SN~Ia rates (the fiducial RSS and its
1$\sigma$ upper and lower errors). For the CC~SN rates, the RSS is
fixed at the adopted fiducial value ($-$0.38). One can see that the
choice of the RSS has a significant effect on the two-component
model. As the RSS for the SN~Ia rates becomes bigger, the correlation
between the SN~Ia rates and the CC~SN rates becomes weaker, as
indicated by the larger tardy component (``$a$"), the smaller
coefficient and significance for the CC~SN rate fraction (``$b$"), and
the smaller reduced $\chi^2$ for a constant fit.

The last three entries of Table 8 and the lower panel of Figure 18
show the correlation for the galaxies with different sizes. When the
galaxy size becomes bigger, the significance of the correlation does
not change (as indicated by the same reduced $\chi^2$ for a constant
fit).  However, the tardy component becomes smaller, and the
coefficient for the CC~SN rate fraction becomes bigger. This can be
understood by multiplying Eq. (8) by $(M/M_0)^{-0.25}$, which yields
the following:

\begin{equation}
\begin{array}{ll}
 {\rm SNuM(Ia,} M) = &0.046\,(M/M_0)^{-0.25} \\
   & +\,\,  0.248\,M^{0.13} \,{\rm SNuM(CC,} M).
\end{array}
\end{equation}
\noindent
In other words, the tardy component varies with galaxy mass because of
the rate-size relation, while the CC~SN rate fraction changes with
galaxy mass because the RSSs for the SN~Ia and CC~SN rates are
different (by 0.13, though with a low significance level).

We have investigated how the two-component model fit results are
affected by different choices of parameters in the rate calculations,
such as the normalisation (SNuB, SNuK, or SNuM), the RSSs (with or
without), and the construction of the rate $(X,Y)$ pairs (using rates
in the different Hubble types or $ B - K$ colours).  The results are
listed in Table 9 and plotted in Figure 19. For Table 9, Column 1
(``Src") shows how the rate $(X,Y)$ pairs are constructed: ``H-type"
means the rates in the different Hubble types are used, while ``$B -
K$" means the rates in the different $B - K$ colours are used. Column
2 (``Rate") shows the normalisation.  The next two blocks of columns
show the fit parameters for the model, with and without RSSs.

Inspection of Table 9 and Figure 19 reveals the following. 

\begin{enumerate}

\item{ The normalisation has a rather significant effect. Using SNuB,
  for example, yields a reverse trend as expected from the
  two-component model (the SN~Ia rate decreases with increasing CC~SN
  rate), although with a low significance level as indicated by the
  small reduced $\chi^2$ for a constant fit. This fact serves as a
  reminder that we have not yet found an ideal normalisation to
  measure the rates for all types of SNe. The blue luminosity, for
  example, is dominated by contributions from very massive stars (a
  small minority of all stars), and is thus a relatively poor gauge of
  the stellar population responsible for the production of SNe~Ia;
  few, if any, SNe~Ia arise from stars having $M \apgt
  8$\,M$_\odot$. The $K$-band luminosity, on the other hand, can be
  used to derive the mass, especially in conjunction with the $B - K$
  colours; it arises from a combination of both young and old
  populations. The discussion of the two-component model should take
  into account the limitations of our current knowledge of the ideal
  normalisation, and the associated pitfalls. It is likely, for
  example, that the rate cannot be quantified by a single
  normalisation parameter, as witnessed by the existence of the
  rate-size relation. Other more subtle effects such as environmental
  influences (metallicity, active galactic nuclei, radio jets, etc.)
  may become more obvious in future studies with larger and more
  complete samples. The SNuB correlations will {\it not} be considered
  hereafter unless explicitly expressed otherwise. }

\item{The rate-size relation affects the results of the two-component
  model fit.  The correlation between the SN~Ia rates and the CC~SN
  rates in general becomes weaker after the rate-size correction is
  applied, as indicated by the fit parameters: the tardy component
  becomes larger (comparing ``$a_2$" to ``$a_1$" in Table 9), the CC~SN
  rate fraction and significance become smaller (compare ``$b_2$" to
  ``$b_1$"), and the reduced $\chi^2$ for a constant fit becomes
  smaller (compare ``$\chi^2(c)_2$" to ``$\chi^2(c)_1$").}


\item{The construction of the rate $(X,Y)$ pairs has a significant
  effect on the two-component model, suggesting that there is not a
  one-to-one correlation between the SN~Ia and the CC~SN rates. For
  the cases both with and without the rate-size corrections, the rates
  from the galaxy $B - K$ colours display a more significant
  correlation than the rates from the galaxy Hubble types. In
  particular, we note that after the rate-size relation is considered,
  the SN~Ia rates in different Hubble types are consistent with being
  a constant (i.e., no correlation with the CC~SN rates), as discussed
  in \S 3.5 and demonstrated by the small $\chi^2(c)_1$ value in Table
  9.  This is disconcerting, and suggests that there could be no
  correlation or a strong correlation, depending on how the galaxies
  are grouped to calculate the rates. It is of course dangerous to
  define, {\it a posteriori},``optimal" ways to group the galaxies for
  the purpose of the two-component model analysis. }

\end{enumerate}

To further explore the correlation between the SN~Ia and the CC~SN
rates, we attempt to split the SN~Ia rates into two components: the
contribution from (1) old and (2) young {\it stellar populations} in
galaxies (hereafter, the ``old-s" and ``young-s" components,
respectively).  Note that this approach is different from the
two-component model for the SN~Ia rates by M05 and Scannapieco \&
Bildsten (2005), where the SNe~Ia are split into an
old/tardy and a young/prompt component.  In other words, the old/tardy
component in the two-component model is proportional to {\it the total
  mass} of a galaxy, while the ``old-s" component in our approach is
proportional to {\it the mass of the old stellar population} in a
galaxy.  It is generally accepted that early-type galaxies (E/S0) are
predominantly made of old stellar populations, while late-type
galaxies (Sc/Scd) consist of mostly young stellar populations, so we
adopt a toy model in which the fraction of the ``old-s" SN~Ia
component decreases from 100\% in E galaxies, to 83.3\% in S0, 66.7\%
in Sab, 50.0\% in Sb, 33.3\% in Sbc, 16.7\% in Sc, and 0\% in Scd
galaxies\footnote{Thus, our toy model naively assumes that the number
  sequence 1--7 for the E--Scd galaxies represents a linear decrease
  of the fraction of the old stellar population.}. Our goal is to
study whether there is a significant correlation between the
``young-s" SN~Ia rate and the CC~SN rate.

Figure 20 shows the results for the SNuM rates for a fiducial galaxy.
The left panel shows the conventional (M05; Scannapieco \& Bildsten
2005) two-component model fit (dashed line) for the total SN~Ia rates
(solid dots) as follows:

\begin{equation}
\begin{array}{ll}
 {\rm SNuM(Ia,} M_0) = &(0.116 \pm 0.012) \\
  & +\,\,  (0.051 \pm 0.032) \, {\rm SNuM(CC,} M_0).
 \end{array}
\end{equation}
\noindent
As discussed above, the SN~Ia rates can be well fit by a constant, and
the correlation with the CC~SN rates is not significant (at only the
$\sim 1.5\sigma$ level). Also shown in the panel is our adopted
``old-s" SN~Ia component (dash-dotted line): it accounts for 100\% of
the SNe~Ia in E galaxies and 0\% in Scd galaxies. The ``young-s"
component, the difference between the total rate and the ``old-s"
component, is plotted in the right-hand panel, together with a
two-component model fit as follows:

\begin{equation}
\begin{array}{ll}
 {\rm SNuM[Ia(young),} M_0] = &(0.001 \pm 0.005) \\
  & +\,\,  (0.187 \pm 0.027) \\
  & {\rm SNuM(CC,} M_0).
 \end{array}
\end{equation} 
\noindent
Not surprisingly, the tardy component is consistent with being
zero. There is also a strong correlation between the ``young-s"
component of the SN~Ia rate and the CC~SN rate (at the $\sim 7\sigma$
level).

This exercise suggests that the fundamental idea of the two-component
model for the SN~Ia rates as proposed by M05 and Scannapieco \&
Bildsten (2005) is correct. The only required modification to the
model is that the ``tardy/delayed" component is related to the mass of
the old stellar population, rather than to the total mass, of the
galaxies. However, we caution that the treatment of the young/old
stellar populations in our toy model is {\it ad hoc}; there is clear
observational evidence indicating that early-type galaxies do harbor
some young stellar populations, and that late-type galaxies do also
contain an old component (Mannucci et al. 2008, and references
therein).

A more sophisticated analysis will only become possible if methods are
developed to properly reconstruct the age distributions of stellar
populations in galaxies and identify the SNe~Ia associated with
different populations (Neill et al. 2009; Brandt et al. 2010; Maoz et
al. 2011). While we have some clues (e.g., SN 1991bg-like objects
probably come from an old stellar population while SN 1991T-like
objects from a young population), we do not have a clear picture for
normal SNe~Ia, which are two-thirds of the total SN~Ia population
(Paper II) and occur in galaxies of all Hubble types. It is thus
impossible to directly measure the SN~Ia rates in stellar populations
of different ages. Brandt et al. (2010) and Maoz et al. (2011)
developed a recovery method to constrain the DTD of SNe~Ia in
different stellar populations.  In particular, Maoz et al. (2011)
found evidence for a population of SNe~Ia in both ``young'' (age $<
420$~Myr) and ``old'' (age $>2.4$~Gyr) stellar populations, which they
called the ``prompt" and ``delayed" components.  We note, however,
that their ``delayed" component refers to SNe~Ia that occur in old
(age $>2.4$~Gyr) stellar populations, so in essence it is the ``old-s"
component we discussed above, not the delayed component discussed in
the original two-component model of M05 and Scannapieco \& Bildsten
(2005), which is proportional to the total mass (including young
stellar populations) of a galaxy.


In summary, the correlation between the SN~Ia and the CC~SN rates is
affected by the normalisation and the way the galaxies are
grouped. Whether there is a {\it physical connection} between the
rates hinges on finding the ideal normalisation and an optimal way to
group the galaxies. It is also found that the rate-size relation plays
a significant role in the two-component model. While the cause of the
rate-size relation is not clear (see the discussion in the previous
section), the fact remains that for galaxies having the same size, the
correlation between the SN~Ia and CC~SN rates becomes rather weak
(e.g., for the rates from different $B - K$ colours), or nonexistent
(e.g., for the rates from different Hubble types). We also find that
the SN~Ia rate for the young stellar population in galaxies might have
a significant correlation with the CC~SN rate even after applying the
rate-size corrections.


While recent studies provide indisputable evidence of a ``weak"
bimodality (Mannucci 2008) --- that SNe~Ia come from stellar
populations that are both young and old (e.g., M05; Maoz et al. 2011)
--- it is unclear whether the two progenitor populations are well
separated (the so-called ``strong" bimodality; e.g., Mannucci et
al. 2006; Scannapieco \& Bildsten 2005; our toy model above) or form a
continuous distribution. Models of binary-star evolution exists that
produce both bimodalities (e.g., Greggio 2005; Hachisu et al. 2008;
Lipunov et al. 2010).  Due to uncertainties in the two-component
models, fitting the SN~Ia rate evolution with redshift based on the
two-component models and the star-formation history becomes highly
uncertain until we understand the origin of the rate-size relation and
properly parameterise the two-component models.

\subsection{The Volumetric SN Rates}

Supernova rates at different redshifts provide important information
on the evolution of a number of physical processes over cosmic time,
such as the cosmic SFR and the DTD for the explosion of SNe~Ia. All of
the published rates at moderate to high redshifts are expressed as a
volumetric rate in units of SNe Mpc$^{-3}$ yr$^{-1}$; thus, in this
section, we attempt to derive a volumetric rate in the local Universe
from our dataset.

As discussed in Paper I, our galaxy sample is not complete, even for
the very nearby volume within $D < 60$~Mpc, so we cannot directly
measure a volumetric rate using the control-time method. To convert
our rates for galaxies of different Hubble types into a volumetric
rate, we require knowledge of the local luminosity density for
galaxies of different Hubble types. A further complication is the
presence of the rate-size relation; we need to know the distribution
of the sizes for the galaxies (i.e., the galaxy luminosity function).

Unfortunately, our combined knowledge of the galaxy luminosity
function and local density for different Hubble types and colours is
still rather limited. We were only able to find a complete set of
measurements in the literature with galaxies split into broad
early-type and late-type bins. In our calculation, we make use of the
$K$-band galaxy luminosity function and density published by Kochanek
et al. (2001). In particular, we adopt the standard model in their
Table 3 for the early-type and late-type galaxies. With our adopted
Hubble constant, this means local $K$-band luminosity densities of
$j_{\rm early} = (2.25 \pm 0.36) \times 10^8\, {\rm L}_\odot$
Mpc$^{-3}$ and $j_{\rm late} = (2.96 \pm 0.42) \times 10^8\, {\rm
  L}_\odot$ Mpc$^{-3}$.

Our volumetric rates are derived with the following steps. We first
calculate SNuK for a fiducial galaxy for the early-type (E-S0) and
late-type (Sa--Irr) galaxies using the ``full-optimal" SN sample.
These rates are reported in the first two rows in Table 10 for the
different SN types\footnote{We note that our definition of the
  early-type and late-type galaxies is somewhat different from that
  adopted by Kochanek et al. (2001), but the ratio of the integrated
  total $K$-band luminosity between the early- and late-type galaxies
  in our sample, 1.27, is consistent with that reported by Kochanek et
  al. ($1.17 \pm 0.12$). }.  The Kochanek et al. (2001) luminosity
functions for the early- and late-type galaxies are then used to
derive the number distribution for the galaxies having different
luminosities, and the RSSs as reported in Table 2 are used to
calculate the rates for different luminosities according to the
rate-size relation. The average values of SNuK, weighted by the number
distributions of the LFs, are reported in the third and forth rows in
Table 10. These SNuK values are multiplied by the corresponding
luminosity densities as reported above, and the contributions from the
early- and late-type galaxies are summed to yield the final volumetric
rates reported in the last row of Table 10.

The evolution of the volumetric rate versus redshift (up to $z \approx
0.5$) for SNe~Ia is shown in Figure 21. The published rates (all
converted to our adopted Hubble constant) include those of C99 (Botticella et al. 2008), Hardin
et al. (2000), Madgwick et al. (2003), Tonry et al. (2003), Blanc et
al. (2004), Dahlen et al. (2004), Barris \& Tonry (2006), Neill et
al. (2006), Neill et al. (2007), Dilday et al. (2008), Botticella et
al. (2008), and Horesh et al. (2008). Our volumetric rate is plotted
with the statistical error only (half-solid circle, displaced for
clarity at $z = -0.01$), and then with the statistical error and
systematic error added in quadrature (solid circle).  We note that our
measurement with the total uncertainty has roughly the same precision
as some of the other measurements, despite the fact that we have used
more SNe in our calculations. The explanation is twofold.  First, the
precision of our volumetric rate is limited by the precision of the
local luminosity density. Second, we take an aggressive approach to
calculating the systematic errors (as discussed in \S 3.4), and hence
may overestimate the total errors. As can be seen, our measurement
including only the statistical error is the most precise among all the
points.

Our rate is consistent with the C99 measurement at the same redshift
to within the uncertainties\footnote{Note that the differences between
  our volumetric rates and the C99 measurements are caused by a
  combination of several factors: the difference in the rate numbers,
  the use of RSSs and galaxy LFs in our calculation (see also Mannucci
  et al. 2006), and the difference in the adopted luminosity density
  for the galaxies.}.  The SN~Ia rates are consistent with being a
constant from $z = 0$ to $\sim0.3$, followed by a rise toward higher
redshifts; however, a gentle rising behaviour from $z = 0$ to 0.5
cannot be ruled out, as indicated by the lower dash-dotted line, which
is evaluated at the 1$\sigma$ lower error bar of the C99 measurement
and follows a rate $\propto (1 + z)^{3.6}$, a functional form that is
the same as the derived SFR history from Hopkins \& Beacom (2006). The
dashed and the upper dash-dotted lines follow the same functional form
but are evaluated at our measurement, and at the 1$\sigma$ upper error
bar of our measurement, respectively, and they do not provide a
satisfactory fit to the ensemble of the measurements. We also plot the
expected SN~Ia rate from the SFR history study by Mannucci et
al. (2007; dotted line).  Detailed discussions of the redshift
evolution of the SN~Ia rate, the comparison to SFR history, and
constraints on the DTD are beyond the scope of this paper.

Figure 22 shows the redshift evolution of the volumetric rates of the
CC~SNe.  The published rates (all converted to our adopted value of
the Hubble constant) include C99, Dahlen et al. (2004), Cappellaro et
al. (2005), Botticella et al. (2008), and Bazin et al. (2009).  Our
volumetric rate, obtained by summing the SN~Ibc and SN~II rates in
Table 10, is plotted with the statistical error only (half-solid circle,
displaced at $z = -0.03$ for clarity), and then with the total error
(solid circle). Our rate is consistent with the C99 measurement at the
same redshift to within uncertainties, though our number is nominally
higher (by 65\%).  The dashed line again gives a rate $\propto (1 +
z)^{3.6}$, while the dotted line follows the SFR history from Mannucci
et al. (2007).  Both curves are evaluated at our measurement and offer
excellent fits to all of the published results. Taken at face value,
it would appear that the CC~SN rate closely follows the SFR history,
though we caution that most of the measurements have rather large
uncertainties. Detailed discussions of the CC~SN rate redshift
evolution and the various renditions of the SFR are beyond the scope
of this paper.

We further note that to investigate the rate evolution at different
redshifts, one needs to consider potential galaxy-size evolution
(i.e., a luminosity function change) at different redshifts because of
the rate-size relation.

\subsection{The SN Rates in the Milky Way}

From our measured rates reported in Table 4, we can determine the
expected SN rates in the Milky Way Galaxy (MW, hereafter) and compare
them with values obtained from other sources.  To achieve this, we
require knowledge of the size and the Hubble type of the MW.  We
assume the MW to be of Hubble type Sbc (e.g., van den Bergh \& McClure
1994).  The total $B$-band luminosity of the MW is quite uncertain; we
adopt $(2.0 \pm 0.6) \times 10^{10}$\,L$_\odot$ (van der Kruit 1987)
and $(2.6 \pm 0.6) \times 10^{10}$\,L$_\odot$ (van den Bergh
1988). Alternatively, we can assume that the MW has a size similar to
that of the Andromeda galaxy (M31), as they are the largest galaxies
in the Local Group and generally thought to be similar in many
ways. For M31, the $B$-band magnitude, 3.36, is adopted from RC3 and
is corrected for both internal (due to inclination) and Milky Way
extinction. The $K$-band magnitude (0.875) is adopted from the 2MASS
extended source catalog and corrected for extinction as well. The
distance to M31 ($D = 0.778 \pm 0.017$ Mpc) is calculated from 17
Cepheid measurements archived in NED. Finally, we assume that the MW
has the average size of the Sbc galaxies in the ``optimal" LOSS galaxy
sample.

Table 11 lists all of our rate estimates (in SNe per century). The
rates for a fiducial Sbc galaxy in Table 4 are corrected by the
rate-size relation according to the size of the MW (with the RSSs in
Table 2).  The uncertainties for the individual measurements are not
reported, as they are much smaller than the scatter among the
different measurements.  The average rates are given in the last row
together with the 1$\sigma$ scatter. Considering the uncertainties for
the Hubble type and size of the MW, these rates may have a systematic
uncertainty of a factor of $\sim 2$.  In particular, we note that the
MW rate in SNe per century is proportional to size$^{1 + {\rm RSS}}$
(\S 2.2), so even if the MW size is off by a factor of 10, the SN rate
is erroneous only by a factor of 2.8--5.9 (depending on the SN type
and the normalisation).

Our fiducial estimate of $2.84 \pm 0.60$ SNe per century is in good
agreement with published results of 1.4--5.8 SNe per century based on
different techniques, including direct star counting, pulsar birth
rates, the number of radio SN remnants, and historical SN records (van
den Bergh 1991; van den Bergh \& Tammann 1991; Cappellaro et al. 1993;
van den Bergh \& McClure 1994).  Our CC~SN rate estimate of $2.30 \pm
0.48$ SNe per century is also consistent with published values
(1.9--2.6 per century) based on observations of gamma-ray emission
from radioactive $^{26}$Al within the MW (Timmes et al. 1997; Diehl et
al. 2006).

\subsection{The Rate Ratio as a Function of Galaxy Mass} 

We report the rate-size relation in \S 2.2 and offer more discussion
of the possible causes in \S 4.2. One interesting question is whether
the relative fractions of SNe also change with galaxy size.  To
investigate this, we divide the galaxies into different size bins for
the ``full-optimal" sample, calculate the SNuM rate ratios relative to
the CC~SN rates, and show them in Figure 23. We have included only
spiral galaxies (Hubble type = 3--7) in this analysis because CC~SN
rates are negligible in E/S0 galaxies and very uncertain in Irr
galaxies.  Only the statistical errors are used to derive the ratios.

The ratio of the SN~Ia to the CC~SN rates shows a marginal ($\sim
1.5\sigma$) trend from the least to the most massive galaxies. The
SN~Ia rate is about 25\% of the CC~SN rate when the galaxy mass is
smaller than $\sim 3 \times 10^{10}$\,L$_\odot$, and about 40\% for
the large galaxies. There are two likely causes for this trend.  (a)
The result of the different RSSs in the rate-size relation for the
SNe~Ia and the CC~SNe. As SNuM(Ia) $\propto M^{-0.50}$ and SNuM(CC)
$\propto M^{-0.55}$, the ratio SNuM(Ia)/SNuM(CC) should increase with
mass in proportion to $M^{0.05}$. (b) More massive galaxies have, on
average, an earlier spiral Hubble type. As SNuM(Ia) is nearly constant
in different galaxy Hubble types, and SNuM(CC) decreases significantly
in earlier spiral galaxies, the SNuM(Ia)/SNuM(CC) ratio should
increase in earlier spiral or higher mass galaxies. Because of the
uncertainties of the ratios, it is difficult to disentangle the
relative contributions of these two causes.

The ratio of the SN~Ibc to the overall CC~SN rates is $\sim 35$\% for
galaxies with $M > 1.0 \times 10^{10}$\,L$_\odot$, and then declines
to $\sim 10$\% at $M = 0.15 \times 10^{10}$\,L$_\odot$. In other
words, for the least massive galaxies, there are fewer SNe~Ibc
relative to the total population of CC~SNe. This is consistent with
the host-galaxy properties of the LF SNe as discussed in Paper II,
where we found that the SN~Ibc hosts are skewed toward more massive
galaxies than the SN~II hosts, possibly indicating a metallicity
effect. However, we note again that the trend is of low significance
($\sim 2\sigma$). Arcavi et al. (2010) reported that the SN~Ibc
fraction among the CC~SNe does not change in dwarf galaxies, though
with small-number statistics.

\section{Conclusions and Future Improvements} 

This is the third installment of a series of papers aiming to derive
the SN rates in the local Universe from the Lick Observatory SN Search
(LOSS).  The goal of this paper is to put together all of the
ingredients from Papers I (Leaman et al. 2011) and II (Li et al. 2011)
to derive the final rates for the most common types of SNe: Ia, Ibc,
and II.  We refer readers to Paper I for an outline of the series, and
a detailed list of improvements of our rate calculation over past
work. Section 1.1 of the current paper also has a short summary of the
first two papers in the series. The conclusions that are specific to
this paper are as follows.

\begin{enumerate}

\item{The control-time calculations for the galaxies are done
  numerically due to the combined complexity of the SN light curves,
  the limiting magnitudes, and the detection-efficiency curves. For
  each SN type, the luminosity functions are considered separately for
  two broad Hubble-type bins. }

\item{ SN rates are traditionally expressed in units which are
  linearly normalised by the host-galaxy mass or luminosity
  (SNuM/SNuB/SNuK), but we find this to be an inadequate description
  of the data. The rates calculated in SNu units all demonstrate a
  correlation with galaxy sizes such that galaxies of smaller mass or
  luminosity have higher SN rates.  As the result of this rate-size
  relation, our rates are derived for galaxies having a fiducial size,
  and correction factors with a rate-size power-law slope (RSS) are
  used to evaluate the rates for any given galaxy size. Another
  implication is that the SN frequency (SNe per year) for a galaxy is
  not linearly proportional to its size, but rather to size$^{1 + {\rm
  RSS}}$.}

\item{The RSSs are found to have a strong dependence on the
  normalisation, the SN types, and the galaxy grouping methods.  The
  RSSs for the two types of core-collapse SNe (Ibc and II) are in
  general consistent with each other. No apparent dependence on the
  galaxy Hubble types or colours is found, but this may be due to the
  limitation of the precision of our RSS measurements.  }

\item{We have tested the robustness of our rate-calculation pipeline
  in several ways.  The SN rates using different SN subsamples are in
  general consistent with each other. No systematic trend is found for
  the rates in different distance bins, or in galaxies of different
  angular sizes. The SN~Ia rates are insensitive to the input SN
  luminosity function, but the core-collapse SNe are. When the SNe in
  the highly inclined galaxies ($i > 75^\circ$) are excluded from the
  rate calculations, the SN~Ia and SN~Ibc rates do not show a
  significant difference in the different inclination bins. The
  SNe~II, however, exhibit a potential difference in the late-type
  spiral galaxies or galaxies having different $B - K$ colours. No
  inclination correction factor is used in our calculation, and the
  implication on the uncertainty is discussed. }

\item{We use Poisson statistics to calculate the statistical
  uncertainty in the rates. For the systematic uncertainty, we
  consider several sources: the different SN samples, the different SN
  input luminosity functions, the uncertainties of the RSSs, the
  uncertainty caused by the treatment of the inclination correction
  factors, and a universal miscellaneous uncertainty. The systematic
  uncertainty is comparable to the statistical uncertainty in most
  cases ($\sim 20$\%), but can be as high as 80\% of the measurements
  for the SN~II rates in late-type spiral galaxies due to the
  uncertain inclination corrections.}

\item{The SN~Ia rate in a galaxy of the fiducial size with the
  $B$-band luminosity normalisation (SNuB) declines from the early- to
  the late-type galaxies, and from the red to the blue galaxies,
  likely due to the increasing influence of very massive stars in the
  total $B$-band luminosity in the blue, late-type galaxies. The
  core-collapse SN rates are small in the early-type (E--S0) and red
  galaxies, and increase toward late-type and blue galaxies. }

\item{For the rates in a galaxy of the fiducial size with the $K$-band
  luminosity or the mass as the normalisation (SNuK or SNuM), the
  SN~Ia rate is nearly constant among different Hubble types.  The
  core-collapse SN rates in general have an increasing trend from
  early-type to late-type, and red to blue, galaxies. However, the
  SN~Ibc rate may decline for the bluest galaxy bins or from Sc to
  Irr galaxies. }

\item{Our average SN rates for galaxies of different Hubble types or
  colours agree with the published results to within uncertainties when
  the rates are calculated in the same manner (in particular, without
  adopting the rate-size relation in our rate calculations).}

\item{While the rate-size relation for the core-collapse SNe may be
  linked to the connection between the specific star-formation rate
  and the galaxy sizes, it is not clear that such a link can be
  established for the SNe~Ia.  It is important to investigate whether
  the RSSs are universal in galaxies of different properties such as
  Hubble type, colours, and specific star-formation
  rates. Numerically, the rate-size relation indicates that the SN
  rates cannot be adequately parameterised by a single parameter using
  galaxy Hubble types or $B - K$ colours.}

\item{We attempt to fit the SN~Ia rates with the two-component model
  of Mannucci et al.  (2005) and Scannapieco \& Bildsten (2005). We
  find that the model is affected by the choice of the normalisation
  for the rates, the rate-size relation, and the way the galaxies are
  grouped (Hubble type or colour). There may not be a one-to-one
  correlation or physical connection between the SN~Ia and the
  core-collapse SN rates. The SN Ia rates in young stellar populations 
  may have a strong correlation with the core-collapse SN rates. }

\item{We derive a local volumetric rate of $0.301 \pm 0.062$, $0.258
  \pm 0.072$, and $0.447 \pm 0.139$ for SNe~Ia, Ibc, and II,
  respectively (in units of $10^{-4}$ SN Mpc$^{-3}$ yr$^{-1}$).  The
  uncertainties of these rates are dominated by the uncertainties in
  the galaxy luminosity density used to convert our per-galaxy rates to volumetric
  rates. }

\item{We derive a SN rate of $2.84 \pm 0.60$ per century for the Milky
  Way (to within a systematic factor of $\sim 2$, dominated by the
  uncertainty in the properties of the Galaxy), consistent with
  previous estimates.}

\item{The ratio of the SN~Ibc rate to the total core-collapse SN rate
  declines for the least massive galaxies, perhaps indicating a
  metallicity effect on the binary evolution of massive stars.}

\end{enumerate}

While the first three papers in this series conclude our investigation
for the rates of the most common SN types (Ia, Ibc, and II) in
galaxies of different Hubble types and $B - K$ colours, more analysis
is underway to determine our rates for the ``known unknowns" --- the
rare and peculiar transients and SNe in our search.  We also plan to
investigate the SN rates in additional categories of galaxies, such as
radio and other active galaxies, interacting galaxies, and cluster
versus field galaxies. In addition, we are in the process of deriving
more physical parameters for our sample galaxies, such as the
star-formation rates. One important study, for example, is to check
whether there is a rate-size relation when the galaxies are grouped
among different (specific) star-formation rates.

We are continuing our SN search in order to decrease the statistical
uncertainties. As discussed throughout this series, even though we
have a large number of SNe in the rate calculations, the measurements
of the rates and certain parameters will benefit from a even larger
sample of SNe. In particular, improved precision on the RSSs will
provide information on whether they are insensitive to the galaxy
properties and thus are universal, which in turn will constrain the
origin of the rate-size relation; improved precision on the rates will
help determine their dependence on the galaxy inclinations; more SNe
discovered in the Irr galaxies will improve the precision of the
rather poor measurements reported in this series. We plan to improve
our rate-calculation pipeline so that the rates may be easily updated
with new SN discoveries and monitoring history information.

\section*{Acknowledgments}

We thank the referee, Enrico Cappellaro, for useful comments and
suggestions which helped improve the paper.  We are grateful to the
many students, postdocs, and other collaborators who have contributed
to the Katzman Automatic Imaging Telescope and the Lick Observatory
Supernova Search over the past two decades, and to discussions
concerning the determination of supernova rates --- especially Ryan
Foley, Saurabh Jha, Maryam Modjaz, Frank Serduke, Jeffrey Silverman,
Nathan Smith, Thea Steele, and Richard Treffers.  Assaf Horesh
helped make Figure 21.  We thank the Lick Observatory staff for their
assistance with the operation of KAIT.  LOSS, conducted by A.V.F.'s
group, has been supported by many grants from the US National Science
Foundation (NSF; most recently AST-0607485 and AST-0908886), the
TABASGO Foundation, US Department of Energy SciDAC grant
DE-FC02-06ER41453, and US Department of Energy grant
DE-FG02-08ER41563. KAIT and its ongoing operation were made possible
by donations from Sun Microsystems, Inc., the Hewlett-Packard Company,
AutoScope Corporation, Lick Observatory, the NSF, the University of
California, the Sylvia \& Jim Katzman Foundation, and the TABASGO
Foundation.  We give particular thanks to Russell M. Genet, who made
KAIT possible with his initial special gift; former Lick Director
Joseph S. Miller, who allowed KAIT to be placed at Lick Observatory
and provided staff support; and the TABASGO Foundation, without which
this work would not have been completed.  J.L. is grateful for a
fellowship from the NASA Postdoctoral Program.  D.P. is supported by
an Einstein Fellowship.  X.W.  acknowledges NSFC grants (10673007,
11073013) and the China-973 Program 2009CB824800.  We made use of the
NASA/IPAC Extragalactic Database (NED), which is operated by the Jet
Propulsion Laboratory, California Institute of Technology, under
contract with NASA. We acknowledge use of the HyperLeda database
(http://leda.univ-lyon1.fr).

\newpage

\newpage

\begin{figure*}
\includegraphics[scale=1.0,angle=270,trim=0 80 0 0]{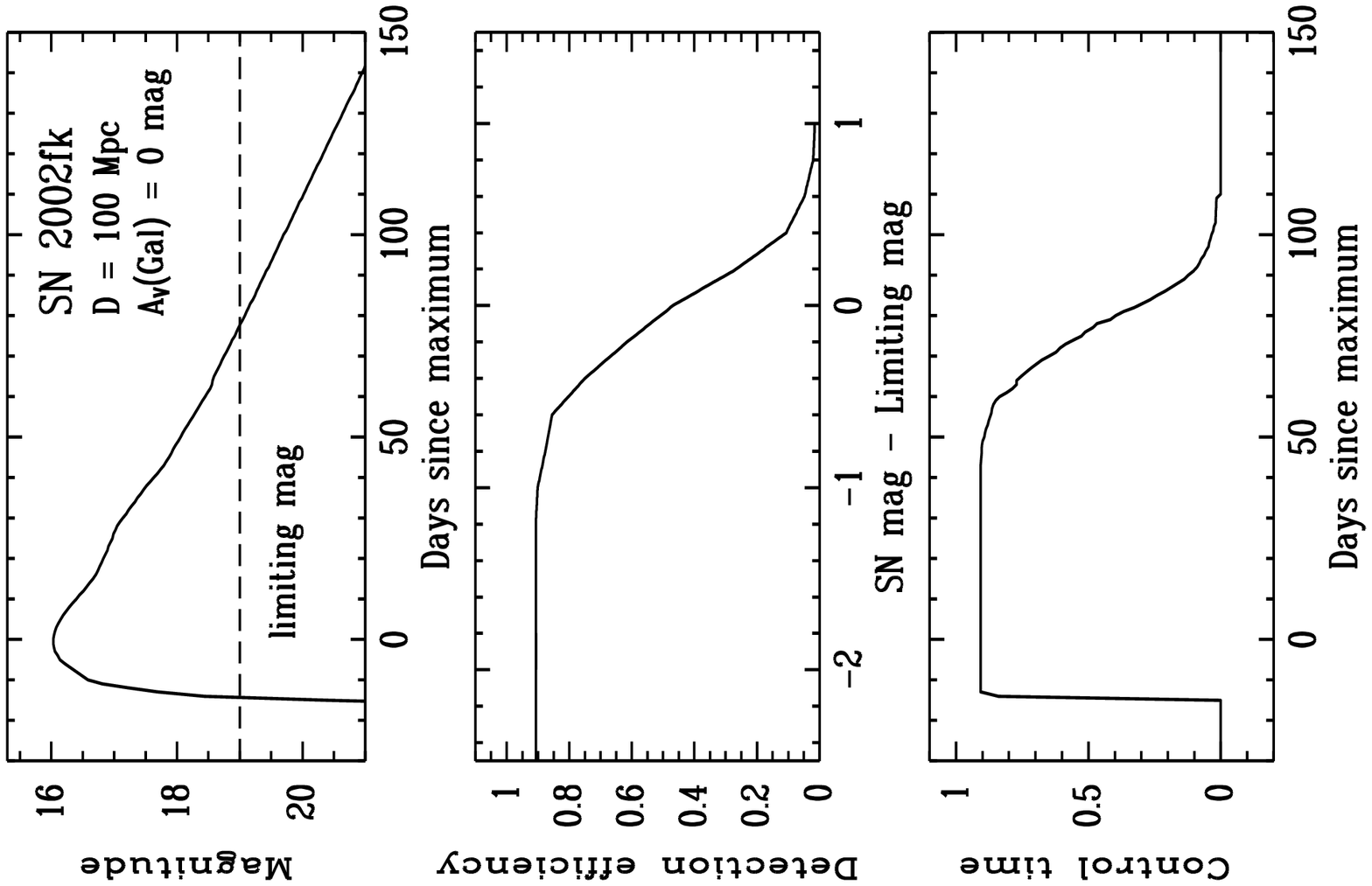}
\caption[] { An example of the control-time calculation. {\it Top
    panel:} The apparent light curve of SN 2002fk in a galaxy at
  100~Mpc, with no Milky Way extinction. The limiting magnitude (19)
  is marked by the dashed line. {\it Middle panel:} The
  detection-efficiency curve for the galaxy (assumed to be of type
  Sb--Sbc), as adopted from Paper I.  {\it Bottom panel:} The
  control-time curve. The curve generally has a rising, a constant,
  and a declining portion. }
\label{1}
\end{figure*}

\clearpage

\begin{figure*}
\includegraphics[scale=1.0,angle=270,trim=0 80 0 0]{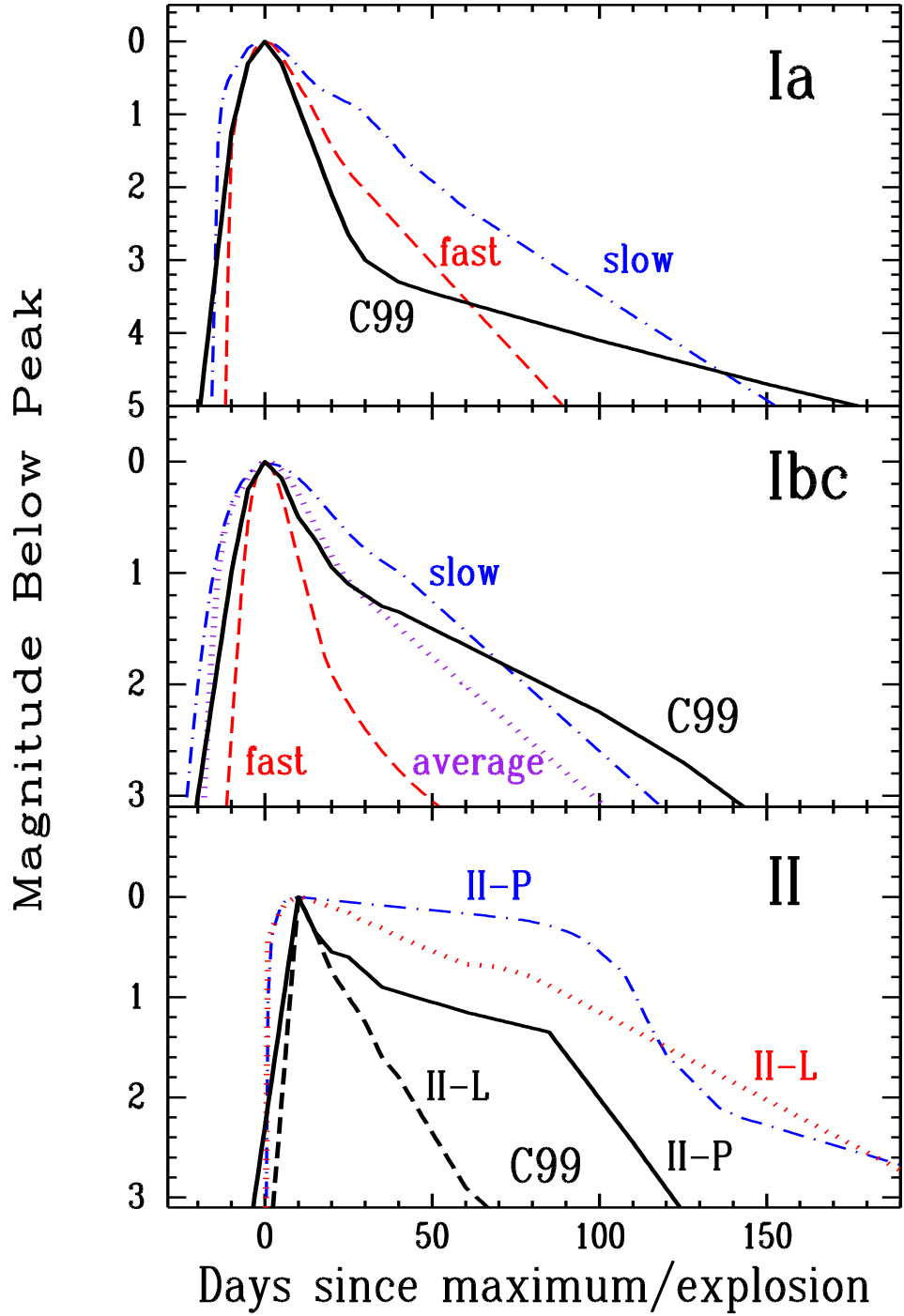}
\caption[] { A comparison of the light curves adopted in our rate
  calculation and those used by C99. The C99 light curves are the
  thick solid or dashed lines. There is a dramatic difference between
  the two light-curve sets due to the different passbands used in the
  surveys, and a direct comparison is not very meaningful. }
\label{2}
\end{figure*}

\clearpage

\begin{figure*}
\includegraphics[scale=0.7,angle=270,trim=0 0 0 0]{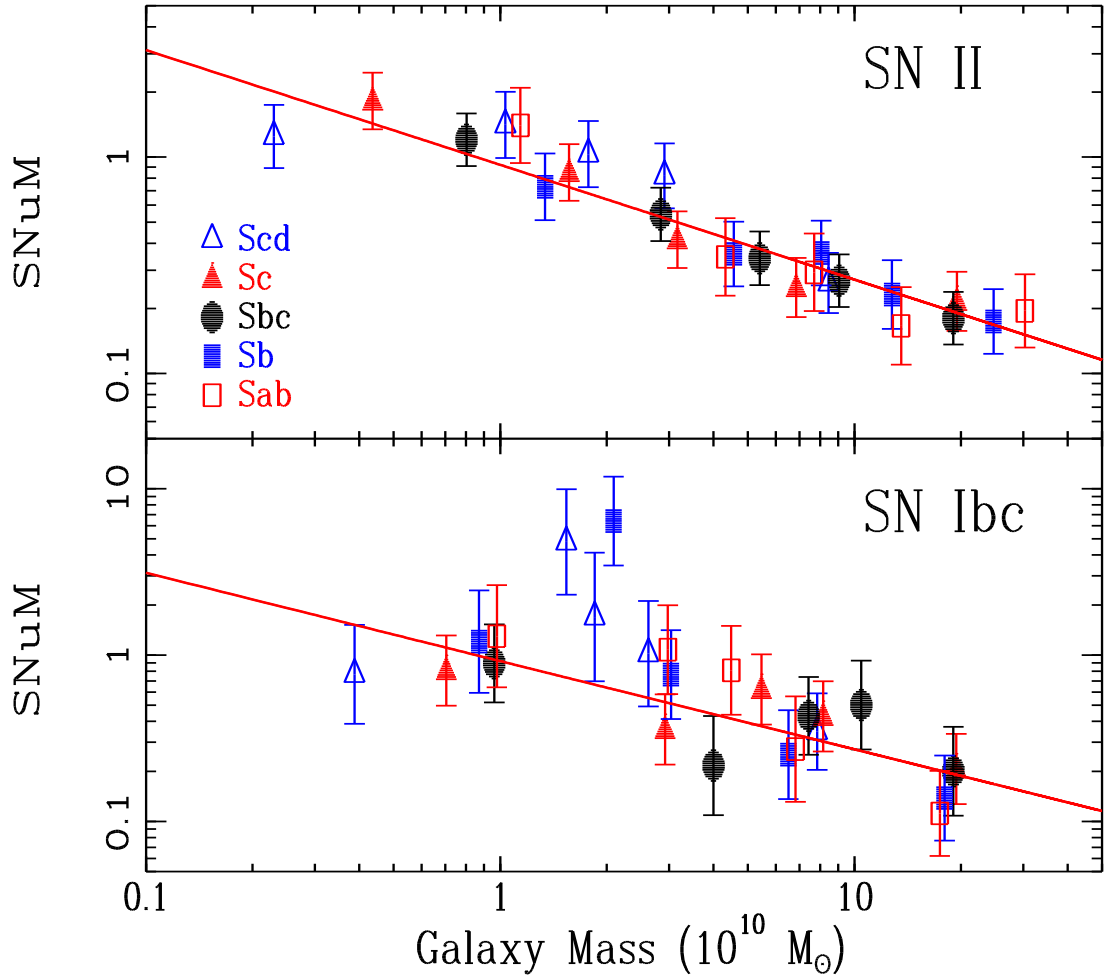}
\caption[] { The core-collapse SN rates (in SNuM) in galaxies of
  different masses. {\it Top panel:} The SNe~II in Sab--Scd galaxies
  are split into 7 bins according to the mass of their host galaxies,
  and the rates (SNuM) are calculated for each mass bin. A
  $\chi^2$-minimizing technique is used to scale and fit the rates
  with the solid curve (using the rates in the Sbc galaxy bins as the
  anchor points), which has a power-law index of $-0.55$.  {\it Bottom
    panel:} The same as the top panel, but for the rates of SNe~Ibc.
  The rates in different galaxy Hubble types are scaled to be fit by
  the linear curve derived from the SN~II rates as shown in the top
  panel.  A similar relation exists between the SNuK rates and galaxy
  $L_K$, and the SNuB rates and galaxy $L_B$, but it is not shown here
  for clarity.  }
\label{4}
\end{figure*}

\clearpage

\begin{figure*}
\includegraphics[scale=0.7,angle=270,trim=0 0 0 0]{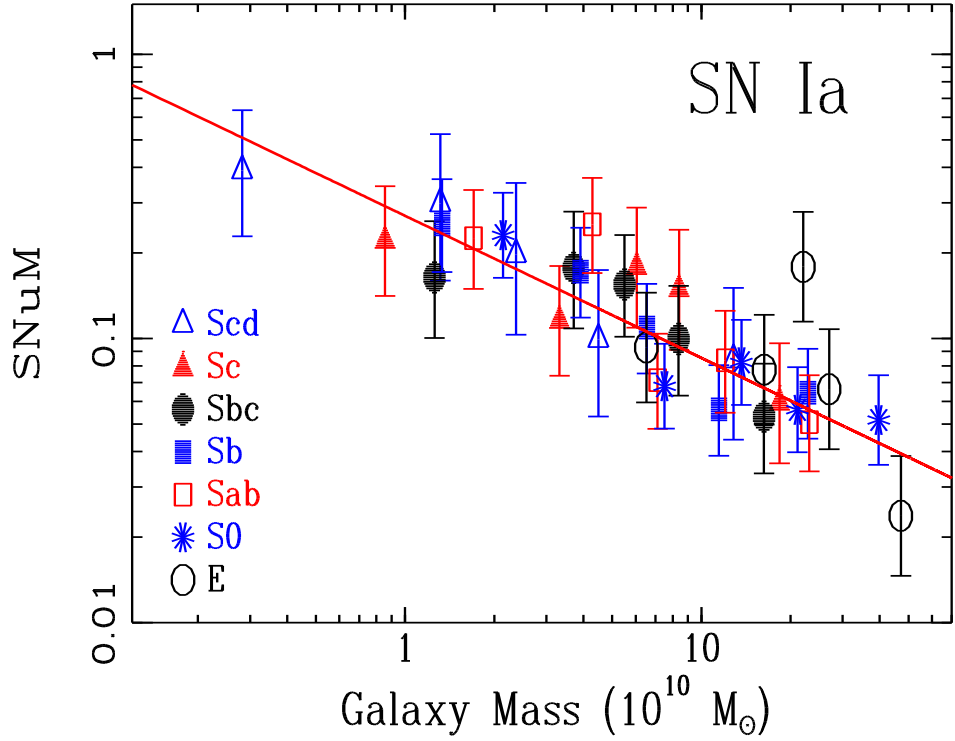}
\caption[] {Same as the top panel of Figure 3, but for the SNuM rates
  of SNe~Ia. The rates for the different Hubble types have been scaled
  up and down to match the normalisation for the Sb galaxies, which
  have been used as anchor points for the fit. The solid curve has a
  power-law index of $-0.50$. The SNuK and SNuB rates have similar
  dependences on galaxy $L_K$ and $L_B$, respectively, but are not
  shown here for clarity.  }
\label{5}
\end{figure*}

\clearpage

\begin{figure*}
\includegraphics[scale=0.7,angle=270,trim=0 0 0 0]{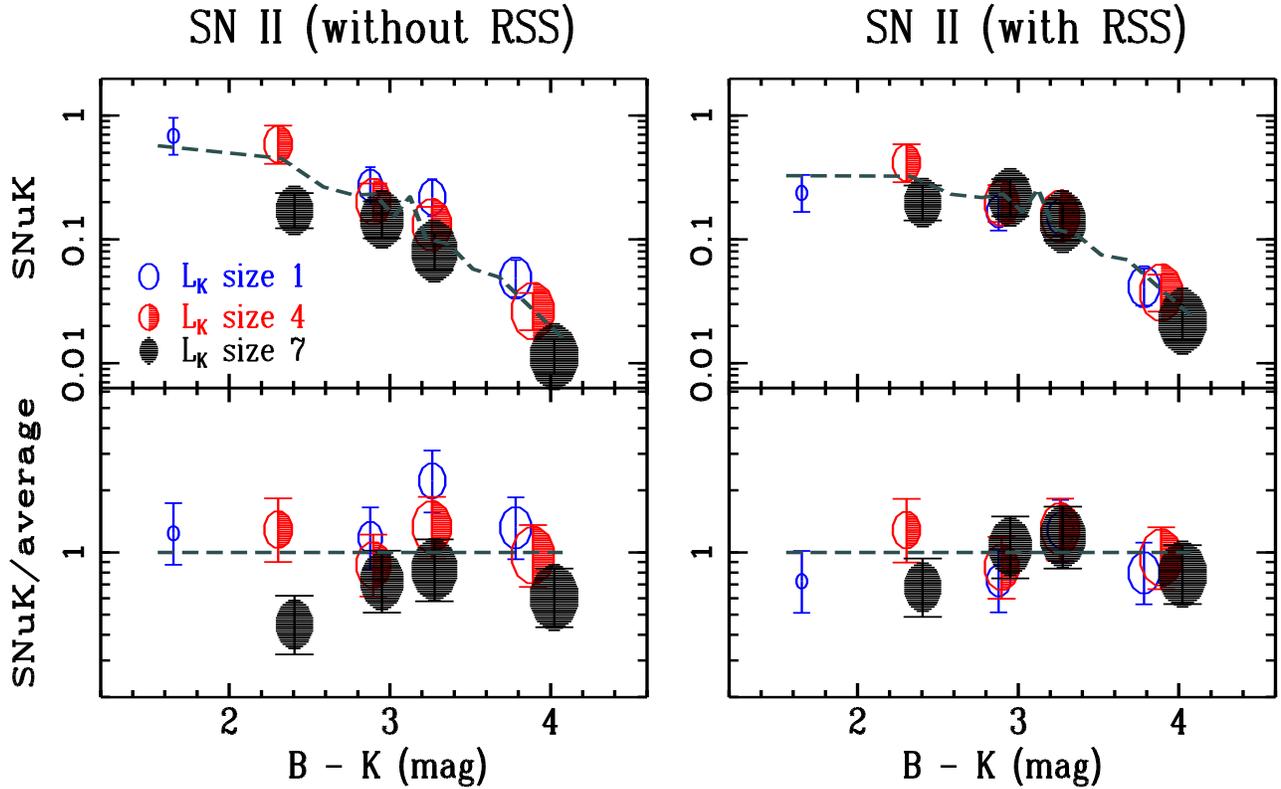}
\caption[] { The rate-size relation for the SN rates with different
  galaxy $B - K$ colours.  The SNuK rates of SNe~II are used as an
  example (the SNuB and SNuM rates of SNe~II, and the SN~Ia rates, all
  exhibit a similar relation). The left and right panels show the SNuK
  rates as a function of galaxy $B - K$ colour without and after
  considering the rate-size relation, respectively, while the top and
  bottom panels display the original rates and the rates after being
  normalised by the average measurements (the dashed lines),
  respectively. For each panel, the galaxies are first divided into 4
  $B - K$ colour groups (same symbol, left to right). For each colour
  group, the galaxies are then divided into 7 $L_K$ bins (different
  symbols). For clarity, only 3 size bins are shown from the smallest
  (i.e., size bin 1, open circles) to the intermediate (size bin 4,
  half-solid circles) to the largest (size bin 7, solid circles).  The
  size of the symbol is proportional to the logarithm of the average
  $L_K$ value of each bin. The left panels exhibit a systematic trend 
  in which the less massive galaxy bins have higher rates (for the same
  colour group). After considering the rate-size relation (with RSS =
  $-0.38$), this trend is gone (the right panels). }
\label{5}
\end{figure*}

\clearpage

\begin{figure*}
\includegraphics[scale=0.7,angle=270,trim=0 0 0 0]{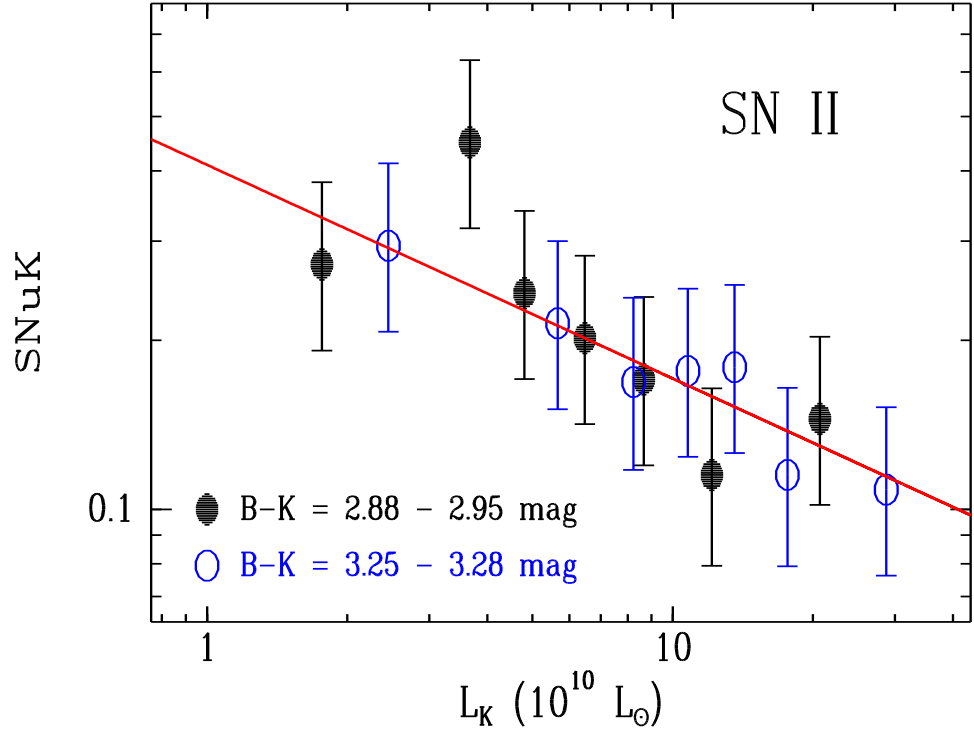}
\caption[] { A further demonstration of the rate-size relation for the
  SNuK rates of SNe~II in two narrow ranges of galaxy $B - K$ colour:
  the solid circles are for the galaxies with average $B - K$ colour
  of 2.88--2.95 mag (used as the anchor points), and the open circles
  are for $B - K$ = 3.25--3.28 mag. Since there are only minimal
  differences in the $B - K$ colours within the same colour group, the
  correlation between SNuK and $L_K$ is close to the intrinsic
  rate-size relation, shown as the solid line (with RSS = $-0.38$).  }
\label{5}
\end{figure*}

\clearpage

\begin{figure*}
\includegraphics[scale=0.7,angle=270,trim=0 -20  0 0]{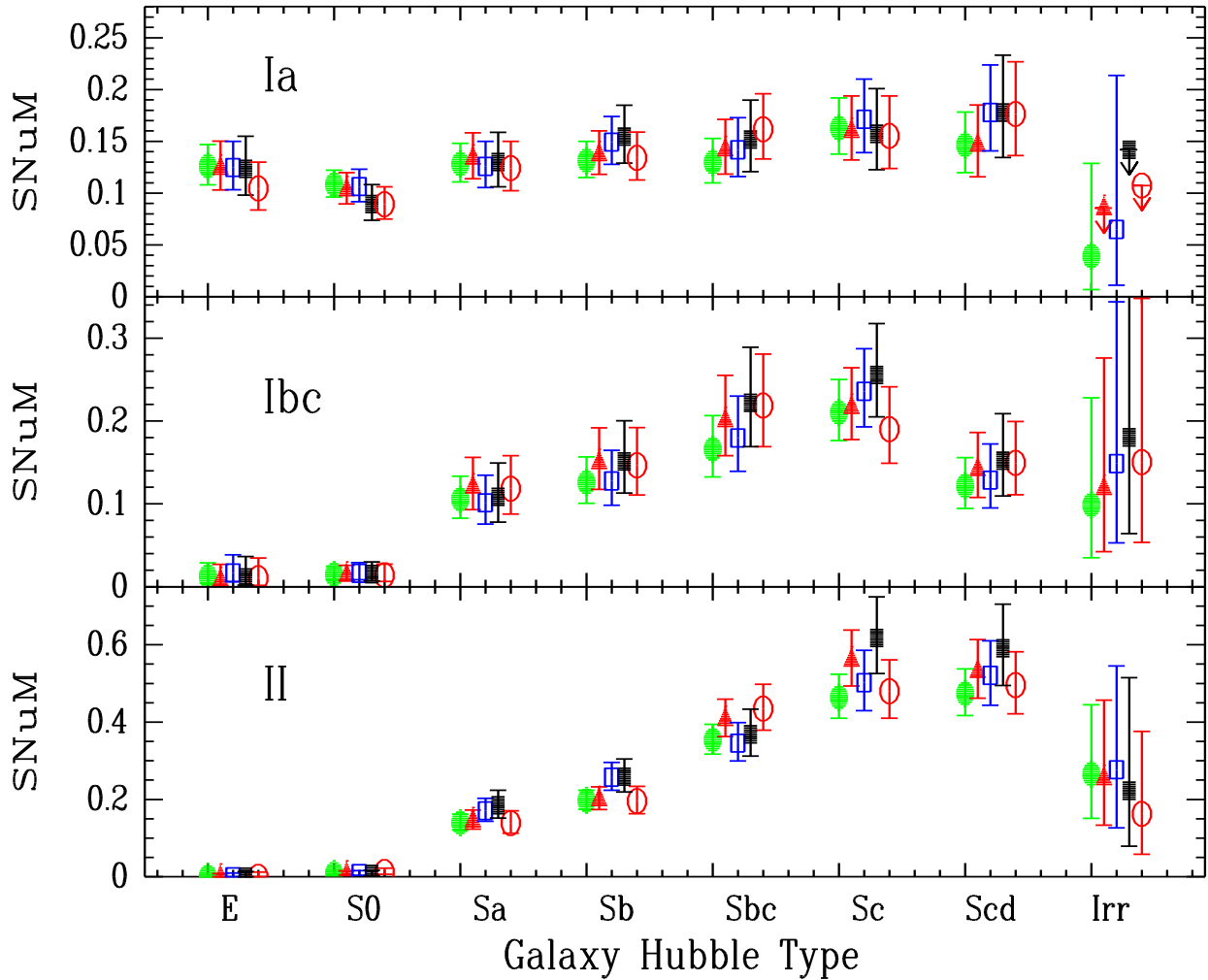}
\caption[] { The SN rates (for a galaxy of the fiducial size) from
  different SN samples.  Solid dots: the ``full" sample (total number
  of SNe = 929).  Triangles: the ``full-optimal" sample (total = 726).
  Open squares: the ``season" sample (total = 656). Solid squares: the
  ``season-optimal" sample (total = 499). Open circles: same as the
  ``full-optimal" sample but only with SNe discovered before the end
  of 2006.  }
\label{8}
\end{figure*}

\clearpage

\begin{figure*}
\includegraphics[scale=0.7,angle=270,trim=0 -20 0 0]{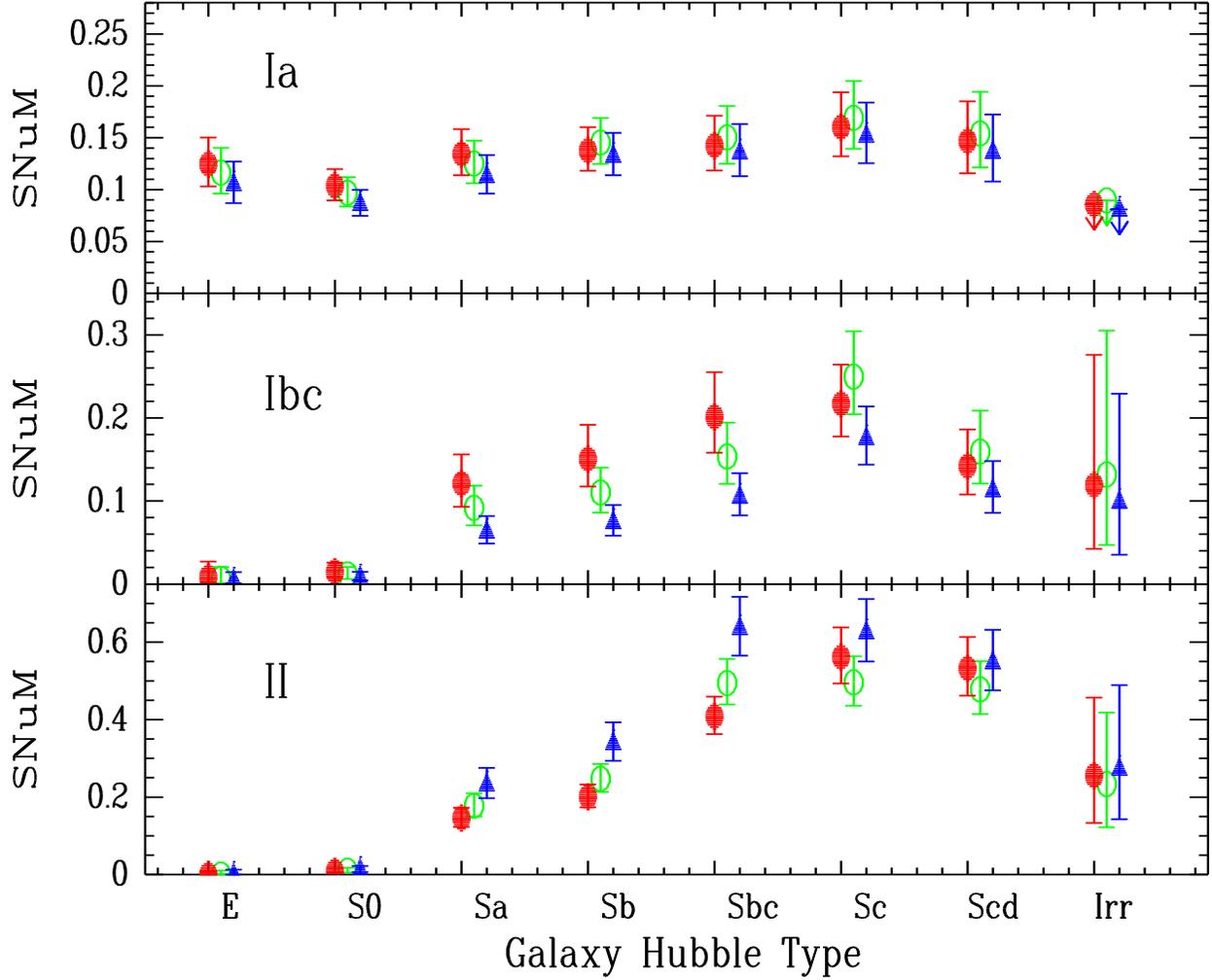}
\caption[] { The SN rates (for a galaxy of the fiducial size) using
  different input LFs and light-curve shapes. Solid dots: two LFs are
  used for each type of SN. Open circles: a single LF is used for each
  SN type. Triangles: the C99 light curves (i.e., no LF) are used.  }
\label{9}
\end{figure*}

\clearpage

\begin{figure*}
\includegraphics[scale=0.7,angle=270,trim=0  0 0 0]{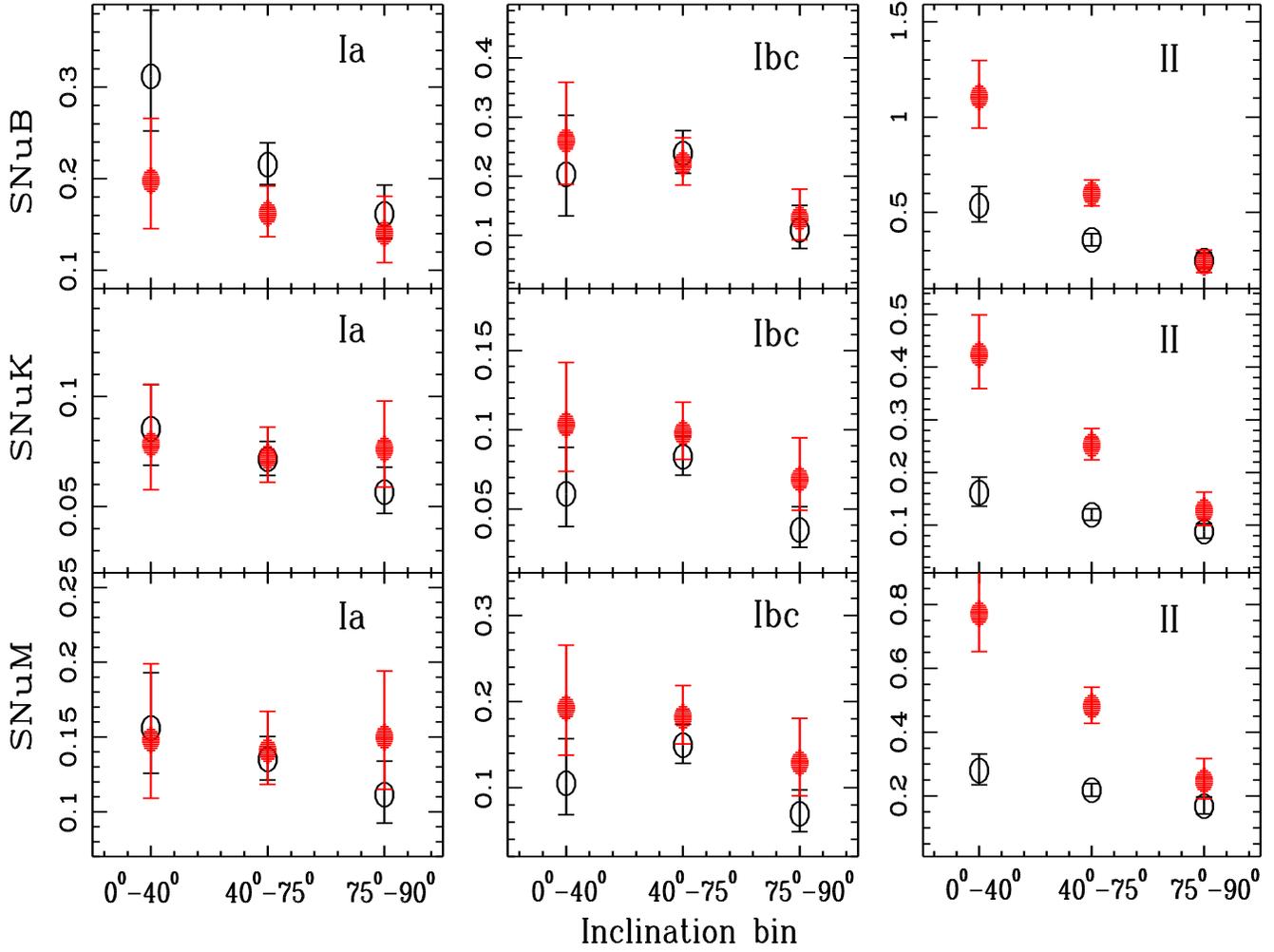}
\caption[] { The SN rates (for a galaxy of the fiducial size) in
  different inclination bins. The solid dots are for the rates in
  late-type spiral galaxies (Sc--Scd), while the open circles are for
  the early-type spirals (Sa--Sbc). From left to right are the rates
  for the SNe~Ia, Ibc, and II, respectively. From top to bottom are
  the rates using different normalisations.  }
\label{13}
\end{figure*}

\clearpage

\begin{figure*}
\includegraphics[scale=0.9,angle=270,trim=0 80 0 0]{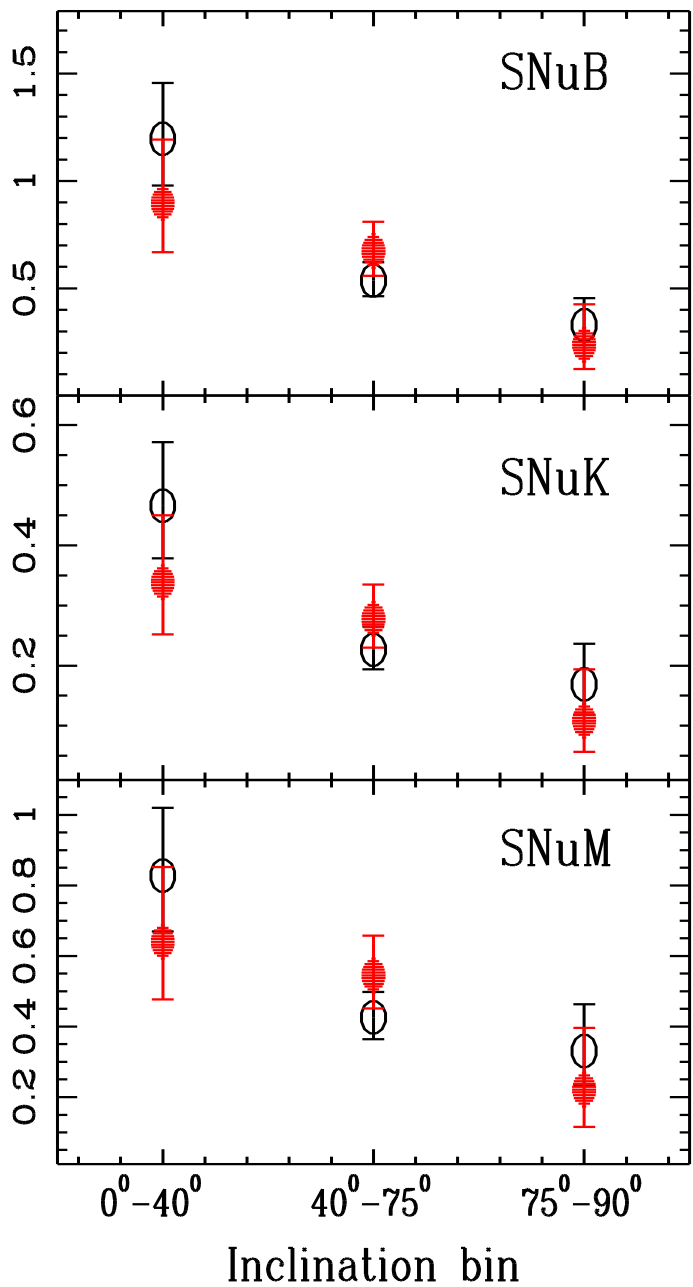}
\caption[] { The SN~II rates (for a galaxy of the fiducial size) in
  the late-type spirals (Sc--Scd) in different inclination bins. The
  open circles are for the rates in the galaxies within 75~Mpc, while
  the solid dots are for the rates in the galaxies more distant than
  75~Mpc.  }
\label{13}
\end{figure*}

\clearpage

\begin{figure*}
\includegraphics[scale=0.7,angle=270,trim=0  0 0 0]{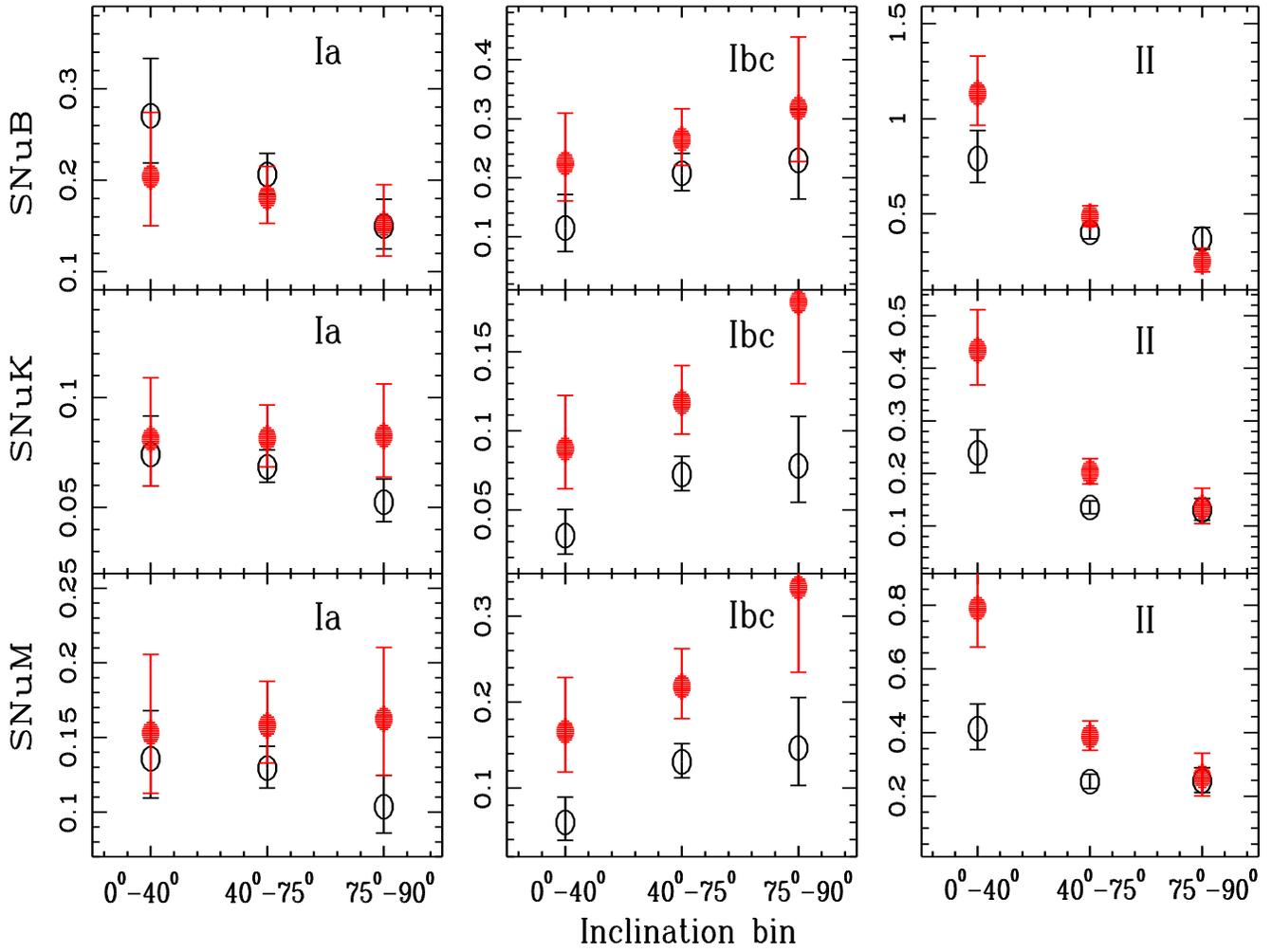}
\caption[] { The same as Figure 9, but with the LF SNe divided into
  different inclination bins when the control time is calculated. In
  other words, the LF is constructed separately for each inclination
  bin, using the SNe in the LF sample whose host-galaxy inclinations
  fall into the same inclination bin.  }
\label{13}
\end{figure*}

\clearpage

\begin{figure*}
\includegraphics[scale=0.7,angle=0,trim=0 0 0 0]{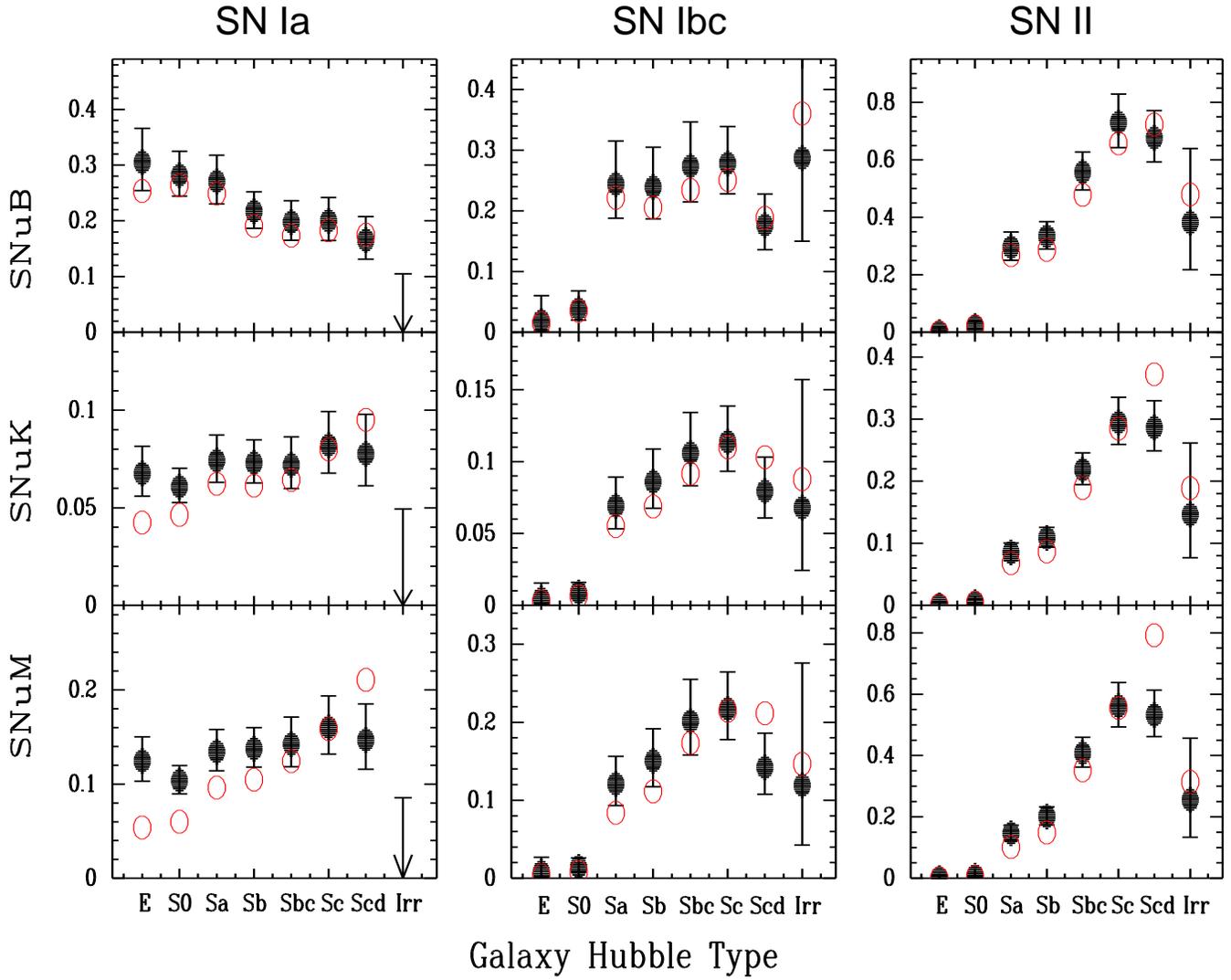}
\caption[] { The SN rates (for a galaxy of the fiducial size) for
  galaxies of different Hubble types (solid circles). The open circles
  without error bars are the rates evaluated at the median galaxy size
  in each Hubble-type bin. }
\label{10}
\end{figure*}

\clearpage

\begin{figure*}
\includegraphics[scale=0.7,angle=0,trim=0 0 0 0]{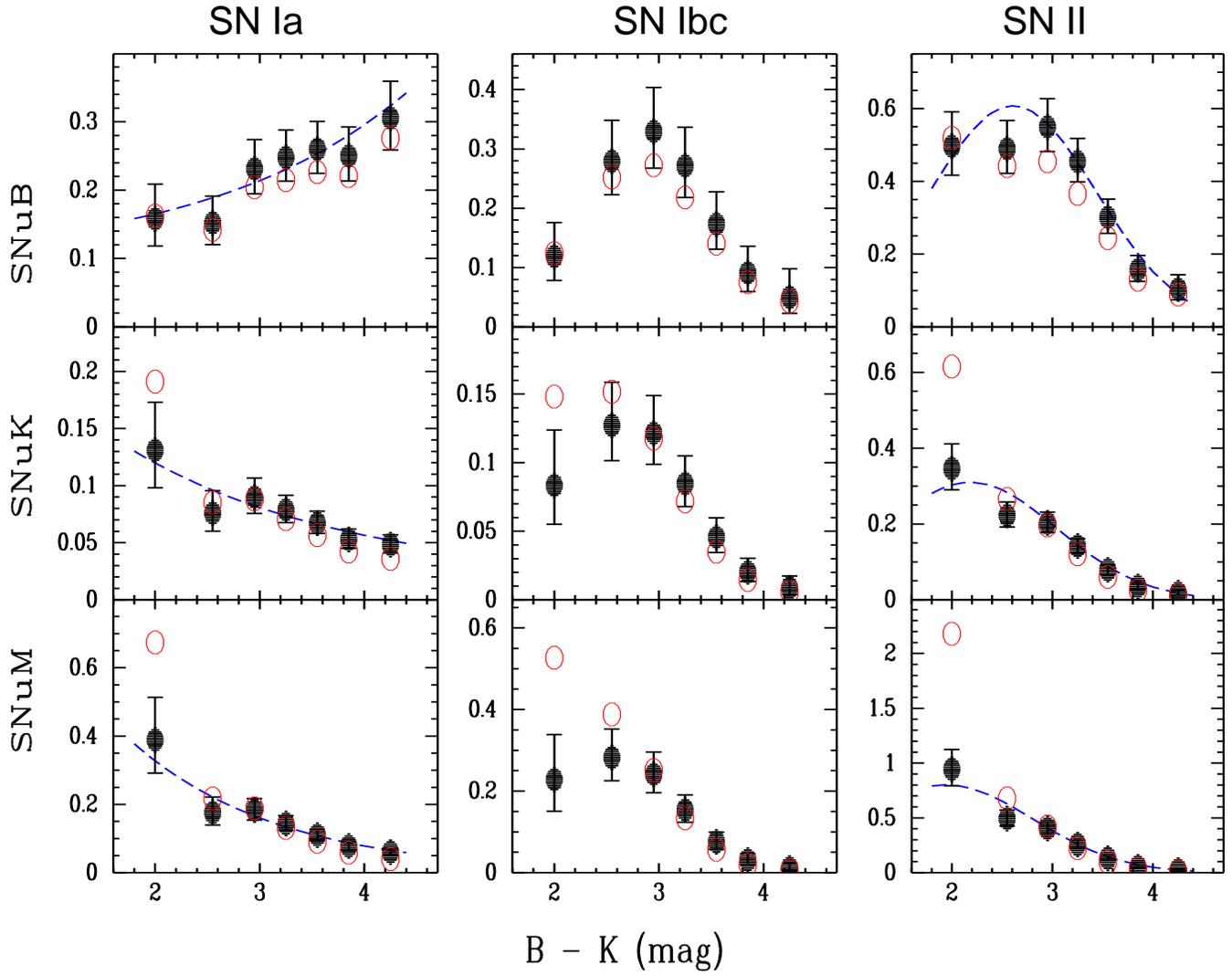}
\caption[] { Same as Figure 12, but for galaxies having different $B -
  K$ colours (solid circles).  The open circles without error bars are
  the rates evaluated at the median galaxy size in each colour
  bin. The dashed lines represent the second-order polynomial fits (as
  a function of $B - K$ colour) for the logarithm of the rates as
  determined during the multivariate linear regression model analysis
  in \S 2.2. }
\label{11}
\end{figure*}

\clearpage

\begin{figure*}
\includegraphics[scale=0.7,angle=0,trim=0 0 0 0]{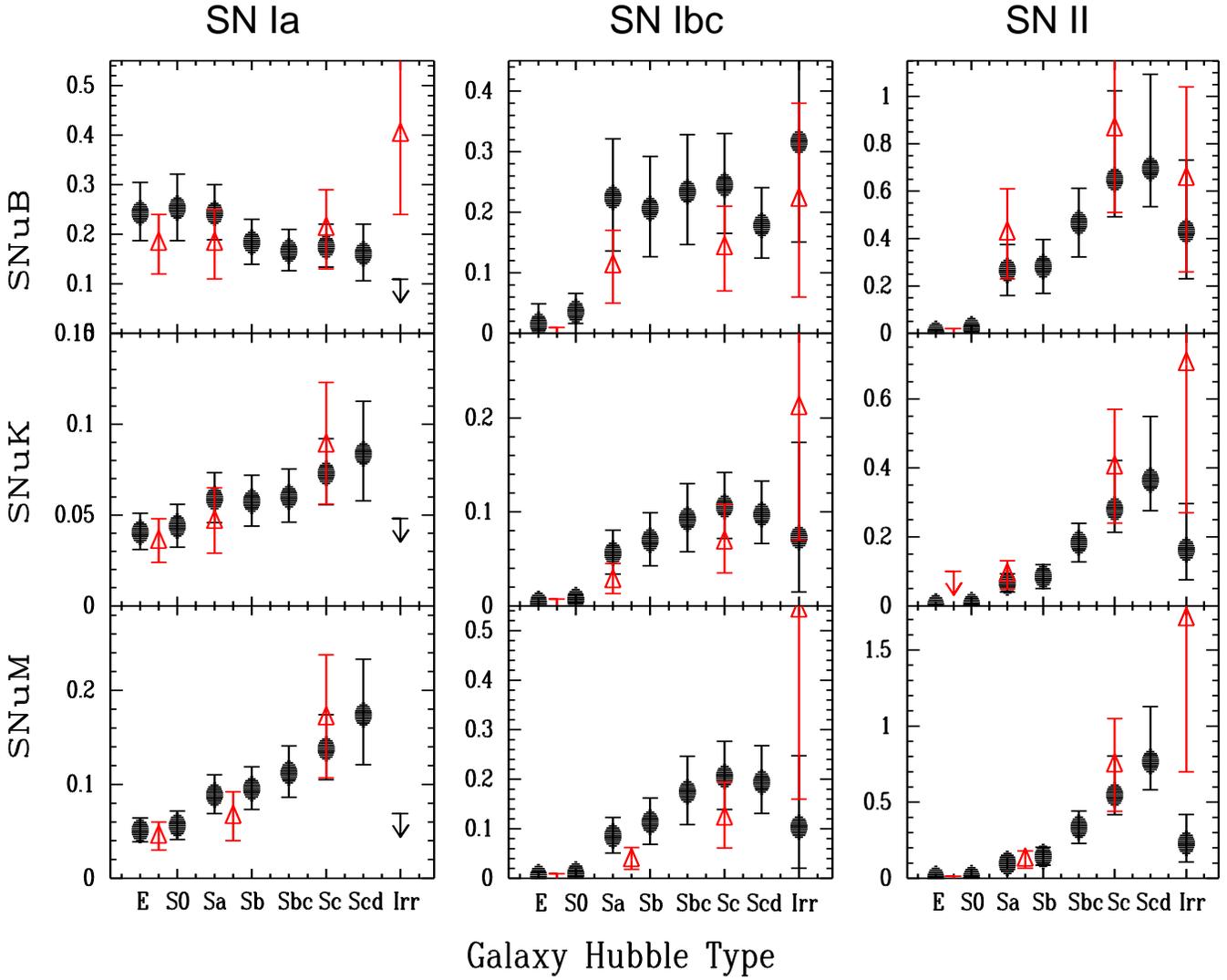}
\caption[] { Comparison with published results. To mimic the past
  calculations, our rates (the solid dots) are calculated without
  using RSSs, so they are the average for the galaxies with different
  sizes.  The published results (open triangles) come from C99 (SNuB)
  and M05 (SNuK and SNuM). The M05 SNuK and SNuM rates for SNe~Ia in
  Irr galaxies are off the scale of the plot (higher than the ordinate
  limit).  }
\label{14}
\end{figure*}

\clearpage

\begin{figure*}
\includegraphics[scale=1.0,angle=270,trim=0 80 0 0]{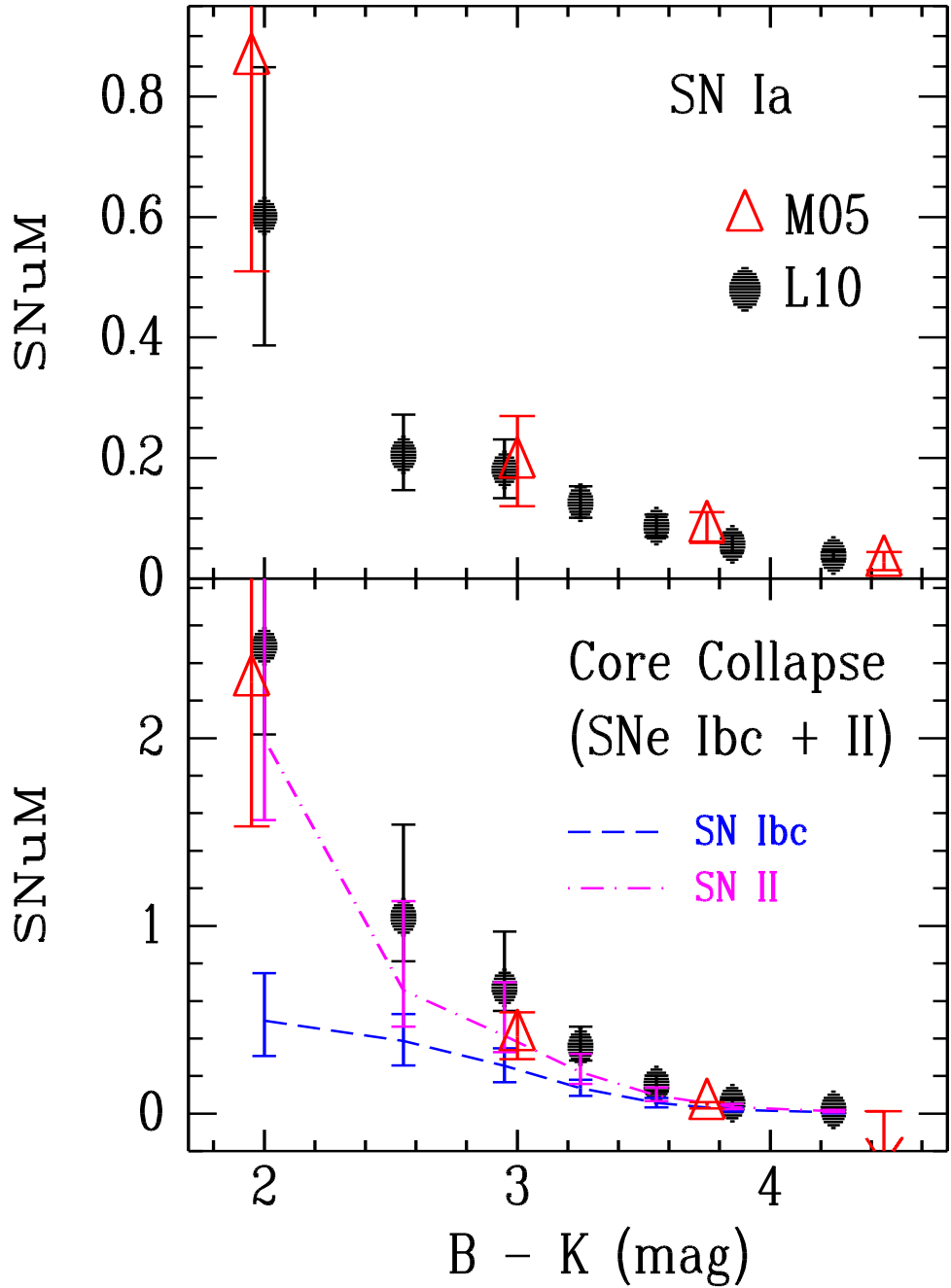}
\caption[] { The same as Figure 14, but for the SNuM rates in galaxies
  having different $B - K$ colours. The dashed and the dash-dotted lines
  in the bottom panel show the contribution of the SN~Ibc and SN~II
  rates to the total core-collapse SN rates, respectively. }
\label{15}
\end{figure*}

\clearpage

\begin{figure*}
\includegraphics[scale=0.7,angle=90,trim=0 0 0 0]{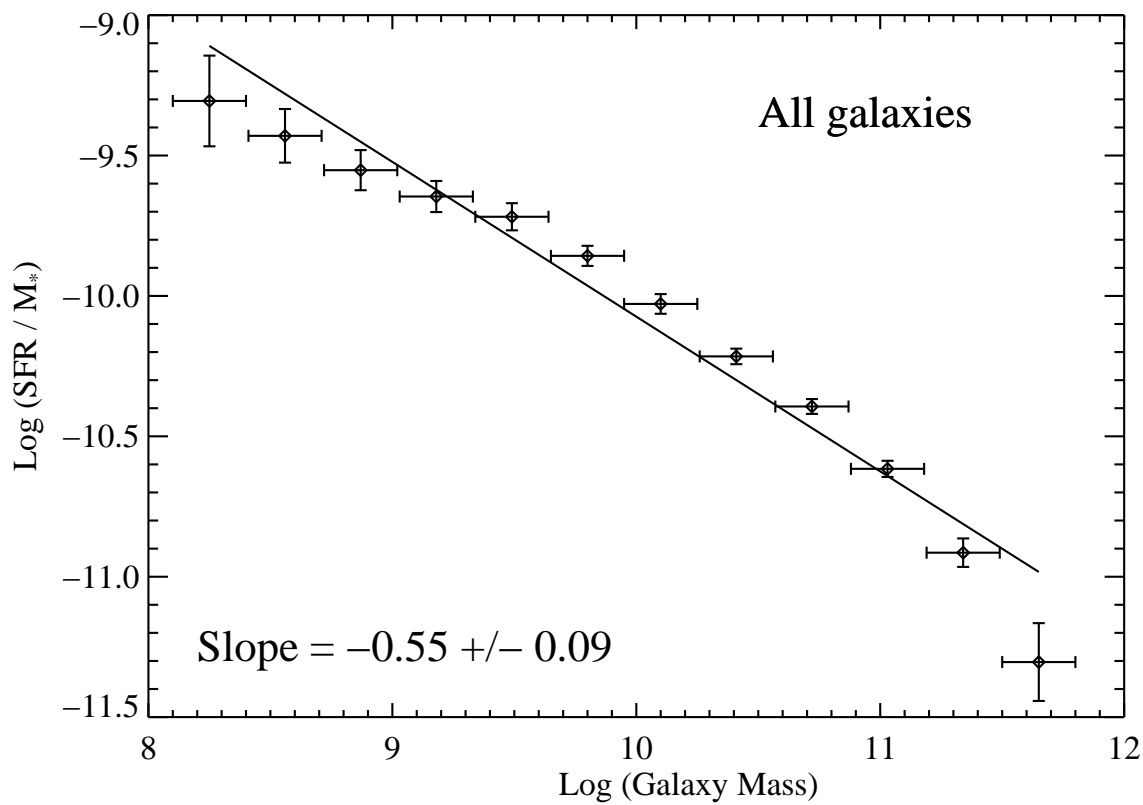}
\caption[] { The correlation between the specific star-formation rate
  and the galaxy stellar mass. This is an integration of the published
  contour map in Figure 7 of Schiminovich et al. (2007). The linear
  fit (the solid line) gives a power-law index of $-0.55 \pm 0.09$,
  similar to that of the rate-size relation for the core-collapse SNe. }
\label{12}
\end{figure*}

\clearpage

\begin{figure*}
\includegraphics[scale=0.7,angle=270,trim=0  0 0 0]{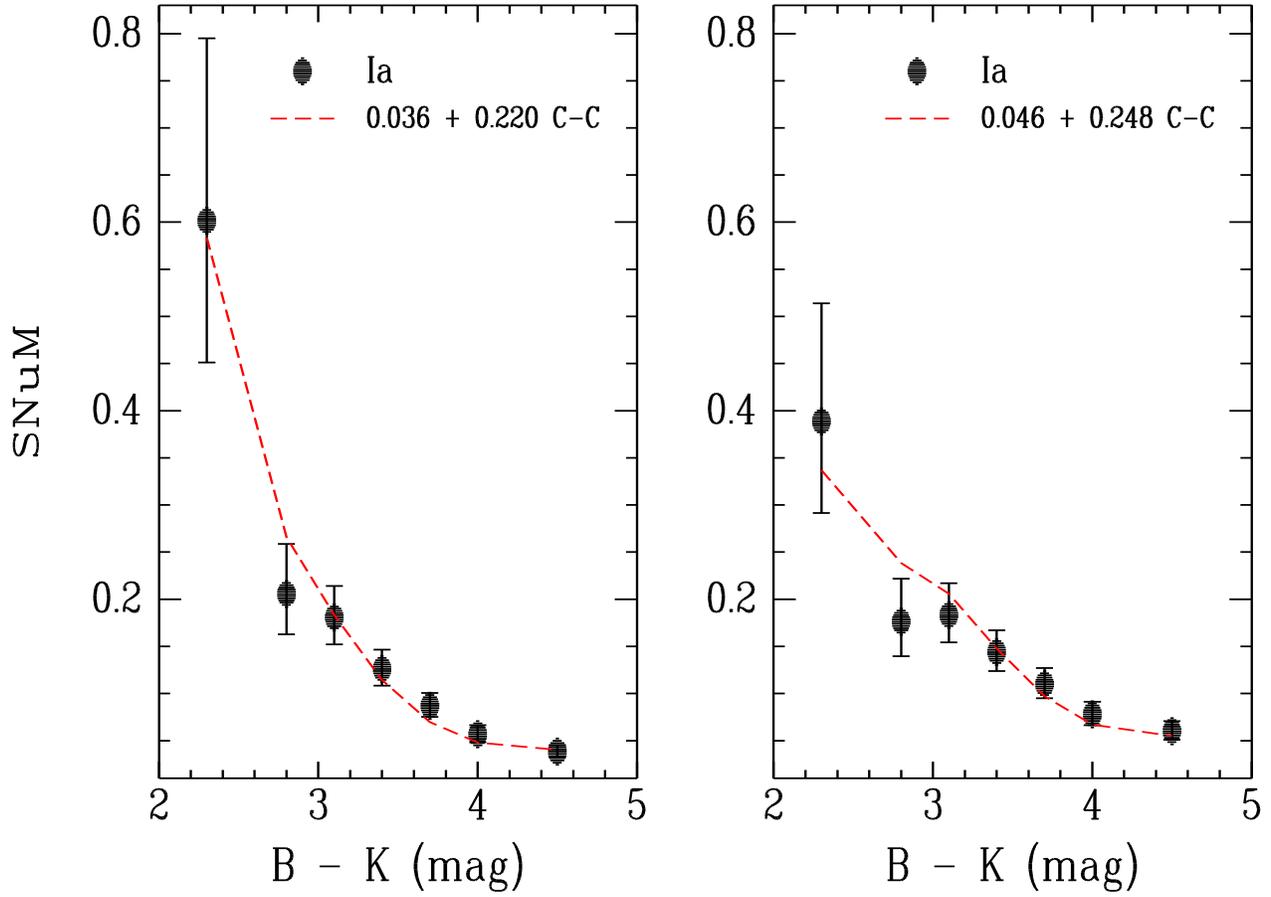}
\caption[] { The two-component model fits for the SN~Ia rates when
  SNuM versus galaxy $B-K$ colour are considered. The left panel shows
  the model fit for the average galaxy size (i.e., no RSSs are used),
  while the right panel shows the fit for the fiducial galaxy size
  (i.e., RSSs are used).  Both fits have $\chi^2/{\rm DOF} < 1.0$.  }
\label{17}
\end{figure*}

\clearpage

\begin{figure*}
\includegraphics[scale=0.8,angle=270,trim=0 40 0 0]{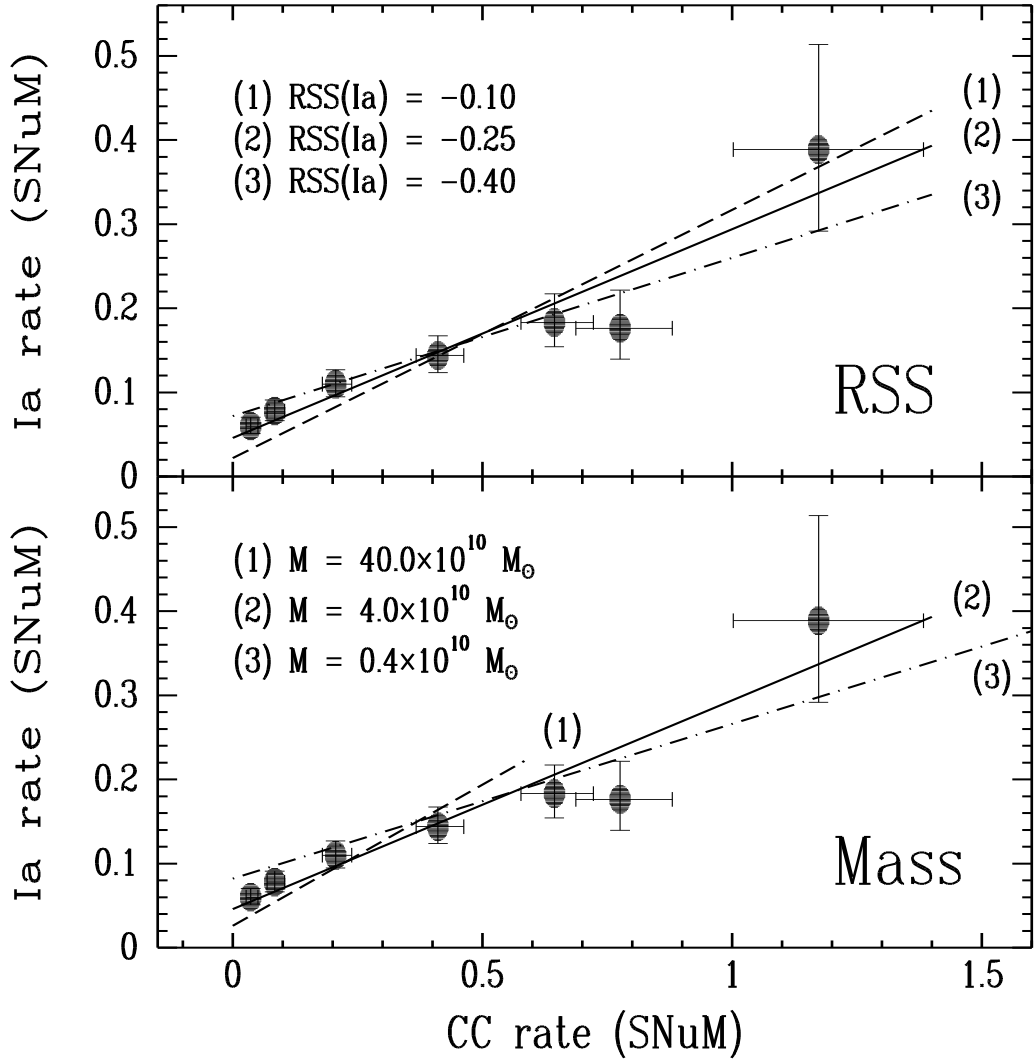}
\caption[] { The effect of different RSSs (top panel) and galaxy
  masses (bottom panel) for the two-component model fits for the SN~Ia
  rates.  For clarity, only the data points (and their error bars) for
  fit (1) (top panel) and fit (2) (bottom panel) are shown,
  respectively.  All fits have reduced $\chi^2 < 1.0$.  }
\label{17}
\end{figure*}

\clearpage

\begin{figure*}
\includegraphics[scale=0.8,angle=270,trim=0 40 0 0]{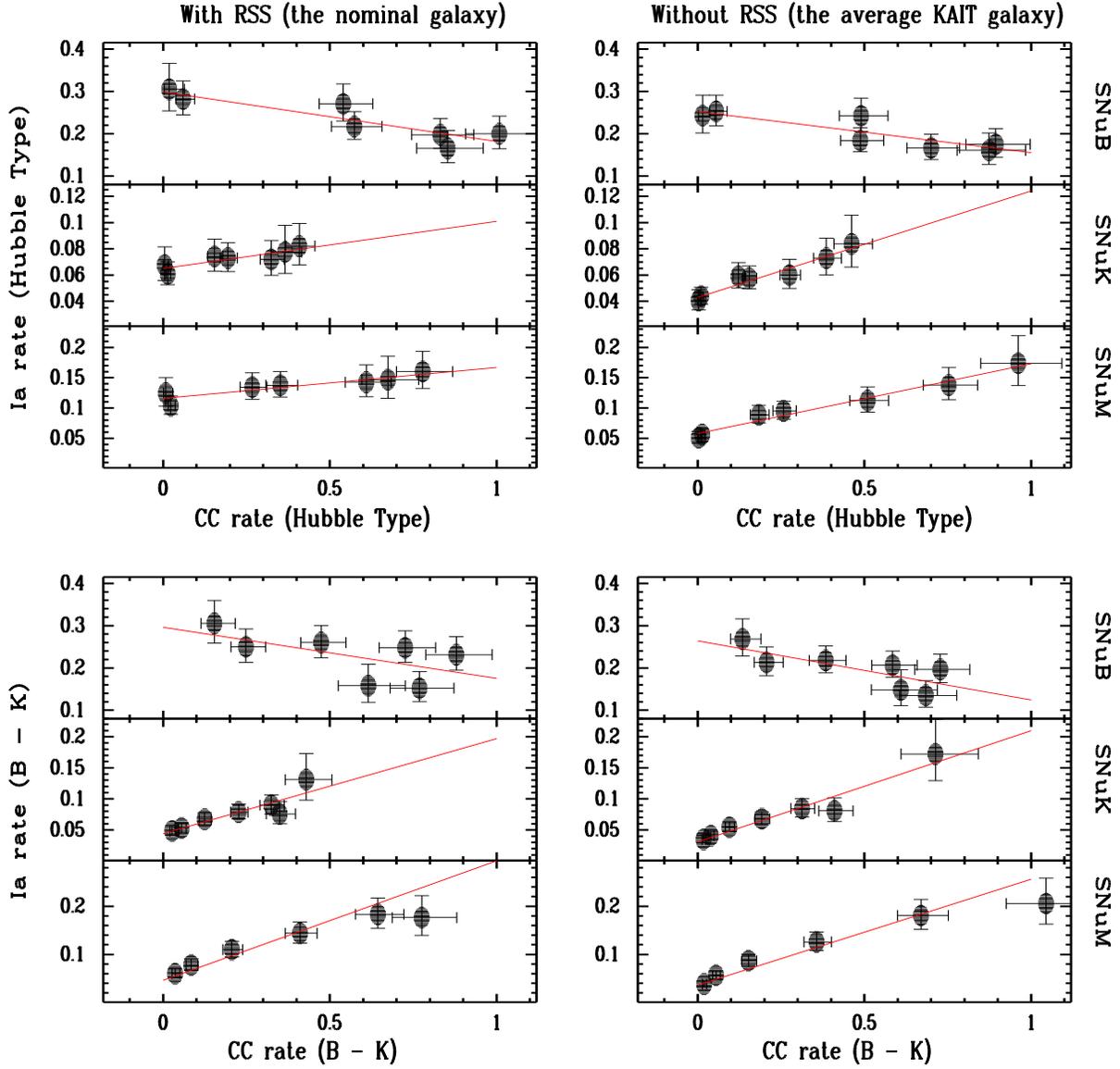}
\caption[] { The two-component model fits for the SN~Ia rates are
  affected by several factors: different normalisations (marked to the
  right), with or without the use of RSSs (left and right panels,
  respectively), and different grouping methods for the galaxies (top
  panel, using galaxy Hubble types; bottom panel, using galaxy $B - K$
  colour).  }
\label{17}
\end{figure*}

\clearpage

\begin{figure*}
\includegraphics[scale=0.7,angle=270,trim=0 0 0 0]{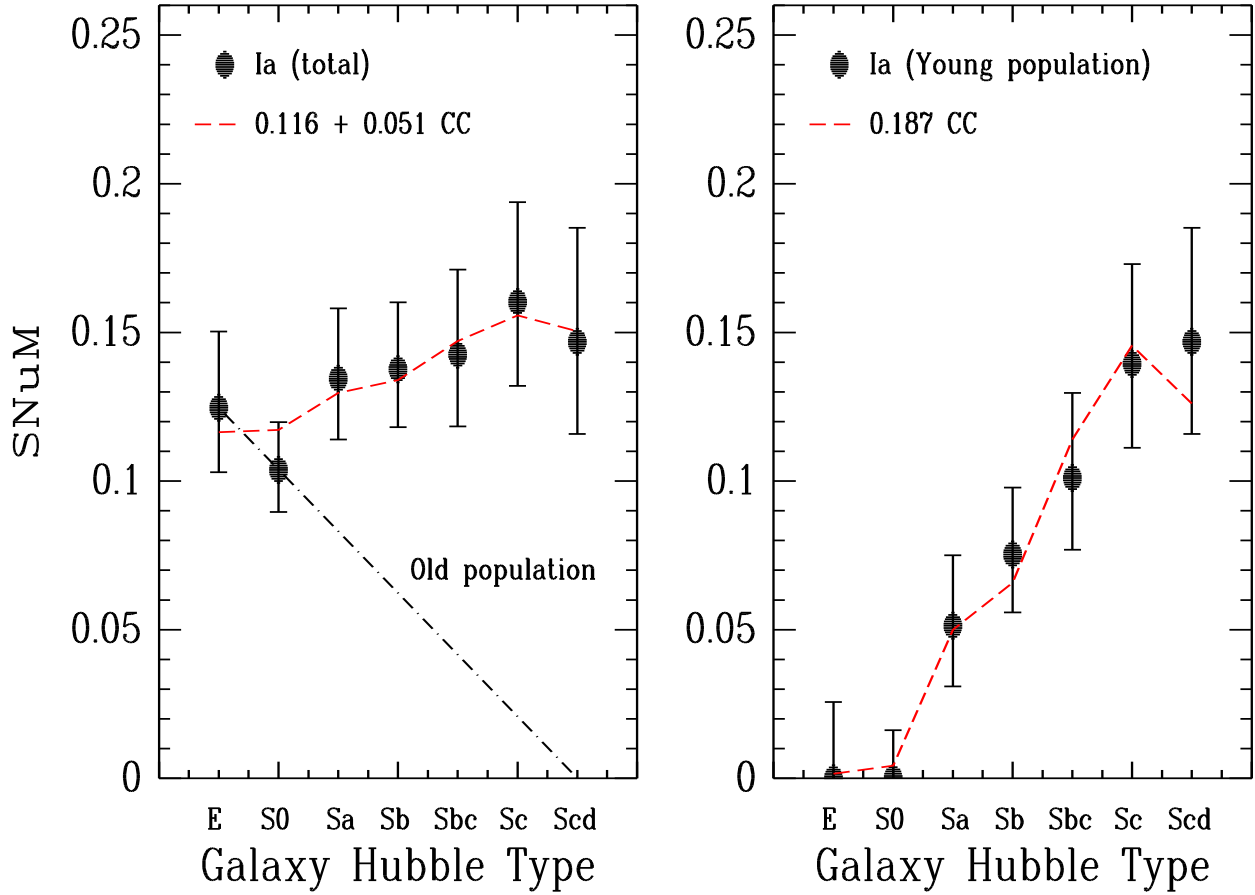}
\caption[] {The effect of considering old and young populations for
  the two-component model for the SN~Ia rates. Here the SNuM rates for
  a fiducial galaxy size are considered for different Hubble types.
  {\it Left panel:} The total SN~Ia rates (solid dots) are fit with
  the two-component model (dashed line). The dash-dotted line shows a
  toy model for the SN~Ia rates in old populations.  {\it Right
    panel:} After subtracting the contribution from old populations,
  the SN~Ia rates in young populations (solid dots) are fit with the
  two-component model (dashed line).  }
\label{17}
\end{figure*}

\clearpage

\begin{figure*}
\includegraphics[scale=0.8,angle=270,trim=0  0 0 0]{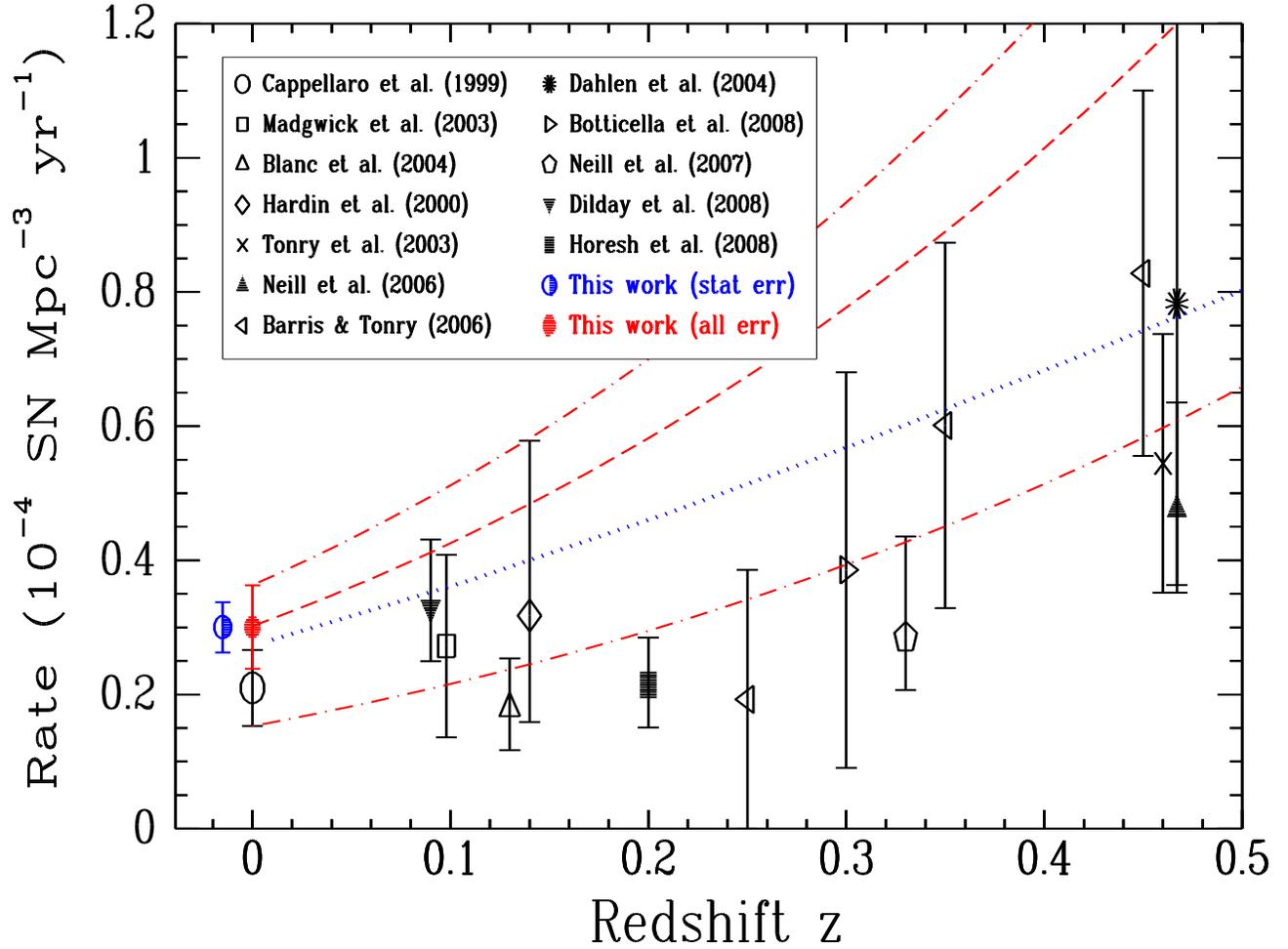}
\caption[] { The volumetric rate of SNe~Ia at different redshifts. Our
  rate is marked with only the statistical uncertainty (half-filled
  circle), and with the total uncertainty (solid circle). The rest of
  the rates are adopted from Horesh et al. (2008).  The dashed line is
  evaluated at our rate with a functional form of ($1 + z$)$^{3.6}$,
  and is the star-formation rate history from Hopkins \& Beacom
  (2006). The upper and lower dash-dotted lines follow the same
  functional form and are evaluated at the 1$\sigma$ upper error bar
  of our measurement, and the 1$\sigma$ lower error bar of the C99
  measurement, respectively. The dotted line is the expected
  SN~Ia rate from the SFR history study by Mannucci et al. (2007).
  }
\label{16}
\end{figure*}

\clearpage

\begin{figure*}
\includegraphics[scale=0.7,angle=270,trim=0 -20 0 0]{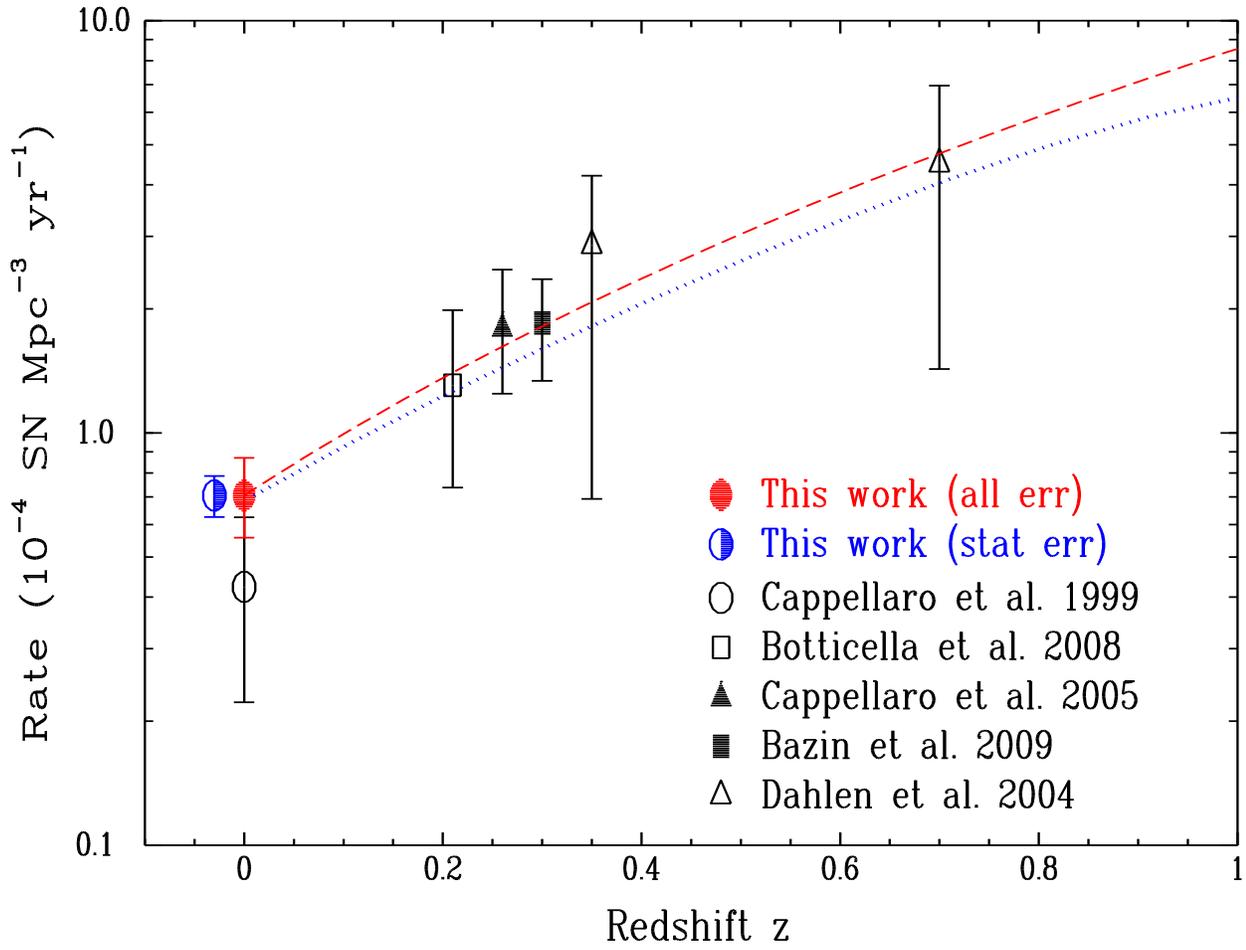}
\caption[] { The same as Figure 21, but for the core-collapse SN
  rates.  }
\label{16}
\end{figure*}

\clearpage

\begin{figure*}
\includegraphics[scale=0.8,angle=270,trim=0  0 0 0]{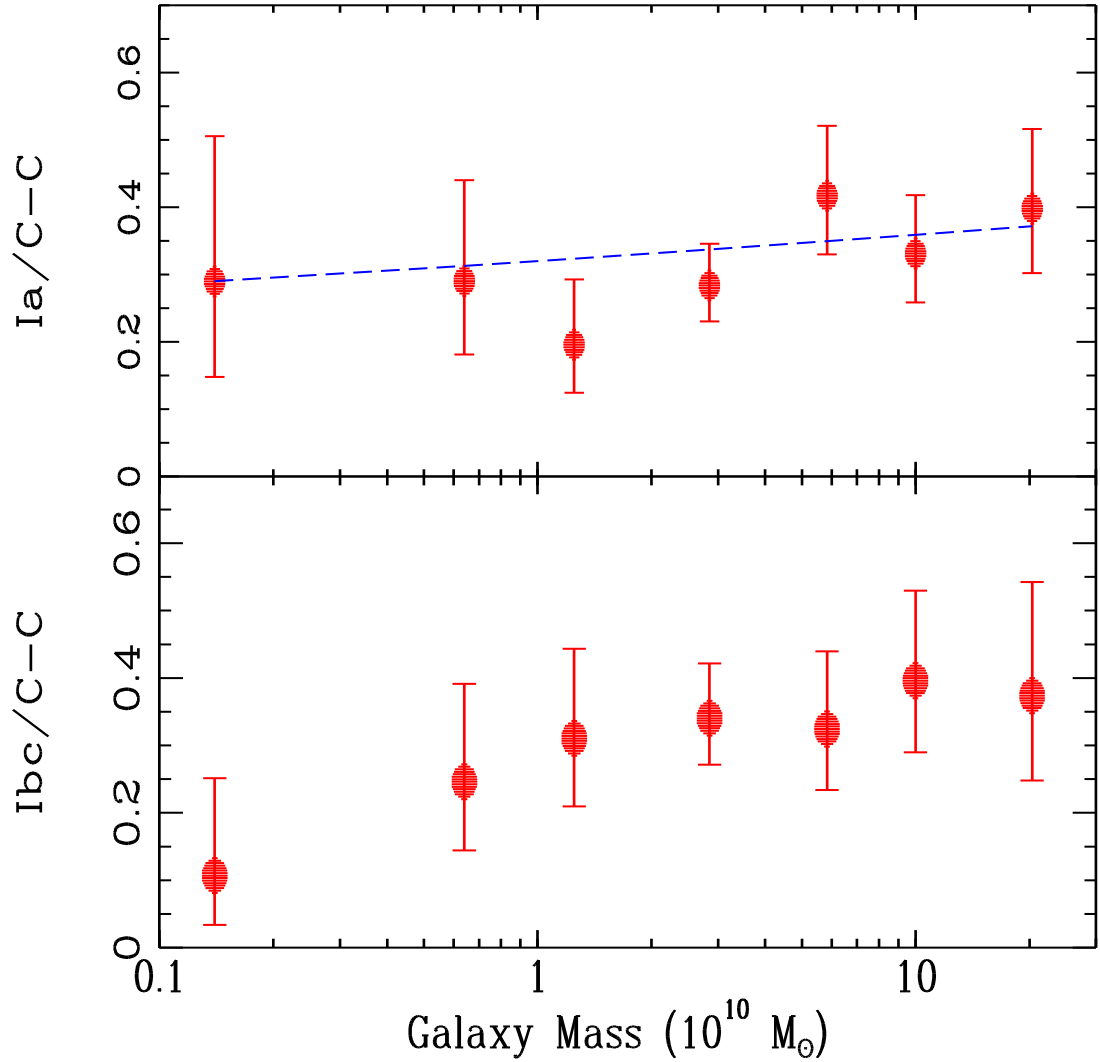}
\caption[] { The rate ratios for different galaxy masses. The top
  panel shows the ratio of the SN~Ia rate to the core-collapse SN
  rate, while the bottom panel gives the ratio of the SN~Ibc rate to
  the core-collapse SN rate. The dashed line in the top panel follows
  $M^{0.05}$, which is the expected SN~Ia to CC~SN rate ratio due to
  different RSSs in the rate-size relations.  }
\label{19}
\end{figure*}

\newpage
\clearpage

\renewcommand{\baselinestretch}{1.5}
\scriptsize

\begin{table*}
\caption{The rate-size correction factors.$^{a}$}
\begin{tabular}{lcc}
\hline
\hline
Hubble Type&{RSS(SN~Ia)} &RSS(SN~II)  \\
\hline
E & $-0.513 \pm 0.316$ & \nodata \\ 
S0 & $-0.503 \pm 0.158$ & \nodata \\ 
Sa & $-0.637 \pm 0.199$ & $-0.653 \pm 0.167$ \\ 
Sb & $-0.555 \pm 0.171$ & $-0.498\pm 0.165$ \\ 
Sbc & $-0.443 \pm 0.241$ & $-0.628 \pm 0.121$ \\ 
Sc & $-0.329 \pm 0.201$ & $-0.626 \pm 0.111$ \\ 
Scd & $-0.435 \pm 0.195$ & $-0.437 \pm 0.128$ \\ 
\hline
\hline
\end{tabular}

\medskip
$^{a}$SNuM rates for SNe~Ia and II in the ``full-optimal" sample are
used.
\end{table*}

\begin{table*}
\caption{Adopted correction factors and fiducial galaxy sizes.}
\begin{tabular}{clccl}
\hline
\hline
{Rate} &{Galaxy groups} & {RSS(Ia)} &{RSS(Ibc, II)}
&{Fiducial size} \\ 
\hline
SNuB & Hubble type & RSS$_B = -0.23\pm0.20$ & RSS$_B = -0.27\pm0.10$ & $L_{B0} = 2\times10^{10}\,{\rm L}_\odot$ \\
SNuK & Hubble type & RSS$_K = -0.35\pm0.10$ & RSS$_K = -0.45\pm0.10$ & $L_{K0} = 7\times10^{10}\,{\rm L}_\odot$ \\
SNuM & Hubble type & RSS$_M = -0.50\pm0.10$ & RSS$_M = -0.55\pm0.10$ & $M_0 = 4\times10^{10}\,{\rm M}_\odot$ \\
SNuB & $B - K$     & RSS$_B = -0.25\pm0.15$ & RSS$_B = -0.38\pm0.10$ & $L_{B0} = 2\times10^{10}\,{\rm L}_\odot$ \\
SNuK & $B - K$     & RSS$_K = -0.25\pm0.15$ & RSS$_K = -0.38\pm0.10$ & $L_{K0} =  7\times10^{10}\,{\rm L}_\odot$ \\
SNuM & $B - K$     & RSS$_M = -0.25\pm0.15$ & RSS$_M = -0.38\pm0.10$ & $M_0 = 4\times10^{10}\,{\rm M}_\odot$ \\
\hline
\hline
\end{tabular}
\end{table*}

\begin{table*}
\caption{SN rates in different galaxy inclination bins.}
\begin{minipage}{180mm}
\begin{tabular}{lllllllrrr}
\hline
\hline
{Rate} &{Gal} &{SN} &{$r0$(0--40)$^{a}$} 
&{$r1$(40--75)$^{a}$} &{$r2$(75--90)$^{a}$} 
&{$r3$(0--75)$^{a}$}
&{$r0/r1 - 1$} &{$r0/r2 - 1$}&{$r0/r3 - 1$} \\
\hline
\hline
SNuB&Sa--Sbc&Ia&0.312(0.066)&0.215(0.023)&0.162(0.029)&0.231(0.022)&0.45(0.34)&0.93(0.54)&0.35(0.31)\\
SNuB&Sc--Scd&Ia&0.198(0.060)&0.162(0.028)&0.140(0.036)&0.170(0.025)&0.22(0.43)&0.41(0.56)&0.16(0.39)\\
SNuB&Sa--Sbc&Ibc&0.203(0.085)&0.238(0.036)&0.109(0.036)&0.233(0.033)&$-$0.15(0.38)&0.86(0.99)&$-$0.13(0.39)\\
SNuB&Sc--Scd&Ibc&0.260(0.086)&0.222(0.040)&0.129(0.043)&0.230(0.036)&0.17(0.44)&1.02(0.95)&0.13(0.41)\\
SNuB&Sa--Sbc&II&0.535(0.093)&0.356(0.032)&0.247(0.038)&0.385(0.030)&0.51(0.29)&1.17(0.50)&0.39(0.26)\\
SNuB&Sc--Scd&II&1.107(0.177)&0.599(0.068)&0.236(0.059)&0.710(0.065)&0.85(0.36)&3.68(1.39)&0.56(0.29)\\
\hline
SNuK&Sa--Sbc&Ia&0.085(0.018)&0.072(0.008)&0.056(0.011)&0.074(0.007)&0.19(0.29)&0.51(0.43)&0.15(0.27)\\
SNuK&Sc--Scd&Ia&0.078(0.024)&0.072(0.013)&0.076(0.020)&0.074(0.011)&0.08(0.38)&0.03(0.41)&0.06(0.36)\\
SNuK&Sa--Sbc&Ibc&0.060(0.025)&0.083(0.012)&0.037(0.013)&0.079(0.011)&$-$0.28(0.32)&0.62(0.88)&$-$0.24(0.33)\\
SNuK&Sc--Scd&Ibc&0.103(0.034)&0.098(0.018)&0.069(0.023)&0.099(0.016)&0.06(0.40)&0.50(0.70)&0.04(0.38)\\
SNuK&Sa--Sbc&II&0.162(0.028)&0.120(0.011)&0.088(0.014)&0.127(0.010)&0.35(0.26)&0.84(0.43)&0.27(0.24)\\
SNuK&Sc--Scd&II&0.424(0.070)&0.252(0.030)&0.128(0.032)&0.294(0.028)&0.68(0.34)&2.32(0.99)&0.44(0.27)\\
\hline
SNuM&Sa--Sbc&Ia&0.156(0.034)&0.135(0.015)&0.111(0.021)&0.139(0.013)&0.16(0.28)&0.40(0.40)&0.12(0.26)\\
SNuM&Sc--Scd&Ia&0.148(0.045)&0.141(0.024)&0.150(0.040)&0.143(0.021)&0.05(0.37)&$-$0.01(0.40)&0.04(0.35)\\
SNuM&Sa--Sbc&Ibc&0.105(0.044)&0.149(0.023)&0.070(0.024)&0.141(0.020)&$-$0.30(0.31)&0.51(0.82)&$-$0.26(0.33)\\
SNuM&Sc--Scd&Ibc&0.193(0.064)&0.182(0.034)&0.129(0.045)&0.184(0.029)&0.06(0.40)&0.49(0.72)&0.04(0.38)\\
SNuM&Sa--Sbc&II&0.280(0.048)&0.218(0.020)&0.168(0.027)&0.229(0.018)&0.28(0.25)&0.66(0.39)&0.22(0.23)\\
SNuM&Sc--Scd&II&0.771(0.128)&0.481(0.057)&0.247(0.063)&0.553(0.052)&0.60(0.33)&2.12(0.96)&0.40(0.27)\\
\hline
\hline
\end{tabular}

\medskip
$^{a}$SN rates for the galaxies with inclination in the range $0^\circ
- 40^\circ$, $40^\circ - 75^\circ$, $75^\circ - 90^\circ$, and
$0^\circ - 75^\circ$, respectively.

\end{minipage}
\end{table*}

\clearpage

\begin{table*}
\vbox to330mm{\vfil Landscape Table 4 to go here.
\caption{SN rates in fiducial galaxies of different Hubble types.}
\vfil}
\label{landtable}
\end{table*}

\begin{table*}
\vbox to330mm{\vfil Landscape Table 5 to go here.
\caption{SN rates in fiducial galaxies of different $B - K$ colours.}
\vfil}
\label{landtable}
\end{table*}

\begin{table*}
\caption{SN rates in average galaxies of different Hubble types.$^{a}$}
\begin{minipage}{180mm}
\begin{tabular}{llcrccrccr}
\hline
\hline

{Hub.} &{SN} &{SNuB$^{b}$} &{$N_B^{c}$}
&{\,\,\,\,} &{SNuK$^{b}$} &{$N_K^{c}$} 
&{\,\,\,\,} &{SNuM$^{b}$} &{$N_M^{c}$} \\

\hline
E   &SN Ia&0.243$^{+0.048}_{-0.041}$(0.038)& 35.0&&0.041$^{+0.008}_{-0.007}$(0.006)& 33.0&&0.051$^{+0.010}_{-0.009}$(0.008)& 33.0\\
S0  &SN Ia&0.253$^{+0.038}_{-0.034}$(0.057)& 56.0&&0.044$^{+0.007}_{-0.006}$(0.010)& 54.0&&0.056$^{+0.009}_{-0.008}$(0.013)& 54.0\\
Sab &SN Ia&0.242$^{+0.042}_{-0.036}$(0.039)& 44.3&&0.059$^{+0.010}_{-0.009}$(0.010)& 43.3&&0.089$^{+0.016}_{-0.013}$(0.015)& 43.3\\
Sb  &SN Ia&0.184$^{+0.030}_{-0.026}$(0.036)& 50.2&&0.058$^{+0.009}_{-0.008}$(0.011)& 50.2&&0.095$^{+0.016}_{-0.014}$(0.018)& 49.2\\
Sbc &SN Ia&0.166$^{+0.032}_{-0.027}$(0.029)& 36.6&&0.060$^{+0.012}_{-0.010}$(0.010)& 34.6&&0.112$^{+0.023}_{-0.019}$(0.018)& 34.6\\
Sc  &SN Ia&0.175$^{+0.037}_{-0.031}$(0.027)& 32.0&&0.073$^{+0.015}_{-0.013}$(0.012)& 32.0&&0.138$^{+0.029}_{-0.024}$(0.022)& 32.0\\
Scd &SN Ia&0.161$^{+0.041}_{-0.033}$(0.043)& 23.0&&0.084$^{+0.022}_{-0.018}$(0.019)& 22.0&&0.174$^{+0.045}_{-0.037}$(0.038)& 22.0\\
Irr &SN Ia&0.000$^{+0.109}_{-0.000}$( $-$ )&  0.0&&0.000$^{+0.048}_{-0.000}$( $-$ )&  0.0&&0.000$^{+0.069}_{-0.000}$( $-$ )&  0.0\\
&&&&&&&&&\\ 
E   &SN Ibc&0.015$^{+0.034}_{-0.012}$(0.007)&  1.0&&0.003$^{+0.006}_{-0.002}$(0.001)&  1.0&&0.004$^{+0.008}_{-0.003}$(0.002)&  1.0\\
S0  &SN Ibc&0.036$^{+0.028}_{-0.017}$(0.010)&  4.0&&0.007$^{+0.005}_{-0.003}$(0.002)&  4.0&&0.009$^{+0.007}_{-0.004}$(0.003)&  4.0\\
Sab &SN Ibc&0.224$^{+0.065}_{-0.052}$(0.072)& 18.5&&0.056$^{+0.016}_{-0.013}$(0.018)& 18.5&&0.086$^{+0.025}_{-0.020}$(0.028)& 18.5\\
Sb  &SN Ibc&0.206$^{+0.056}_{-0.045}$(0.065)& 20.5&&0.070$^{+0.019}_{-0.015}$(0.023)& 21.5&&0.113$^{+0.031}_{-0.025}$(0.037)& 20.5\\
Sbc &SN Ibc&0.234$^{+0.062}_{-0.050}$(0.071)& 21.3&&0.092$^{+0.025}_{-0.020}$(0.029)& 21.3&&0.175$^{+0.047}_{-0.038}$(0.055)& 21.3\\
Sc  &SN Ibc&0.245$^{+0.053}_{-0.045}$(0.066)& 30.0&&0.106$^{+0.023}_{-0.019}$(0.028)& 30.0&&0.206$^{+0.045}_{-0.037}$(0.055)& 30.0\\
Scd &SN Ibc&0.178$^{+0.052}_{-0.041}$(0.035)& 18.7&&0.097$^{+0.029}_{-0.023}$(0.021)& 17.7&&0.194$^{+0.060}_{-0.047}$(0.042)& 16.7\\
Irr &SN Ibc&0.316$^{+0.249}_{-0.151}$(0.068)&  4.0&&0.073$^{+0.095}_{-0.047}$(0.034)&  2.0&&0.103$^{+0.136}_{-0.067}$(0.049)&  2.0\\
&&&&&&&&&\\
E   &SN II&0.000$^{+0.014}_{-0.000}$( $-$ )&  0.0&&0.000$^{+0.003}_{-0.000}$( $-$ )&  0.0&&0.000$^{+0.003}_{-0.000}$( $-$ )&  0.0\\
S0  &SN II&0.020$^{+0.015}_{-0.009}$(0.006)&  4.0&&0.004$^{+0.003}_{-0.002}$(0.001)&  4.0&&0.005$^{+0.004}_{-0.002}$(0.001)&  4.0\\
Sab &SN II&0.266$^{+0.047}_{-0.041}$(0.098)& 42.2&&0.066$^{+0.012}_{-0.010}$(0.024)& 42.2&&0.098$^{+0.018}_{-0.015}$(0.035)& 41.2\\
Sb  &SN II&0.282$^{+0.043}_{-0.037}$(0.106)& 56.3&&0.085$^{+0.013}_{-0.012}$(0.032)& 53.3&&0.144$^{+0.023}_{-0.020}$(0.055)& 53.3\\
Sbc &SN II&0.466$^{+0.058}_{-0.052}$(0.134)& 80.1&&0.183$^{+0.023}_{-0.020}$(0.052)& 81.1&&0.335$^{+0.042}_{-0.038}$(0.098)& 79.1\\
Sc  &SN II&0.649$^{+0.088}_{-0.078}(^{+0.364}_{-0.137})$& 69.0&&0.280$^{+0.038}_{-0.034}(^{+0.136}_{-0.058})$& 68.0&&0.547$^{+0.075}_{-0.066}(^{+0.245}_{-0.112})$& 68.0\\
Scd &SN II&0.695$^{+0.097}_{-0.086}(^{+0.386}_{-0.135})$& 65.3&&0.364$^{+0.055}_{-0.048}(^{+0.176}_{-0.075})$& 57.3&&0.767$^{+0.116}_{-0.102}(^{+0.342}_{-0.154})$& 56.3\\
Irr &SN II&0.431$^{+0.291}_{-0.186}$(0.074)&  5.0&&0.162$^{+0.128}_{-0.078}$(0.039)&  4.0&&0.230$^{+0.181}_{-0.110}$(0.054)&  4.0\\
\hline
\hline
\end{tabular}

$^{a}$Uncertainties are ordered as statistical and systematic (in
parentheses).

$^{b}$The rate for the average galaxy size.

$^{c}$The number of SNe used in the rate calculation. 

\end{minipage}
\end{table*}

\clearpage

\begin{table*}
\caption{SN rates in average galaxies of different $B - K$ colours.$^{a}$}
\begin{minipage}{180mm}
\begin{tabular}{llcrccrccr}
\hline
\hline

{$B - K$} &{SN} &{SNuB$^{b}$} &{$N_B^{c}$}
&{\,\,\,\,} &{SNuK$^{b}$} &{$N_K^{c}$} 
&{\,\,\,\,} &{SNuM$^{b}$} &{$N_M^{c}$} \\

\hline
$<$2.3   &SN Ia&0.148$^{+0.048}_{-0.037}$(0.038)& 15.6&&0.172$^{+0.056}_{-0.043}$(0.044)& 15.6&&0.601$^{+0.194}_{-0.150}$(0.153)& 15.6\\
2.3$-$2.8&SN Ia&0.135$^{+0.035}_{-0.028}$(0.026)& 22.6&&0.081$^{+0.021}_{-0.017}$(0.016)& 22.6&&0.206$^{+0.053}_{-0.043}$(0.040)& 22.6\\
2.8$-$3.1&SN Ia&0.196$^{+0.036}_{-0.031}$(0.041)& 39.8&&0.084$^{+0.016}_{-0.013}$(0.018)& 39.8&&0.181$^{+0.033}_{-0.029}$(0.038)& 39.8\\
3.1$-$3.4&SN Ia&0.206$^{+0.034}_{-0.029}$(0.030)& 50.2&&0.068$^{+0.011}_{-0.010}$(0.010)& 50.2&&0.126$^{+0.020}_{-0.018}$(0.018)& 50.2\\
3.4$-$3.7&SN Ia&0.218$^{+0.034}_{-0.030}$(0.034)& 53.6&&0.054$^{+0.009}_{-0.007}$(0.008)& 53.6&&0.087$^{+0.014}_{-0.012}$(0.014)& 53.6\\
3.7$-$4.0&SN Ia&0.213$^{+0.036}_{-0.031}$(0.041)& 46.0&&0.041$^{+0.007}_{-0.006}$(0.008)& 46.0&&0.057$^{+0.010}_{-0.008}$(0.011)& 46.0\\
$>$4.0   &SN Ia&0.269$^{+0.047}_{-0.041}$(0.045)& 43.0&&0.035$^{+0.006}_{-0.005}$(0.006)& 43.0&&0.038$^{+0.007}_{-0.006}$(0.006)& 43.0\\
&&&&&&&&&\\
$<$2.3   &SN Ibc&0.120$^{+0.058}_{-0.041}$(0.020)&  8.2&&0.141$^{+0.068}_{-0.048}$(0.024)&  8.2&&0.495$^{+0.239}_{-0.169}$(0.084)&  8.2\\
2.3$-$2.8&SN Ibc&0.253$^{+0.063}_{-0.051}$(0.069)& 24.1&&0.152$^{+0.038}_{-0.031}$(0.041)& 24.1&&0.388$^{+0.096}_{-0.078}$(0.105)& 24.1\\
2.8$-$3.1&SN Ibc&0.277$^{+0.062}_{-0.052}$(0.079)& 28.3&&0.119$^{+0.027}_{-0.022}$(0.034)& 28.3&&0.255$^{+0.057}_{-0.048}$(0.073)& 28.3\\
3.1$-$3.4&SN Ibc&0.222$^{+0.053}_{-0.044}$(0.051)& 25.5&&0.073$^{+0.018}_{-0.014}$(0.017)& 25.5&&0.135$^{+0.032}_{-0.027}$(0.031)& 25.5\\
3.4$-$3.7&SN Ibc&0.143$^{+0.045}_{-0.035}$(0.050)& 16.3&&0.036$^{+0.011}_{-0.009}$(0.013)& 16.3&&0.057$^{+0.018}_{-0.014}$(0.020)& 16.3\\
3.7$-$4.0&SN Ibc&0.078$^{+0.038}_{-0.027}$(0.025)&  8.0&&0.015$^{+0.007}_{-0.005}$(0.005)&  8.0&&0.021$^{+0.010}_{-0.007}$(0.007)&  8.0\\
$>$4.0   &SN Ibc&0.045$^{+0.043}_{-0.024}$(0.019)&  3.0&&0.006$^{+0.006}_{-0.003}$(0.002)&  3.0&&0.006$^{+0.006}_{-0.004}$(0.003)&  3.0\\
&&&&&&&&&\\
$<$2.3   &SN II&0.490$^{+0.093}_{-0.079}(^{+0.376}_{-0.070})$& 38.1&&0.572$^{+0.108}_{-0.092}(^{+0.422}_{-0.082})$& 38.1&&1.994$^{+0.378}_{-0.321}(^{+1.332}_{-0.286})$& 38.1\\
2.3$-$2.8&SN II&0.432$^{+0.068}_{-0.060}(^{+0.344}_{-0.113})$& 52.3&&0.258$^{+0.041}_{-0.036}(^{+0.199}_{-0.068})$& 52.3&&0.657$^{+0.104}_{-0.090}(^{+0.462}_{-0.172})$& 52.3\\
2.8$-$3.1&SN II&0.451$^{+0.063}_{-0.056}(^{+0.348}_{-0.076})$& 64.8&&0.194$^{+0.027}_{-0.024}(^{+0.144}_{-0.033})$& 64.8&&0.415$^{+0.058}_{-0.051}(^{+0.280}_{-0.070})$& 64.8\\
3.1$-$3.4&SN II&0.363$^{+0.051}_{-0.045}(^{+0.174}_{-0.092})$& 65.3&&0.119$^{+0.017}_{-0.015}(^{+0.051}_{-0.030})$& 65.3&&0.221$^{+0.031}_{-0.027}(^{+0.091}_{-0.056})$& 65.3\\
3.4$-$3.7&SN II&0.241$^{+0.041}_{-0.035}(^{+0.117}_{-0.063})$& 47.1&&0.060$^{+0.010}_{-0.009}(^{+0.026}_{-0.016})$& 47.1&&0.096$^{+0.016}_{-0.014}(^{+0.040}_{-0.025})$& 47.1\\
3.7$-$4.0&SN II&0.129$^{+0.032}_{-0.026}(^{+0.062}_{-0.033})$& 24.0&&0.025$^{+0.006}_{-0.005}(^{+0.011}_{-0.006})$& 24.0&&0.034$^{+0.009}_{-0.007}(^{+0.014}_{-0.009})$& 24.0\\
$>$4.0   &SN II&0.090$^{+0.034}_{-0.025}(^{+0.045}_{-0.025})$& 12.0&&0.012$^{+0.004}_{-0.003}(^{+0.005}_{-0.003})$& 12.0&&0.013$^{+0.005}_{-0.004}(^{+0.006}_{-0.004})$& 12.0\\
\hline
\hline
\end{tabular}

$^{a}$Uncertainties are ordered as statistical and systematic (in
parentheses).

$^{b}$The rate for the average galaxy size.

$^{c}$The number of SNe used in the rate calculation. 

\end{minipage}
\end{table*}

\clearpage

\begin{table*}
\caption{Two-component model fits to the SN~Ia rates.$^{a}$}
\begin{minipage}{75mm}
\begin{tabular}{ccccc}
\hline
\hline
{Mass$^{b}$} &{RSS} &{$a$}
&{$b$} &{$\chi^2(c)^{c}$} \\
\hline
4.0 & $-0.10$ &0.022(0.023)& 0.295(0.083) &10.06 \\
4.0 & $-0.25$ &0.046(0.019)& 0.248(0.071) & 7.13 \\
4.0 & $-0.40$ &0.072(0.016)& 0.188(0.058) & 4.31 \\
&&&&\\
0.4 & $-0.25$ &0.082(0.034)& 0.184(0.053) & 7.13 \\
4.0 & $-0.25$ &0.046(0.019)& 0.248(0.071) & 7.13 \\
40.0 & $-0.25$ &0.026(0.011)& 0.335(0.096)& 7.13 \\
\hline
\hline
\end{tabular}

\medskip

$^{a}${The rates for galaxies of different $B - K$ colours 
are used in the analysis.}

$^{b}${Galaxy mass, in units of $10^{10}\,{\rm L}_\odot$.}

$^{c}${$\chi^2$/DOF for a constant fit to the SN~Ia rates.}

\end{minipage}
\end{table*}

\begin{table*}
\caption{More two-component model fits to the SN~Ia rates.$^{a}$}
\begin{minipage}{180mm}
\begin{tabular}{cccrccrc}
\hline
\hline
{Src} &{Rate}
&{$a_1$(with RSS)} &{$b_1$(with RSS)} &{$\chi^2(c)_1^{b}$} 
&{$a_2$(no RSS)} &{$b_2$(no RSS)} &{$\chi^2(c)_2^{b}$} \\
\hline
H-type&SNuB&0.299(0.031)& $-0.117$(0.044) &1.465&0.252(0.025)& $-0.097$(0.041) &1.277\\
H-type&SNuK&0.065(0.007)& 0.036(0.032) &0.353&0.043(0.005)& 0.081(0.030) &1.787\\
H-type&SNuM&0.116(0.012)& 0.051(0.032) &0.770&0.058(0.008)& 0.115(0.030) &5.184\\
$B - K$&SNuB&0.296(0.039)& $-0.121$(0.063) &1.641&0.264(0.033)& $-0.140$(0.062) &1.339\\
$B - K$&SNuK&0.044(0.008)& 0.153(0.054) &2.477&0.031(0.009)& 0.179(0.058) &4.983\\
$B - K$&SNuM&0.046(0.019)& 0.248(0.071) &7.128&0.036(0.022)& 0.220(0.067) &11.960\\
\hline
\hline
\end{tabular}

\medskip

$^{a}${The correlation is fit as ${\rm rate(SN~Ia)} = a + b \times {\rm rate(SN~CC)}$.}

$^{b}${$\chi^2$/DOF for a constant fit to the SN~Ia rates.}

\end{minipage}
\end{table*}

\begin{table*}
\caption{Volumetric rate.}
\begin{tabular}{lccc}
\hline
\hline
{Rate} & {SN~Ia} 
&{SN~Ibc} & {SN~II}\\
\hline
Early(fiducial; SNuK)&0.064$^{+0.008}_{-0.007}(^{+0.013}_{-0.013})$ & 0.008$^{+0.006}_{-0.004}(^{+0.002}_{-0.002})$ & 0.004$^{+0.003}_{-0.002}(^{+0.001}_{-0.001})$\\
Late(fiducial; SNuK)&0.074$^{+0.006}_{-0.006}(^{+0.012}_{-0.012})$ & 0.096$^{+0.010}_{-0.009}(^{+0.018}_{-0.018})$ & 0.172$^{+0.011}_{-0.011}(^{+0.045}_{-0.036})$\\
Early(LF-average; SNuK)&0.048$^{+0.006}_{-0.005}(^{+0.010}_{-0.010})$ & 0.006$^{+0.004}_{-0.003}(^{+0.002}_{-0.002})$ & 0.003$^{+0.002}_{-0.001}(^{+0.001}_{-0.001})$\\
Late(LF-average; SNuK)&0.065$^{+0.006}_{-0.005}(^{+0.010}_{-0.010})$ & 0.083$^{+0.009}_{-0.008}(^{+0.016}_{-0.016})$ & 0.149$^{+0.010}_{-0.009}(^{+0.039}_{-0.031})$\\
\hline
Vol-rate ($10^{-4}$ SN Mpc$^{-3}$ yr$^{-1}$) &0.301$^{+0.038}_{-0.037}(^{+0.049}_{-0.049})$ & 0.258$^{+0.044}_{-0.042}(^{+0.058}_{-0.058})$ & 0.447$^{+0.068}_{-0.068}(^{+0.131}_{-0.111})$\\
\hline
\hline
\end{tabular}
\end{table*}

\begin{table*}
\caption{Milky Way rate (per century). }
\begin{tabular}{lccccccl}
\hline
\hline
{Normalisation} &{Size$^{a}$} &{SN~Ia} 
&{SN~Ibc} & {SN~II} & {CC~SNe}
&{Total SNe} &{Comments} \\
\hline
$L_B$ &2.0  &0.40    &0.55    &1.11 &1.66  &2.06& Galaxy size from van der Kruit (1987) \\
$L_B$ &2.3  &0.44    &0.61    &1.23 &1.84  &2.28& Galaxy size from van den Bergh (1988) \\
$L_B$ &4.3  &0.71    &0.96    &1.95 &2.91  &3.62& M31 size \\
$L_K$ &6.3  &0.47    &0.70    &1.44 &2.14  &2.61& M31 size \\
Mass  &2.3  &0.43    &0.63    &1.27 &1.90  &2.33& M31 size \\
$L_B$ &3.5  &0.61    &0.82    &1.68 &2.50  &3.11& Average Sbc galaxy size\\
$L_K$ &9.7  &0.62    &0.89    &1.83 &2.72  &3.34& Average Sbc galaxy size\\
Mass  &5.2  &0.65    &0.91    &1.84 &2.75  &3.40& Average Sbc galaxy size\\
\hline
 & & 0.54$\pm$0.12 & 0.76$\pm$0.16 &1.54$\pm$0.32 &2.30$\pm$0.48 & 2.84$\pm$0.60 &\\
\hline
\hline
\end{tabular}

\medskip
\noindent
$^{a}${For $L_{B}$ and $L_{K}$, the units are $10^{10}\,{\rm
  L}_\odot$; for mass, the units are $10^{10}\,{\rm M}_\odot$.}

\end{table*}

\vfill
\clearpage

\appendix

\section{Discovery of the Rate-Size Relation} 

As in \S 2, for the discussion in this Appendix, we use the
``full-optimal" SN sample with 726 objects that occurred in the
``optimal" galaxy sample to compute the SN rates.

We have performed various tests to check the robustness of our
rate-calculation pipeline. One such test yielded an unexpected result,
as shown in Figure A1. Here the SNe are combined in different
Hubble-type bins, and the rates in SNuM are calculated for the
galaxies in different distance bins (with a bin size of 30~Mpc). Only
the statistical uncertainties are shown (see \S 3.4 for more
discussion of how the errors are calculated).  In principle, the rate
for each SN type should remain constant (i.e., no evolution) in the
small redshift range we considered. However, Figure A1 shows a strong
declining trend for the SNe~Ia in E--S0 galaxies, as well as for
SNe~Ibc and SNe~II in Sb--Irr galaxies: the rate in the 0--30 Mpc bin
is a factor of 2--3 higher than that in the 120--150 Mpc bin. The
SN~Ia rates in Sb--Irr galaxies are also consistent with a declining
trend, though with a lower significance.

The search for the possible causes of this trend fundamentally changed
our rate calculations. First, we suspected that the trend could be
caused by a bottom-heavy LF for the SNe --- the LF has many faint
objects that can only be detected in the very nearby galaxies. One
main reason we chose to construct a complete SN sample with the nearby
SNe in Paper II is to quantify the fraction of subluminous SNe after
correction for their incompleteness in our search. As it turns out,
the LF does not solve this problem; the rates shown in Figure A1
have made use of the LFs derived from Paper II.

Another possible cause is the missing fraction of SNe in the nuclei of
galaxies. As we go to larger distances, the average {\it angular} size
of the galaxies becomes smaller, and the central area (with a fixed
radius of a few pixels) which we avoid in the SN search would become a
larger fraction of the total galaxy. This could potentially lead to a
larger missing fraction of SNe in more distant galaxies, resulting in
a lower apparent SN rate. To check this, we divide the galaxies into
bins with different angular sizes and calculate their rates. The
results are shown in Figure A2. Only SNe in Sb--Irr galaxies are
considered, since the early-type galaxies with small angular sizes are
not considered in the final rate calculations.  There is no strong
correlation between the SN rates and the angular galaxy sizes: the
rate is consistent with a constant for all of the three SN types. As
described in \S 4.2.3 of Paper I, the missing fraction of SNe from the
radial distribution and Monte Carlo simulation studies is $\sim 10$\%
of all SNe, not enough to explain the big difference (a factor of
2--3) shown in Figure A1. The lack of a correlation between the SN
rates and the angular sizes of the galaxies is probably due to two
competing factors: while the SN rate can be depressed in galaxies
having smaller angular sizes due to a larger fraction of the missed
SNe in the nuclei, it can also be enhanced if these galaxies have a
smaller average {\it size} (luminosity or mass) as discussed below,
although we note that a smaller angular size does not necessarily
translate into a smaller luminosity or mass.

We also considered whether the trend could be caused by a change of
Hubble-type distribution of the LOSS sample galaxies with distance, as
described in \S 4.1.4 of Paper I. Since the SN rates for galaxies of
different Hubble types are different, a declining trend could be
observed if the more nearby distance bin mainly consists of galaxies
with higher SN rates (i.e., late-type spirals), while the more distant
bin is dominated by galaxies with lower rates (i.e., early-type
spirals). We have calculated the average SN rate for each distance bin
after weighting the SN rate for each galaxy by its total normalised
control time and its average SN rate for its Hubble type (as discussed
in \S 4.1), and find that the effect of changing Hubble-type
distribution over distance is $\simlt 5$\% for SNe~Ia, $\simlt 10$\%
for SNe~Ibc, and $\simlt 20$\% for SNe~II.  Again, this is not enough
to explain the strong rate-distance trend shown in Figure A1.

As discussed in \S 4.1.4 and Figure 5 of Paper I, there are important
changes in the galaxy properties over the 0 to 200 Mpc distance range
of the LOSS sample galaxies due to selection biases. In particular,
the average luminosity of the galaxies increases monotonically with
increasing distance due to a strong Malmquist bias in the current
astronomical databases. If there is a correlation between the galaxy
size (luminosity or mass) and the SN rates, it may explain the observed
declining trend of the SN rates with increasing distance.

At first thought, this may seem unlikely, since the SN rates have
already been {\it linearly} normalised by the galaxy size, as
indicated by Equations (A3) and (A4) of Paper I. However, as shown in
\S 2, a strong correlation between the SN rates and the galaxy size is
found. Consequently, our SN rates are calculated for a fiducial galaxy
size, and the rates for other galaxy sizes are derived using a
rate-size relation, as discussed in more detail in \S 2.

The effect of adopting the rate-size relation in the rate calculations
for the galaxies in different distance bins is illustrated in Figure
A3. Here all of the rates are converted to the fiducial galaxy size. The
rates are consistent with being a constant in different distance bins
for each SN type, suggesting that the rate-size relation is the main
reason for the declining trend in the SN rates over distance.


An alternative way to demonstrate our solution to the rate-distance
trend is shown by the dashed lines in Figure A1. Here, we adopt the
final SNuM for the fiducial galaxy size in each Hubble-type bin (\S
3.5), and calculate the average SNuM for each distance bin. To
calculate this average, the rate for each galaxy is corrected by its
Hubble type and galaxy size, and weighted by its total control
time. In other words, the dashed lines are what we would expect for
the change of the average SNuM for the LOSS galaxies over distance,
after considering the change in the galaxy Hubble-type and size
distribution over distance, and the monitoring history of the
galaxies.  The dashed lines provide an excellent fit to the observed
rate-distance trend.

\clearpage

\setcounter{figure}{0}

\begin{figure*}
\includegraphics[scale=0.7,angle=270,trim=0 0 0 0]{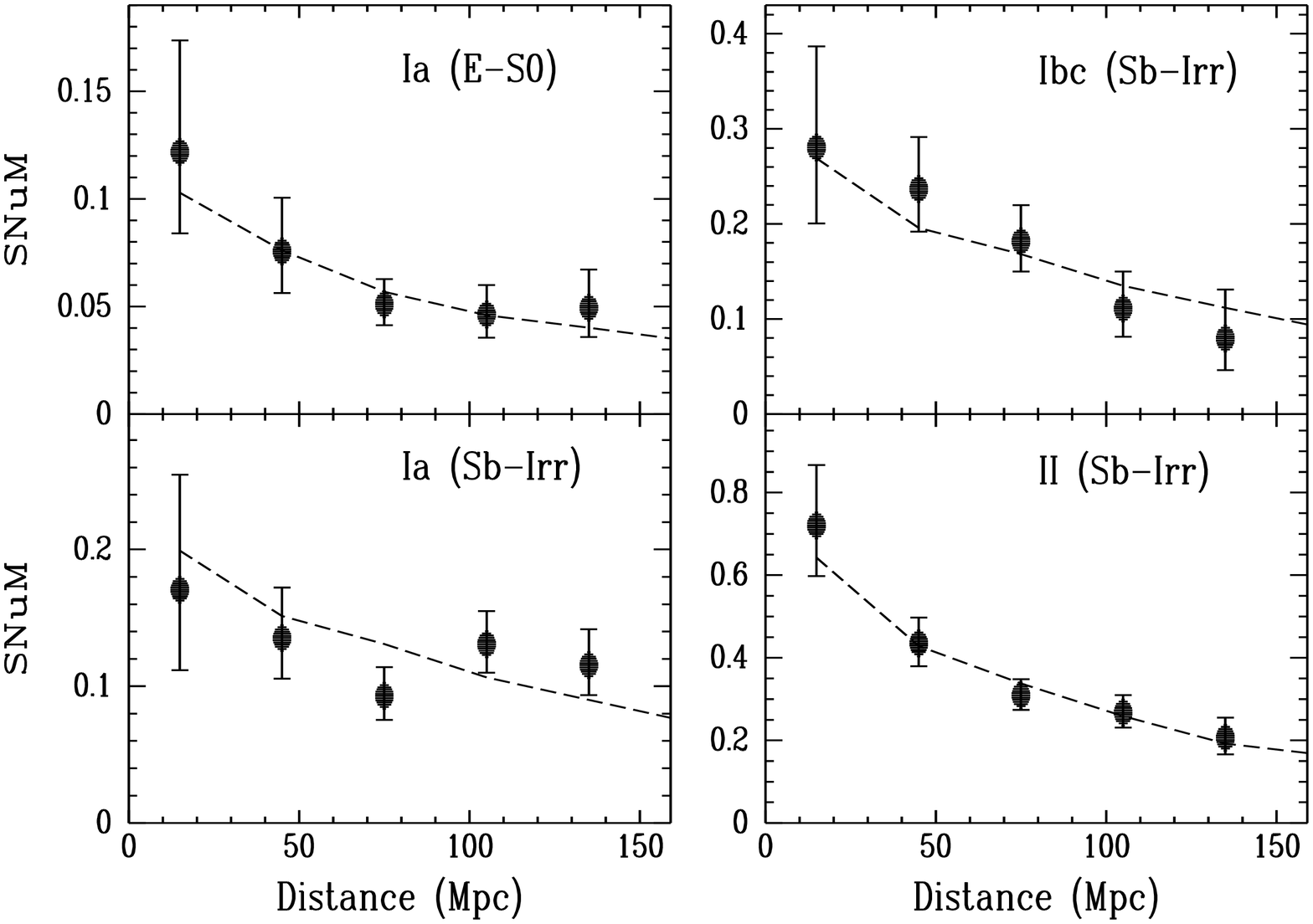}
\caption[] { The SN rates in various Hubble types in different
  distance bins.  There is a significant declining trend for the
  rates. The dashed lines are fits to the trend after considering the
  rate-size relation and the control times of the galaxies. See text
  in the Appendix for more details.  }
\label{3}
\end{figure*}

\begin{figure*}
\includegraphics[scale=0.8,angle=270,trim=0 40 0 0]{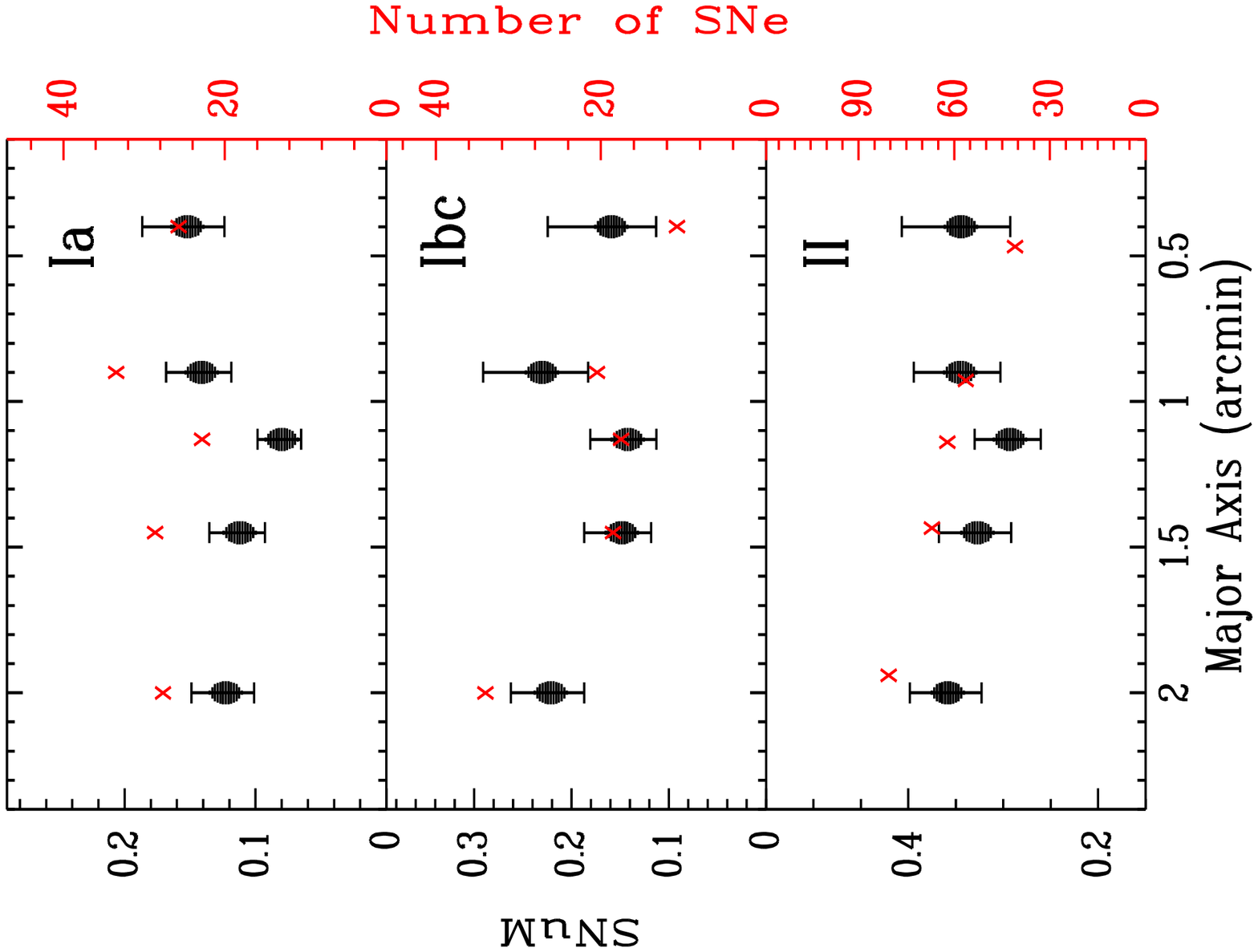}
\caption[] { The SN rates (solid dots) in galaxies of different
  angular sizes.  The number of SNe used to calculate each rate is
  also shown (crosses) with the scale to the right of the plot. No
  obvious trend is found.  }
\label{3}
\end{figure*}

\clearpage

\begin{figure*}
\includegraphics[scale=0.7,angle=270,trim=0  0 0 0]{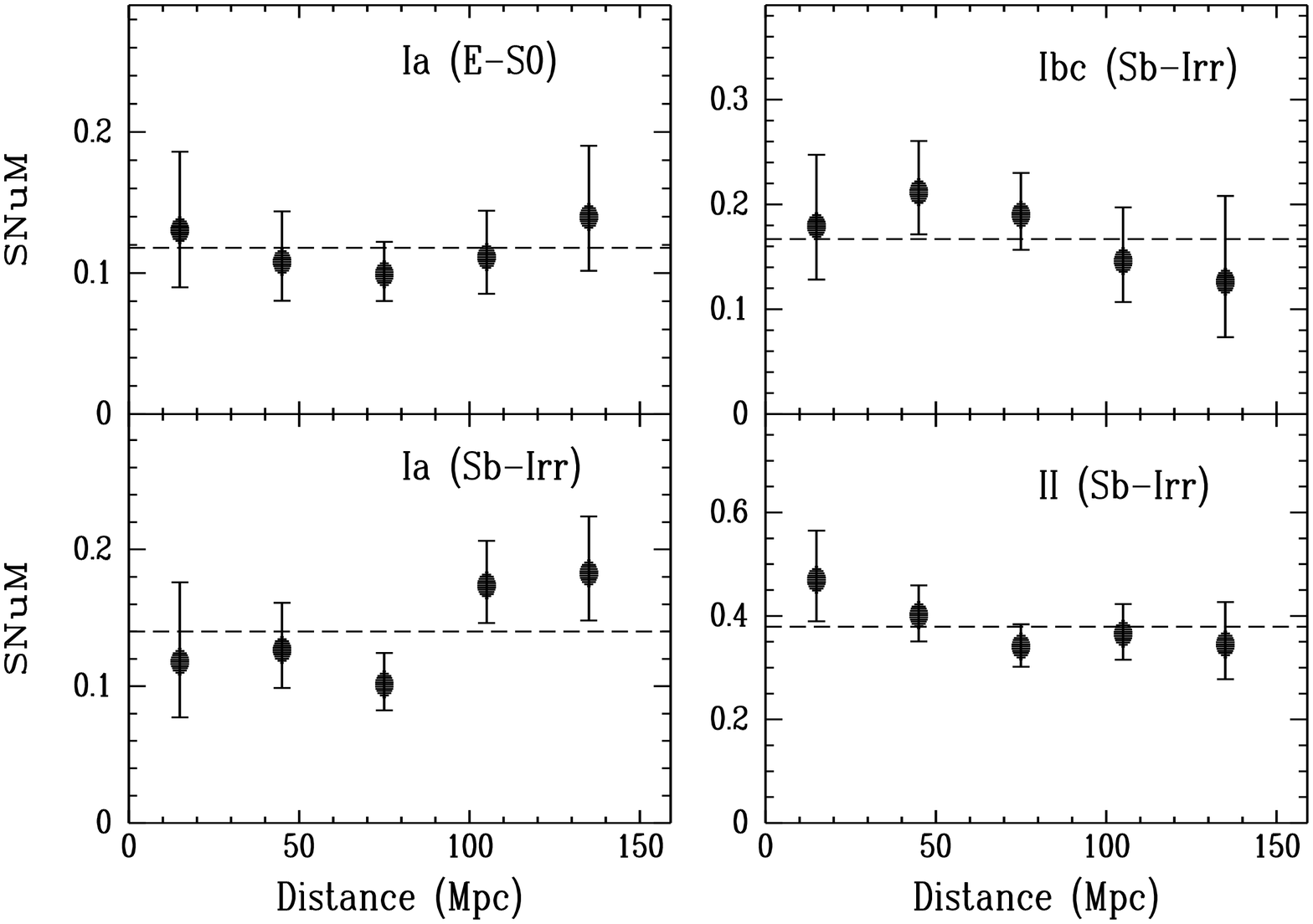}
\caption[] { The SN rates for the same mass ($M = 4.0 \times
  10^{10}\,{\rm M}_\odot$) in different distance bins. The rates are
  consistent with being a constant in different distance bins,
  suggesting that the rate-size relation, combined with the increasing
  Malmquist bias of the galaxy sample, is responsible for the declining
  trend seen in Figure A1.  }
\label{7}
\end{figure*}

\clearpage

\section{Supernova Rate as a Function of Host-Galaxy Properties}

As discussed in \S 4.2, the existence of the rate-size relation
indicates that numerically, the rates cannot be adequately described
by a single parameter using either galaxy Hubble type or $B - K$
colour, and the galaxy size ($L_B$, $L_K$, mass) is used as a second
parameter to quantify the rates (in the form of the rate-size
relation).  We have considered other combinations of parameters to
describe the rates. In particular, here we discuss how the rates can
be parameterised as a function of both galaxy Hubble type and $B - K$
colour. It is generally accepted that there is a correlation between
galaxy size and colour, with smaller galaxies having bluer colours.
Consequently, the empirical rate-size relation can be converted to a
rate-colour relation.

We first investigate the galaxy size and colour correlation using our
own galaxy sample. Figure B1 shows the results for the ``optimal"
sample. A strong correlation between the galaxy $L_K$ (or mass) and $B
- K$ colour is confirmed. The galaxy $L_B$ value shows a relatively
weak (but significant) correlation with $B - K$ colour as well. We
divide the galaxies into ten $B-K$ colour bins and calculate their
average galaxy sizes and $B - K$ colours; the results are plotted
as the dashed lines.  A linear regression fit yields the following
correlations:

\begin{equation}
{\rm log} (L_B) = {\rm constant} + (1.20 \pm 0.27) {\rm log} (B - K), 
\end{equation}

\begin{equation}
{\rm log} (L_K) = {\rm constant} + (3.99 \pm 0.17) {\rm log} (B - K), 
~{\rm and}
\end{equation}

\begin{equation}
{\rm log (Mass)} = {\rm constant} + (5.43 \pm 0.11) {\rm log} (B - K).
\end{equation}
\noindent
In principle, the rate-size relation given by Eqs. (1)--(3) in \S 2.1
can then be written as (using SNuB as an example)

\begin{equation}
\begin{array}{ll}
{\rm log} {\rm SNuB}(L_B) & = {\rm constant} + {\rm RSS}_B \times 
{\rm log} (L_B) \\
              & = {\rm constant}^\prime + {\rm RSS}_B \times 
(1.20 \pm 0.27) {\rm log} (B - K). 
\end{array}
\end{equation} 
\noindent
In other words, the rate-size relation can be converted to a
rate-colour relation,

\begin{equation}
{\rm SNuB}(B - K) = {\rm SNuB}[(B - K)_0] \Bigl[\frac{B - K} 
{(B - K)_0}\Bigr]^{{\rm RCS}_B}, 
\end{equation}
\noindent
where ``RCS" stands for rate-colour slope, and RCS$_B$ = RSS$_B \times
(1.20\pm0.27)$. For completeness, the equations for SNuK and SNuM are
as follows:

\begin{equation}
{\rm SNuK} (B - K) = {\rm SNuK}[(B - K)_0] \Bigl[\frac{B - K} 
{(B - K)_0}\Bigr]^{{\rm RCS}_K}, ~{\rm and}
\end{equation}

\begin{equation}
{\rm SNuM} (B - K) = {\rm SNuM}[(B - K)_0] \Bigl[\frac{B - K} 
{(B - K)_0}\Bigr]^{{\rm RCS}_M}, 
\end{equation}
\noindent
where RCS$_K$ = RSS$_K \times (3.99\pm0.17)$, and RCS$_M$ =
RSS$_M \times (5.43\pm0.11)$. These RCS values are listed
in the last two columns of Table B1 when the RSS$_B$, RSS$_K$, and
RSS$_M$ values in Table 2 are adopted.

The above analysis applies the galaxy size-colour correlation to the
rate-size relation to derive the rate-colour relation. Numerically,
the rate-colour relation can be directly derived from the data, as
shown in Figure B2. The left and right panels illustrate the results
for SNe~Ia and SNe~II, respectively, while the top to bottom panels
show the results for the different normalisations. Consider the SNuB
rate of SNe~Ia (the top-left panel) as an example. For each Hubble
type, the galaxies are sorted according to their $B - K$ colours and
then divided into 5 bins from the bluest to the reddest, and the rates
are calculated for each bin. The rates in Sb galaxies are used as the
anchor points, and those in the other Hubble types are scaled by a
multiplicative constant, so the ensemble of data can be fit by the
dashed line. For the SN~II rates in the right panels, the rates in Sbc
galaxies are used as the anchor points.

The dashed lines in Figure B2 provide good fits to the data, with
$\chi^2$/DOF $<$ 1.0 for all of the cases. This indicates that a
power-law correlation between the rates and $B - K$ colours --- that
is, the rate-colour relation as expressed by Eqs. (B5)--(B7) --- is a
reasonable choice. The power-law indexes (i.e., the RCS values derived
directly from the data) are reported in the second and third columns
of Table B1.

A comparison between the two sets of RCS values for the rate-colour
relation indicates that for the SNuK and SNuM rates, the RCSs are
consistent with each other to within the uncertainties. For the SNuB
rates, however, the rate-colour relation derived directly from the
data shows a trend that is contrary to the expectation from the galaxy
size-colour correlation: galaxies with bluer colours have smaller SNuB
rates. We emphasise that this trend is only marginal: $\sim 2.0\sigma$
for the SN~Ia rates, and $\sim 1.4\sigma$ for SN~II rates. As the RCS
values derived from the galaxy size-colour correlation also have
rather large uncertainties, the two sets of RCSs are only different at
the $\sim 2\sigma$ level. For our final rate-colour relations, we
adopt the RCS values derived directly from the data (Columns 2 and 3
in Table B1).

The existence of the rate-colour relation implies that the galaxy
Hubble-type rates should be evaluated at a fiducial $B - K$ colour, so
the rates at any given colour can be calculated using
Eqs. (B5)--(B7). We adopt $(B - K)_0 = 3.0$ mag in our analysis, a
value that is close to the average colour of the ``optimal" galaxy
sample.

The rates for the fiducial $B - K$ colour can be calculated by scaling
the control time of each galaxy by a factor of [$(B - K)/(B -
  K)_0]^{\rm RCS}$, where RCS is listed in Table B1 and is different
for different normalisations\footnote{When performing this
  calculation, we need to reject $\sim 50$ galaxies with negative $B -
  K$ colours (i.e., very blue galaxies) because otherwise it is
  numerically impossible to calculate the power-law value.}.  This is
similar to the treatment of the rate-size relation as described in \S
2.3. These rates are listed in Table B2.

An interesting question is whether the rate for a galaxy with the
fiducial $B - K$ colour is consistent with the rate for a galaxy with
the fiducial size. In Figure B3, we compare the rates for the fiducial
$B - K$ colour (open circles, from Table B2) to those for the fiducial
sizes (solid circles, from Table 4).  As the values for the fiducial
$B - K$ colour and size (in $L_B$, $L_K$, and mass) are chosen quite
arbitrarily, the rates can be scaled by a multiplicative constant
(which simply means adopting a different fiducial
value). Nevertheless, because the fiducial values are chosen to be
close to the average of the KAIT sample galaxies, the two sets of
rates are in good agreement for SNuK and SNuM. The SNuB rates for the
fiducial $B - K$ colours need to be scaled by a factor of 0.91 to
achieve better agreement with those for the fiducial sizes.

The rate-colour relation discussed in this section offers an
alternative to the rate-size relation employed in our rate
calculations. There are advantages and disadvantages to adopting
either relation: while it may be easier to physically understand a
correlation between the rates and colours (which are often tied to the
star-formation rate), the two relations likely share the same physical
origin because of the tight correlation between galaxy sizes and
colours discussed here.  Numerically, both relations are empirically
derived from the data, so either relation can be used to parameterise
the rates.  We do not adopt the rate-colour relation in our rate
calculations for the following reasons: (a) the relation for the SNuB
rates is contrary to expectations, though with a low significance;
(b) numerically, it is impossible to calculate rates for galaxies with
$B - K < 0$ mag for SNuK and SNuM; and (c) it is difficult to publish
the rates in galaxies of different $B - K$ colours using the
rate-colour relation, which requires the rates to be described by a
single $B - K$ colour parameter (and we have demonstrated in \S 2.2
that this is not the case).

\setcounter{table}{0}

\begin{table*}
\caption{Correction factors using $B - K$ colours for the Hubble-type rates.}
\begin{minipage}{140mm}
\begin{tabular}{cccccc}
\hline
\hline
{Rate} & {RCS(Ia)} &{RCS(Ibc, II)}
&{Fiducial $B - K$} &{RCS(Ia)(exp)$^{a}$} &
{RCS(Ibc, II)(exp)$^{a}$} \\
\hline
SNuB & $  0.73\pm0.36$ & $  0.38\pm0.28$ & 3.0 & $-0.28\pm0.25$ & $-0.32\pm0.14$\\
SNuK & $ -1.46\pm0.36$ & $ -1.55\pm0.26$ & 3.0 & $-1.40\pm0.40$ & $-1.80\pm0.41$\\
SNuM & $ -2.96\pm0.36$ & $ -2.77\pm0.28$ & 3.0 & $-2.72\pm0.55$ & $-2.99\pm0.55$\\
\hline
\hline
\end{tabular}

\medskip

$^{a}${These RCS values are derived from the RSS values in
  the rate-size relation (as reported in Table 2) and the correlation
  between size and $B - K$ colour (Eqs. B1--B3).}

\end{minipage}  
\end{table*}

\begin{table*}
\vbox to330mm{\vfil Landscape Table B2 to go here.
\caption{SN rates in galaxies of different Hubble type.}
\vfil}
\label{landtable}
\end{table*}

\setcounter{figure}{0}

\begin{figure*}
\includegraphics[scale=0.7,angle=270,trim=0 0 0 0]{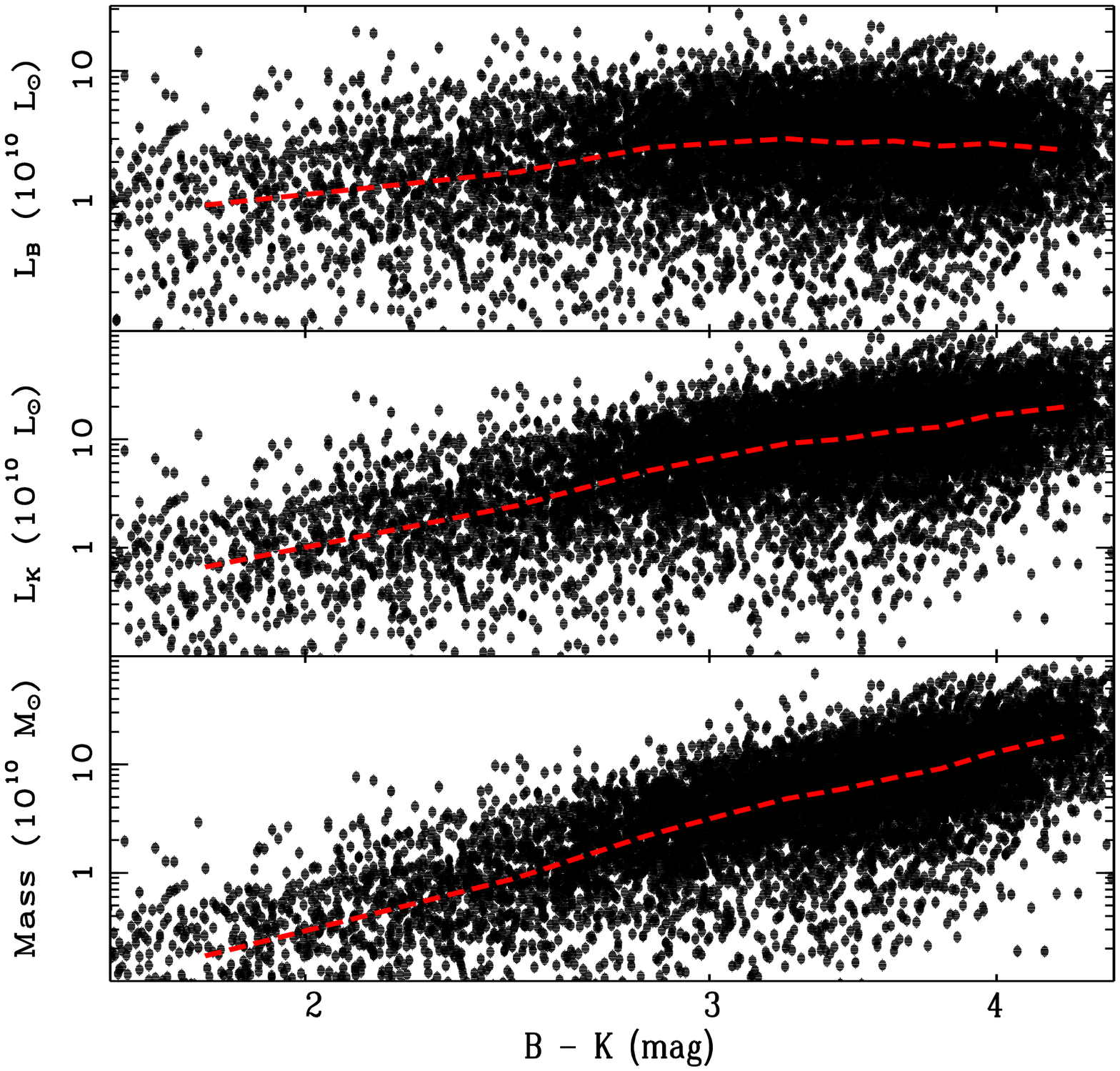}
\caption[] { The correlation between the galaxy size ($L_B$, top
  panel; $L_K$, middle panel; mass, bottom panel) and $B - K$ colour. }
\label{12}
\end{figure*}
\clearpage

\begin{figure*}
\includegraphics[scale=0.7,angle=0,trim=0 0 0 0]{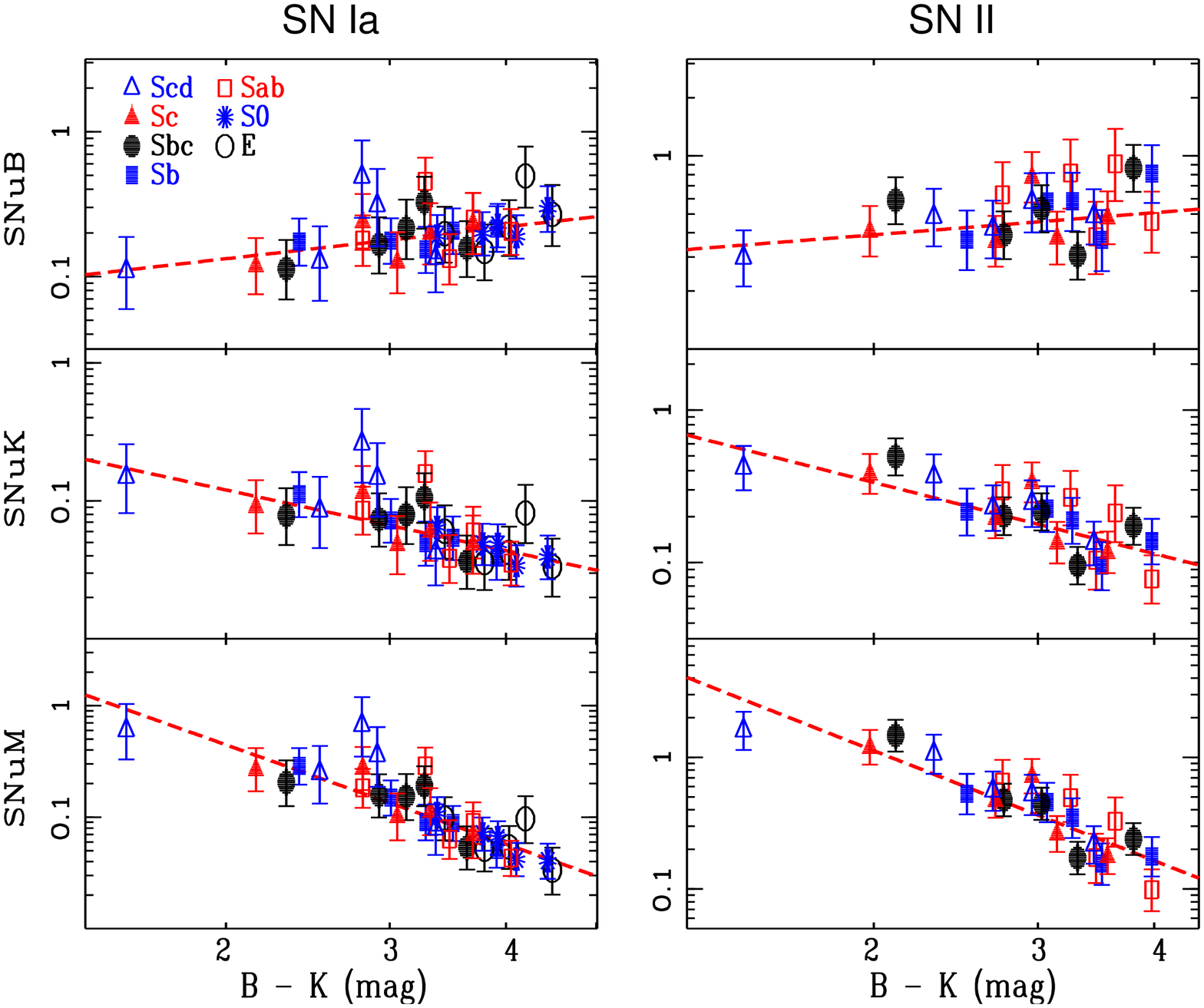}
\caption[] { The Hubble-type SN rates in galaxies of different $B - K$
  colours.  For each Hubble type, the galaxies are divided into five
  $B - K$ colour bins, and their rates are derived. The left panels
  show the results for SNe~Ia, where the rates for the Sb galaxies are
  used as the anchor points and the rates in other Hubble types are
  scaled by a multiplicative constant, so the ensemble of data points
  can be fit by the dashed lines. The right panels show the results
  for SNe~II, where the rates for the Sbc galaxies are used as the
  anchor points.  }
\label{12}
\end{figure*}
\clearpage

\begin{figure*}
\includegraphics[scale=0.7,angle=0,trim=0 0 0 0]{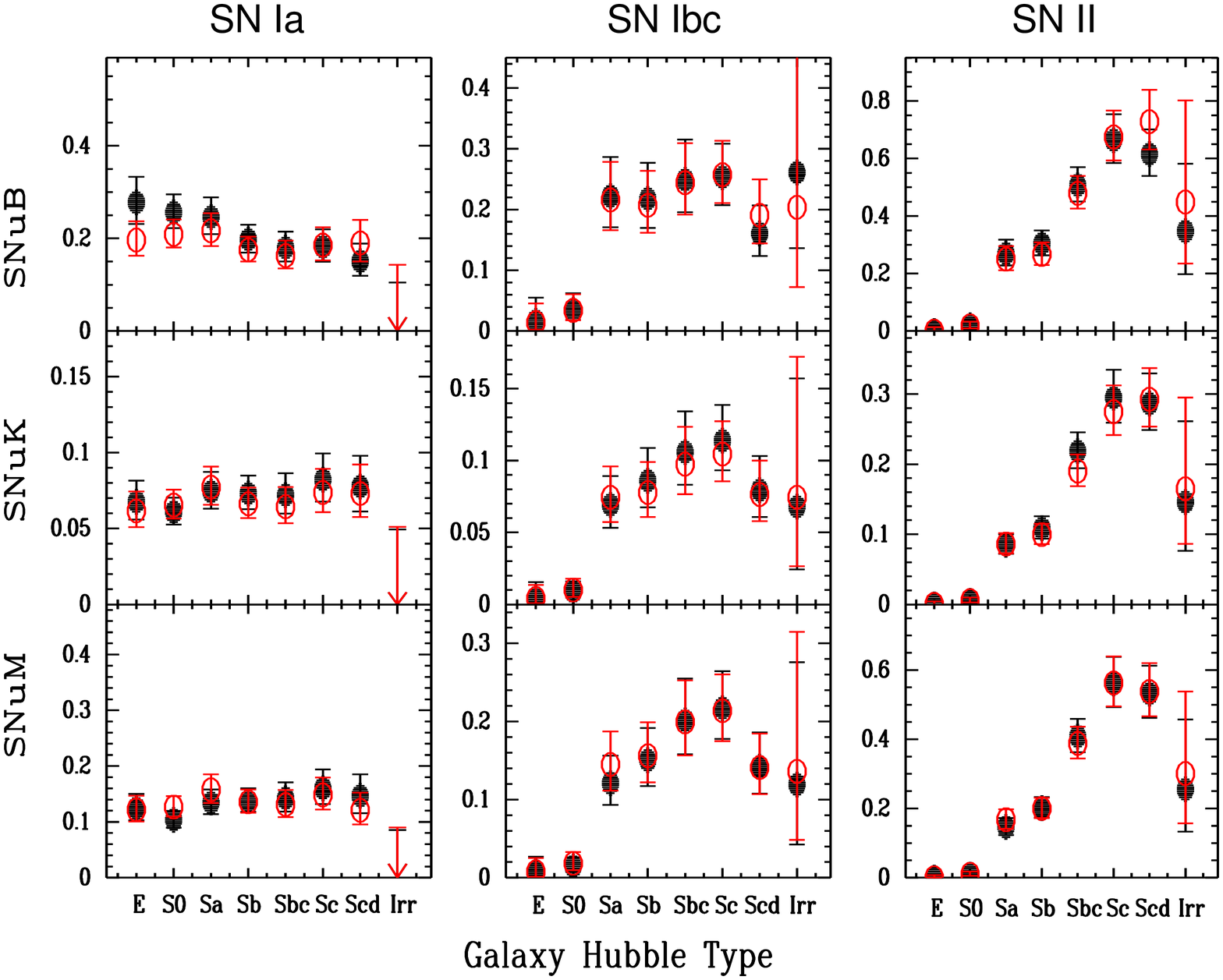}
\caption[] { The SN rates for galaxies of different Hubble types.  The
  open circles indicate rates for a galaxy with the fiducial $B - K$
  colour, while the solid circles give rates for a galaxy with the
  fiducial size. The SNuB rates for the fiducial $B - K$ colour are
  scaled by a factor of 0.91. See text in the Appendix (\S B) for
  more details. }

\label{12}
\end{figure*}
\clearpage

\end{document}